\documentclass[twocolumn]{aastex63}

\usepackage{amsmath,amssymb}
\usepackage{bm}
\usepackage{mathtools}
\usepackage{algorithmic}
\usepackage{algorithm}
\usepackage{textcomp}
\usepackage[nooneline]{subfigure}
\subfiguretopcaptrue
\newcommand{\argmin}{\mathop{\rm argmin}\limits}
\newcommand{\argmax}{\mathop{\rm argmax}\limits}
\newcommand{\minimize}{\mathop{\rm minimize\ }\limits}
\newcommand{\subjectto}{\mathop{\rm \  subject\, to\ }\limits}

\DeclareMathOperator{\prox}{prox}
\DeclareMathOperator{\dom}{dom}
\DeclareMathOperator{\diag}{diag}
\DeclareMathOperator{\MRSA}{MRSA}

\usepackage{amsthm}
\newtheorem{thm}{Theorem}
\newtheorem{dfn}[thm]{Definition}
\newtheorem{prp}[thm]{Proposition}
\newtheorem{cor}[thm]{Corollary}

\usepackage{color}
\usepackage{ulem}
\newcommand{\add}[1]{\textcolor{black}{#1}}

\usepackage{comment}
\definecolor{red}{rgb}{0.8,0.0,0.0}
\definecolor{blue}{rgb}{0.0,0.0,0.8}
\definecolor{green}{rgb}{0.0,0.5,0.0}

\received{February 15, 2022}
\revised{March 24, 2022}
\accepted{April 4, 2022}
\submitjournal{ApJ}

\shorttitle{Global Composition Mapping Using Sparse Modeling}
\shortauthors{Kuwata et al.}

\begin{document}

%%%%%%%%%%%%%%%%%%%%%%%%%%%%%%
\title{Global Mapping of Surface Composition on an Exo-Earth Using Sparse Modeling}

\correspondingauthor{Atsuki Kuwata}
\email{a.kuwata@astron.s.u-tokyo.ac.jp}

\author[0000-0002-3244-7136]{Atsuki Kuwata}
\affiliation{Department of Astronomy, The University of Tokyo, 7-3-1, Hongo, Bunkyo-ku, Tokyo 113-0033, Japan}
%\nocollaboration{1}

\author[0000-0003-3309-9134]{Hajime Kawahara}
\affiliation{Department of Earth and Planetary Science, The University of Tokyo, 7-3-1, Hongo, Bunkyo-ku, Tokyo 113-0033, Japan}
\affiliation{Research Center for the Early Universe, School of Science, The University of Tokyo, 7-3-1, Hongo, Bunkyo-ku, Tokyo 113-0033, Japan}

\author[0000-0001-8877-4497]{Masataka Aizawa}
\affiliation{Tsung-Dao Lee Institute, Shanghai Jiao Tong University, 800 Dongchuan Road, Shanghai 200240, China}

\author[0000-0001-6181-3142]{Takayuki Kotani}
\affiliation{Astrobiology Center, 2-21-1 Osawa, Mitaka, Tokyo 181-8588, Japan}
\affiliation{National Astronomical Observatory of Japan, 2-21-1 Osawa, Mitaka, Tokyo 181-8588, Japan}
\affiliation{Department of Astronomy, School of Science, The Graduate University for Advanced Studies, SOKENDAI, 2-21-1 Osawa, Mitaka, Tokyo 181-8588, Japan}

\author[0000-0002-6510-0681]{Motohide Tamura}
\affiliation{Department of Astronomy, The University of Tokyo, 7-3-1, Hongo, Bunkyo-ku, Tokyo 113-0033, Japan}
\affiliation{Astrobiology Center, 2-21-1 Osawa, Mitaka, Tokyo 181-8588, Japan}
\affiliation{National Astronomical Observatory of Japan, 2-21-1 Osawa, Mitaka, Tokyo 181-8588, Japan}

%%%%%%%%%%%%%%%%%%%%%%%%%%%%%%

\begin{abstract}
The time series of light reflected from exoplanets by future direct imaging can provide spatial information with respect to the planetary surface. We apply sparse modeling to the retrieval method that disentangles the spatial and spectral information from multi-band reflected light curves termed as spin-orbit unmixing. We use the $\ell_1$-norm and the Total Squared Variation norm as regularization terms for the surface distribution. Applying our technique to a toy model of cloudless Earth, we show that our method can infer sparse and continuous surface distributions and also unmixed spectra without prior knowledge of the planet surface. We also apply the technique to the real Earth data as observed by DSCOVR/EPIC. We determined the representative components that can be interpreted as cloud and ocean. Additionally, we found two components that resembled the distribution of land. One of the components captures the Sahara Desert, and the other roughly corresponds to vegetation\add{ although their spectra are still contaminated by clouds.} Sparse modeling significantly improves the geographic retrieval, in particular, of cloud and leads to higher resolutions for other components when compared with spin-orbit unmixing using Tikhonov regularization.
\end{abstract}

%\maketitle

\section{Introduction}

The photometric variation of directly imaged exoplanets has been considered as an invaluable probe for the environment of habitable planets as well as spectroscopy \citep{ford2001characterization}. A two-dimensional surface distribution can be decoded from the diurnal and seasonal variations in reflected light \citep{kawahara2010global}. To date, this technique, termed as {\it spin-orbit tomography}, has been studied in terms of the regularization of geography \citep{kawahara2011mapping,fujii2012mapping},  Bayesian formulation \citep{2018AJ....156..146F, kawahara2020bayesian}, planet's axial tilt determination \citep{2016MNRAS.457..926S,2018AJ....156..146F}, dynamical mapping \citep{kawahara2020bayesian}, non-Lambertian effect \citep{2021arXiv210306275L}, and its application to real Earth data \citep{2019arXiv190312182L,2019ApJ...882L...1F}. Furthermore, the importance of regularization has been recognized. Tikhonov regularization, which was originally applied by \citet{kawahara2011mapping} to spin-orbit tomography, tends to exhibit smooth \add{solution} as an inferred map. \citet{aizawa2020global} showed that sparse modeling improved the sharpness of the inferred map. \add{Sparse modeling is an optimization technique that assumes the sparsity of the solution, as described in Appendix~\ref{sec:sparse}.} More recently, \citet{2021A&A...646A...4A} introduced a neural-net-based regulator, {\it  learned denoiser}, which also improved map quality. 

Spin-orbit tomography is formulated to retrieve a two-dimensional spatial map of a single component such as single-band photometry, color difference, and a single principle component. On the other hand, multi-color photometric variation contains information on the spectra of individual surface components such as water, soil, vegetation, snow, and clouds. {\it Rotational spectral unmixing} is a blind retrieval of endmember spectra of individual surface components combined with the disentanglement of geography by spin rotation, and it has been explored using EPOXI satellite data \citep{2013ApJ...765L..17C, 2018AJ....156..301L}. However, rotational spectral unmixing suffers from non-uniqueness of the inferred endmember spectra \citep{2017AJ....154..189F}.

Recently, \citet{kawahara2020global} proposed {\it spin-orbit unmixing}, which is a unified scheme of spin-orbit tomography and spectral unmixing that leverages non-negative matrix factorization and simplex volume minimization. The latter technique is essentially regularization of the spectral components proposed in the remote sensing field \citep{craig1994minimum}, which eliminates the ambiguity of the endmember determination. However, the regularization for geography is still a traditional Tikhonov ($\ell_2$-norm) regularization in the current spin-orbit unmixing. In this study, we show that sparse modeling in spin-orbit unmixing improves not only the map quality but also the recovery of the endmember spectra.

The remainder of this paper is organized as follows. We review the basic formulation of spin-orbit unmixing in Section~\ref{chap:method}. In Section~\ref{chap:sparse}, we formulate the spin-orbit unmixing with sparsity. Furthermore, in Section~\ref{chap:experiment}, using a toy model, we verify our technique. In Section~\ref{chap:dscovr}, we demonstrate the
new technique by applying it to real observational data
of the Earth obtained from the Deep Space Climate Observatory (DSCOVR) by \add{\citet{2019ApJ...882L...1F}}. In Section~\ref{chap:conclusion}, we summarize our findings. 

\section{Formulation}\label{chap:method}

\subsection{Spin-Orbit Tomography}\label{sec:SOT} 

The planet flux of reflected light from a solid surface is expressed \citep[Appendix A in][for the derivation]{kawahara2020global} as follows: 
\begin{align}
    f_\mathrm{p}^\mathrm{ref} =\frac{f_\star R_\mathrm{p}^2}{\pi a^2}\int_{S_\mathrm{I}\cap S_\mathrm{V}}\mathrm{d}S R(\vartheta_0,\varphi_0,\vartheta_1,\varphi_1) (\bm{e}_\mathrm{S}\cdot\bm{e}_\mathrm{R}) (\bm{e}_\mathrm{O}\cdot\bm{e}_\mathrm{R}),
\end{align}
where $R_\mathrm{p}$ denotes a planet radius; $a$ is a star--planet distance; $f_\star$ is the stellar flux; $R(\vartheta_0,\varphi_0,\vartheta_1,\varphi_1)$ is the bidirectional reflectance distribution function (BRDF); $\vartheta_0$ and $\varphi_0$ are the incident zenith angle and azimuth angle, respectively; $\vartheta_1$ and $\varphi_1$ are the reflected zenith angle and azimuth angle, respectively; $\bm{e}_\mathrm{O}$, $\bm{e}_\mathrm{S}$, and $\bm{e}_\mathrm{R}$ denote the unit vector from the center of the planet to the observer, primary star, and surface, respectively; and $S_\mathrm{I}\cap S_\mathrm{V}$ denotes the overlapped area of the illuminated region $S_\mathrm{I}$ ( $\bm{e}_\mathrm{S}\cdot\bm{e}_\mathrm{R}>0$ ) and the visible region $S_\mathrm{V}$ ($\bm{e}_\mathrm{O}\cdot\bm{e}_\mathrm{R}>0$).
Assuming isotropic reflection (Lambertian), the local reflectivity on the planet surface is expressed as a function of the spherical coordinate:
\begin{align}
    m(\theta,\phi)\coloneqq R(\vartheta_0,\varphi_0,\vartheta_1,\varphi_1).
\end{align}
Then, we obtain
\begin{align}
    f_\mathrm{p}^\mathrm{ref}(t) &= \int \mathrm{d}S G(t,\theta,\phi) m(\theta,\phi), 
\end{align}
where 
\begin{align}
    G(t,\theta,\phi) &\coloneqq 
    \begin{cases}
    {\displaystyle \frac{f_\star R_\mathrm{p}^2}{\pi a^2} } (\bm{e}_\mathrm{S}\cdot\bm{e}_\mathrm{R}) (\bm{e}_\mathrm{O}\cdot\bm{e}_\mathrm{R}) \\
    \hspace{2.5em} (\bm{e}_\mathrm{S}\cdot\bm{e}_\mathrm{R}>0,\ \bm{e}_\mathrm{O}\cdot\bm{e}_\mathrm{R}>0) \\
    0 \hspace{2em} (\mathrm{otherwise}).
    \end{cases}
\end{align}
The discretization of $t$ and $(\theta,\phi)$ yields
\begin{align}
    f_\mathrm{p}^\mathrm{ref}(t_i) &= \sum_j \Delta S G(t_i,\theta_j,\phi_j) m(\theta_j,\phi_j). \label{eq:ref_flux_descrete}
\end{align}
In this study, we pixelized the planet surface using Hierarchical Equal Area isoLatitude Pixelization (HEALPix) \citep{gorski2005healpix}. Then, Equation~\eqref{eq:ref_flux_descrete} is then written in matrix form as follows:
\begin{align}
    \bm{d}=W\bm{m}, \label{eq:inverse_problem}
\end{align}
where $d_i\coloneqq f_\mathrm{p}^\mathrm{ref}(t_i)$; $m_j\coloneqq m(\theta_j,\phi_j)$; $W_{ij}\coloneqq G(t_i,\theta_j,\phi_j)\Delta S$ $(i=1,\ldots,N_i, j=1,\ldots,N_j)$; $N_i$ and $N_j$ are the number of observation data and pixels of the surface, respectively. 
%%%%%%%%%%%%%%%%%%%
We infer the surface distribution of planet $\bm{m}_\mathrm{est}$ from the observation data $\bm{d}_\mathrm{obs}$. In reality, we must include an observation error $\bm{\varepsilon}$ in the model: 
\begin{align}
    \label{eq:sot}
    \bm{d}_\mathrm{obs}=W\bm{m}+\bm{\varepsilon}.
\end{align}
Spin-orbit tomography infers $\bm{m}$ from $\bm{d}_\mathrm{obs}$. In general, $W$ is a function of \add{an orbital inclination $i$,} an orbital phase at Equinox $\Theta_{\mathrm{eq}}$, and a planet spin vector or equivalently a set of planet obliquity $\zeta$. Additionally, the planet rotation period should be inferred. These nonlinear parameters can be inferred within the framework of spin-orbit tomography \citep{kawahara2010global,2016MNRAS.457..926S,2018AJ....156..146F,kawahara2020bayesian}. In this study, we assume that these nonlinear parameters are known. This type of optimization problem then becomes a linear inverse problem. 

In general, the linear inverse problem requires regularization of the model to suppress the model instability due to noise, namely, to avoid overfitting or overlearning. Regularization can take a variety of formulations such as a bounded model \citep{kawahara2010global}, a non-negative condition \citep{kawahara2020global}, regularization term \citep{kawahara2011mapping, aizawa2020global}, and neural net \citep{2021A&A...646A...4A}. 
Using the regularization term, the inverse problem can be formulated as an optimization problem:
\begin{align}
    \minimize_{\bm{m}} \frac{1}{2}\|\bm{d}-W\bm{m}\|_2^2 +R(\bm{m}). \label{eq:SOT}
\end{align}
\citet{kawahara2011mapping} used Tikhonov regularization as follows:
\begin{align}
    \minimize_{\bm{m}} \frac{1}{2}\|\bm{d}-W\bm{m}\|_2^2 +\lambda_\mathrm{Tik}\|\bm{m}\|_2^2,
\end{align}
where $\lambda_\mathrm{Tik}$ denotes the regularization parameter. Tikhonov regularization tends to provide a smoother \add{solution} than the other regularization methods. To improve the sharpness of the inferred map, \citet{aizawa2020global} introduced a combination of $\ell_1$-norm regularization and Total Squared Variation (TSV) regularization \citep{kuramochi2018superresolution} as a sparse modeling as follows: 
\begin{align}
    \minimize_{\bm{m}} \frac{1}{2}\|\bm{d}-W\bm{m}\|_2^2 +\lambda_{\ell_1}\|\bm{m}\|_1+\lambda_\mathrm{TSV}\|\bm{m}\|_\mathrm{TSV}, \label{eq:SOT_L1TSV}
\end{align}
where $\lambda_{\ell_1}$ and $\lambda_\mathrm{TSV}$ denote the regularization parameters, $\|\cdot\|_1$ and $\|\cdot\|_\mathrm{TSV}$ denote $\ell_1$-norm and TSV norm, respectively.

%Let us denote the prediction data by $\bm{d}_\mathrm{pre}\coloneqq W\bm{m}_\mathrm{est}$ and the prediction error by $\bm{\varepsilon}_\mathrm{pre}\coloneqq \bm{d}_\mathrm{obs}-\bm{d}_\mathrm{pre}$,  

Spin-orbit tomography can retrieve a map of a single component of a light curve. For instance, \cite{kawahara2011mapping} used the color difference between the near-infrared and visible bands to reduce the cloud contribution. The retrieved map accurately represented the land and ocean distribution. \cite{2019ApJ...882L...1F} used a second principle component of the PCA, which also recovered the land/ocean distribution from the real data of Earth obtained via DSCOVR. However, these choices use prior knowledge of the surface of the planet. 

To infer map and surface components, \cite{2013ApJ...765L..17C} applied spectral unmixing to the longitudinal mapping of Earth. However, the retrieved map did not match the real land and ocean distribution. \cite{kawahara2020global} extended their method to a two-dimensional mapping, corresponding to a combination of spectral unmixing and spin-orbit tomography, namely, {\it spin-orbit unmixing}. They also introduced regularization terms for geography and spectra. The detail of spectral unmixing is presented in Appendix~\ref{sec:spectral_unmixing}.

\subsection{Spin-Orbit Unmixing}\label{sec:SOU}

We explain spin-orbit unmixing. Spin-orbit unmixing uses the observation data matrix $D\in\mathbb{R}^{N_i\times N_l}$ along the time and wavelength axes. Let $M\in\mathbb{R}^{N_j\times N_l}$ be the surface distribution defined as $M_{jl}\coloneqq m(\theta_j,\phi_j,\lambda_l)$, and let $A\in\mathbb{R}^{N_j\times N_k}$ and $X\in\mathbb{R}^{N_k\times N_l}$ be the surface distribution matrix and endmember matrix, respectively. Then, the formulation of the spin-orbit unmixing is as follows:
\begin{align}
    D=WAX.
\end{align}
As an optimization problem, the spin-orbit unmixing can be expressed as follows:
\begin{align}
    \minimize_{A,X} \frac{1}{2}\|D-WAX\|_\mathrm{F}^2 +R(A,X) \nonumber \\ 
    \subjectto \add{A_{jk}}\ge 0, X_{kl}\ge 0, \label{eq:SOU}
\end{align}
where $R(A,X)$ denotes a regularization term and $\|\cdot\|_\mathrm{F}$ denotes the Frobenius norm, which can be defined as follows:
\begin{align}
    \|Y\|_\mathrm{F}\coloneqq \sqrt{\sum_i\sum_j Y_{ij}^2}.
\end{align}

\cite{kawahara2020global} used Tikhonov regularization for $A$ and determinant type of the volume regularization for $X$ as follows:
\begin{align}
    \minimize_{A,X} \frac{1}{2}\|D-WAX\|_\mathrm{F}^2 +\frac{\lambda_A}{2}\|A\|_\mathrm{F}^2+\frac{\lambda_X}{2} \det(XX^\top) \nonumber \\
    \subjectto \add{A_{jk}}\ge 0, X_{kl}\ge 0, \label{eq:SOU_L2-VRDet}
\end{align}
where $\lambda_A$ and $\lambda_X$ denote the regularization parameters.
Volume regularization is based on the concept of simplex volume minimization, which was developed in the field of remote sensing \citep{craig1994minimum,fu2015blind,lin2015identifiability,fu2019nonnegative,ang2019algorithms}. With respect to spectral unmixing, it is known that non-negative matrix factorization with the volume regularization term accurately reproduces high-resolution spectrum components from satellite data. Simplex volume minimization is justified under the assumption that the data points are widely spread in the convex hull defined by the endmembers. The true endmembers are then identified by the data-enclosing simplex, whose volume is minimized \citep{craig1994minimum}. An intuitive explanation is provided in Figure 1 in \cite{lin2015identifiability} and Figure 1 in \cite{kawahara2020global}. The choice of $\det{(X X^\top)}$ as a regularization term is based on the following fact\add{: the} volume of an $(N_l-1)$-simplex (or convex hull) in $(N_l-1)$-dimensional space with vertices $\{\bm{x}_1,\bm{x}_2,\ldots,\bm{x}_{N_l}\}$ is $\det(XX^\top)/(N_l!)$, where $\bm{x}_k$ denotes the $k$-th endmember (i.e., the $k$-th column vector) of $X^\top$. 

\section{Spin-Orbit Unmixing with Sparsity} \label{chap:sparse}

In this study, we introduce the sparsity of geography $A$ into spin-orbit unmixing. \add{Most of the elements are zero in a sparse matrix. Sparse optimization is described in Appendix~\ref{sec:sparse}.} The objective function in this study is expressed as follows:
\begin{gather}
    \minimize_{A,X}  Q \subjectto \add{A_{jk}}\ge 0, X_{kl}\ge 0, \label{eq:SOU_problem}\\
    Q\coloneqq\frac{1}{2}\|D-WAX\|_\mathrm{F}^2 +R(A,X). \label{eq:SOU_Q}
\end{gather}
As a regularization of $A$, we consider two types of sparse modeling: $\ell_1$-norm+Total Square Variation (TSV) and trace norm regularization. We also computed Tikhonov regularization for comparison purposes. 

In Equation~\eqref{eq:SOU_problem}, we must optimize $A$ and $X$. Based on \cite{kawahara2020global}, we used the block coordinate descent method, which divides the problem into two separate optimizations for $A$ and $X$ \citep[for example,][]{kim2014algorithms}, 
\begin{align}
    \minimize_{A} & q_A \subjectto \add{A_{jk}}\ge 0, \\
    \minimize_{X} & q_X \subjectto X_{kl}\ge 0,
\end{align}
where $q_A = q_A(A)$ and $q_X = q_X(X)$ are defined \add{by rearranging Equation~\eqref{eq:SOU_Q}} as
\begin{align}
    Q &= q_A(A) + (\mathrm{constant\ for\ } A), \\
    Q &= q_X(X) + (\mathrm{constant\ for\ } X),
\end{align}
and these optimizations are iteratively conducted until convergence.

%%%%%%%%%%%%%%%%
\subsection{Spin-Orbit Unmixing with \texorpdfstring{$\ell_1$}{TEXT}-Norm and TSV Regularization} \label{sec:SOU_L1TSV-VRDet}

In the context of spin-orbit tomography, \citet{aizawa2020global} described a procedure for obtaining a sparse solution using the $\ell_1$-norm and TSV regularization. We extend the optimization method to spin-orbit unmixing. The optimization problem with the $\ell_1$-norm and TSV regularization can be expressed as follows:
\begin{align}
    \minimize_{A,X} &Q_{\ell_1\mathrm{+TSV}} \nonumber\\
    \subjectto &\add{A_{jk}}\ge 0, X_{kl}\ge 0, \label{eq:SOU_L1TSV-VRDet_Q} \\
    Q_{\ell_1\mathrm{+TSV}}\coloneqq& \frac{1}{2}\|D-WAX\|_\mathrm{F}^2 \nonumber \\
    &+ \sum_k\left( \lambda_{\ell_1}\|\bm{a}_k\|_1+\lambda_\mathrm{TSV}\|\bm{a}_k\|_\mathrm{TSV} \right) \nonumber \\
    &+\frac{\lambda_X}{2} \det(XX^\top), \label{eq:Q_L1TSV}
\end{align}
where $\bm{a}_k$ \add{denotes} the $k$-th column vector of $A$, and $\lambda_{\ell_1}$, $\lambda_\mathrm{TSV}$, and $\lambda_X$ denote the regularization parameters. We used the determinant type of volume regularization for $X$ as introduced in Section~\ref{sec:SOU}. 

First, we solve the optimization problem \eqref{eq:SOU_L1TSV-VRDet_Q} for $A$. $Q_{\ell_1\mathrm{+TSV}}$ is rearranged with respect to $\bm{a}_k$ as follows:
\begin{align}
    Q_{\ell_1\mathrm{+TSV}}=& \frac{1}{2} \| \bm{x}_k \|_2^2 \left\| \bm{p}_A - W \bm{a}_k \right\|_2^2 + \lambda_{\ell_1}\|\bm{a}_k\|_1 \nonumber \\ &+\lambda_\mathrm{TSV}\|\bm{a}_k\|_\mathrm{TSV}+(\mathrm{constant\ for\ }\bm{a}_k ), \label{eq:Q_L1TSV_A_sq}
\end{align}
where \add{$\bm{x}_k$ is the $k$-th column vector of $X^\top$,} $\bm{p}_A \coloneqq (1/\| \bm{x}_k \|_2^2) \Delta \bm{x}_k$\add{,} and $\Delta$ is a matrix defined as $\Delta_{il} \coloneqq D_{il} - \sum_j \sum_{s \neq k} W_{ij} A_{js} X_{sl}$. Therefore, the subproblem for solving the optimization problem \eqref{eq:SOU_L1TSV-VRDet_Q} is expressed as follows:
\begin{align}
    \minimize_{\bm{a}_k} &q_{A,\ell_1\mathrm{+TSV}} \nonumber\\
    \subjectto &(\bm{a}_k)_j\ge 0\ (j=1,\ldots,N_j),\label{eq:SOU_L1TSV-VRDet_subproblem_A}  \\
    q_{A,\ell_1\mathrm{+TSV}} \coloneqq &\frac{1}{2}  \left\| \bm{p}_A - W \bm{a}_k \right\|_2^2 \nonumber\\
    &+ \lambda'_{\ell_1}\|\bm{a}_k\|_1+\lambda'_\mathrm{TSV}\|\bm{a}_k\|_\mathrm{TSV},
\end{align}
where $\lambda'_{\ell_1}\coloneqq\lambda_{\ell_1}/\ \|\bm{x}_k\|_2^2$ and $ \lambda'_\mathrm{TSV}\coloneqq\lambda_\mathrm{TSV}/\ \|\bm{x}_k\|_2^2$. Then, we suppose
\begin{align}
    f_{\ell_1\mathrm{+TSV}}(\bm{a}_k)&\coloneqq \frac{1}{2}  \left\| \bm{p}_A - W \bm{a}_k \right\|_2^2 +\lambda'_\mathrm{TSV}\|\bm{a}_k\|_\mathrm{TSV}, \\
    \psi_{\ell_1\mathrm{+TSV}}(\bm{a}_k)&\coloneqq \lambda'_{\ell_1}\|\bm{a}_k\|_1+ \delta_+ (\bm{a}_k),
\end{align}
\add{where $\delta_+$ is defined in Equation~\eqref{eq:nonnegative_indicator}.} It should be noted that $f_{\ell_1\mathrm{+TSV}}(\bm{a}_k)$ is differentiable, and $\psi_{\ell_1\mathrm{+TSV}}(\bm{a}_k)$ is non-differentiable. By adopting the above equation, Equation~\eqref{eq:SOU_L1TSV-VRDet_subproblem_A} is rewritten as follows:
\begin{align}
    \minimize_{\bm{a}_k} f_{\ell_1\mathrm{+TSV}}(\bm{a}_k) +\psi_{\ell_1\mathrm{+TSV}}(\bm{a}_k).
\end{align}
This objective function consists of a combination of two proper convex functions (Definition~\ref{dfn:proper_convex} in Appendix~\ref{chap:mathematics}) that are differentiable and not necessarily differentiable. This type of objective function can be optimized using the proximal gradient method, as explained in Appendix~\ref{sec:prox_grad_method}.

The gradient of $f_{\ell_1\mathrm{+TSV}}(\bm{a}_k) $ and proximal operator of $\psi_{\ell_1\mathrm{+TSV}}(\bm{a}_k)$ can be obtained as follows:
\begin{align}
    \nabla f_{\ell_1\mathrm{+TSV}}(\bm{a}_k) &= \add{-W^\top\left(\bm{p}_A-W\bm{a}_k\right)}+ 2 \lambda'_\mathrm{TSV} \tilde{N}\bm{a}_k, \\
    \prox \left(\bm{w} \mid \gamma\psi_{\ell_1\mathrm{+TSV}}\right) &= \max\{\bm{w}-\gamma\lambda'_{\ell_1}\bm{1}, \bm{0}\}, \label{eq:prox_L1_nonneg}
\end{align}
where $\tilde{N}$ is defined in \add{Equation~\eqref{eq:tilde_N}, $\gamma>0$ is the parameter indicating the step size,} $\bm{1}=(1,1,\ldots,1)^\top$ \add{denotes} a vector such that all entries are one (one vector), and $\max$ \add{denotes} the element-wise maximum. Hence, the update formula of the proximal gradient method for \eqref{eq:SOU_L1TSV-VRDet_subproblem_A} can be expressed as follows:
\begin{align}
    \bm{a}_k^{(i+1)}&= \max\left\{ \bm{a}_k^{(i)} - \gamma \bm{s}_k^{(i)} -\gamma\lambda'_{\ell_1}\bm{1}, \bm{0}\right\}, \\
    \bm{s}_k^{(i)} &\coloneqq \add{-W^\top\left(\bm{p}_A-W\bm{a}_k^{(i)}\right)}+ 2 \lambda'_\mathrm{TSV} \tilde{N}\bm{a}_k^{(i)}.
\end{align}

In this study, we used Monotone FISTA (MFISTA) \citep{beck2009fastIEEE} to solve the subproblem for $\bm{a}_k$ \eqref{eq:SOU_L1TSV-VRDet_subproblem_A}. MFISTA deals with the non-monotonic decreasing nature of \add{Fast Iterative Shrinkage-Thresholding Algorithm (FISTA) \citep{beck2009fastSIAM}}, and \citet{aizawa2020global} used it to solve spin-orbit tomography with the $\ell_1$-norm and TSV regularization. 

Next, we consider the optimization problem \eqref{eq:SOU_L1TSV-VRDet_Q} for $X$. The objective function $Q_{\ell_1\mathrm{+TSV}}$ can be rearranged with respect to $\bm{x}_k$ as follows:
\begin{align}
    Q_{\ell_1\mathrm{+TSV}}=& \frac{1}{2} \bm{x}_k^\top ( \mathcal{L}_X + \mathcal{D}_X ) \bm{x}_k \nonumber\\
    &- \bm{l}_X^\top \bm{x}_k +(\mathrm{constant\ for\ }\bm{x}_k), \label{eq:Q_L1TSV_X_sq}
\end{align}
where $\mathcal{L}_X \coloneqq \| W \bm{a}_k \| _2 ^2 I$, $\mathcal{D}_X \coloneqq \lambda_X \mathrm{det} ( \breve{X}_k \breve{X}_k^\top ) ( I - \breve{X}_k^\top ( \breve{X}_k \breve{X}_k^\top )^{-1} \breve{X}_k )$, $\bm{l}_X \coloneqq  \Delta^\top W \bm{a}_k$, $I$ is an identity matrix, and $\breve{X}_k$ is a submatrix of $X$ \add{with the $k$-th row removed}. The subproblem for solving the optimization problem \eqref{eq:SOU_L1TSV-VRDet_Q} is expressed as follows:
\begin{align}
    \minimize_{\bm{x}_k} &q_{X,\ell_1\mathrm{+TSV}} \nonumber\\
    \subjectto &(\bm{x}_k)_l\ge 0\ (l=1,\ldots,N_l), \label{eq:SOU_L1TSV-VRDet_subproblem_X} \\
    q_{X,\ell_1\mathrm{+TSV}} \coloneqq &\frac{1}{2} \bm{x}_k^\top ( \mathcal{L}_X + \mathcal{D}_X ) \bm{x}_k - \bm{l}_X^\top \bm{x}_k,
\end{align}
Subproblem \eqref{eq:SOU_L1TSV-VRDet_subproblem_X} is an optimization problem with a non-negative constraint for a differentiable objective function. This problem can be optimized using the proximal gradient method as explained in Appendix~\ref{sec:prox_grad_method}. The derivative with respect to $\bm{x}_k$ can be obtained as follows:
\begin{align}
    \nabla q_{X,\ell_1\mathrm{+TSV}}(\bm{x}_k)=( \mathcal{L}_X + \mathcal{D}_X )\bm{x}_k-\bm{l}_X.
\end{align}
Then, the update formula of the proximal gradient method for \eqref{eq:SOU_L1TSV-VRDet_subproblem_X} is written as follows:
\begin{align}
    \bm{x}_k^{(i+1)} &= \max \left\{ \bm{x}_k^{(i)} - \gamma\left(\left( \mathcal{L}_X + \mathcal{D}_X \right)\bm{x}_k^{(i)}-\bm{l}_X \right),\bm{0} \right\}.
\end{align}

\citet{kawahara2020global} optimized the subproblem \eqref{eq:SOU_L1TSV-VRDet_subproblem_X} using \add{FISTA}, which is a proximal gradient method using Nesterov's acceleration method \citep{nesterov2003introductory} with the restart method. Nesterov's acceleration method increased the convergence speed. However, the function value did not necessarily decrease monotonically. The restart method \citep{o2015adaptive} avoids increasing the objective function by restarting Nesterov's acceleration method when the function value does not decrease. 

Consequently, the optimization algorithm for solving \eqref{eq:SOU_L1TSV-VRDet_Q} is summarized as follows:
\begin{algorithm}[H]
    \caption{Spin-orbit unmixing with $\ell_1$-norm and TSV regularization}
    \label{alg:SOU_using_L1TSV}
    \begin{algorithmic}
    \STATE Initialization: non-negative matrices $A_0$ and $X_0$
    \FOR{$n$ in $(1,N_\mathrm{try})$}
    \FOR{$k$ in $(1,N_k)$}
    \STATE Update $\bm{x}_k$ using FISTA with the restart method
    \STATE Update $\bm{a}_k$ using MFISTA
    \ENDFOR
    \ENDFOR
    \end{algorithmic}
\end{algorithm}

%%%%%%%%%%%%%%%%%%%%%%%%%
\subsection{Spin-Orbit Unmixing with Trace Norm Regularization} \label{sec:SOU_Trace-VRDet}

We also consider another type of sparse modeling, termed as trace norm regularization. Trace norm regularization is sparse modeling of a matrix, not a vector. It results in a low-rank matrix. Therefore, vectorization of $A$ is not required as opposed to that in the $\ell_1$-norm and TSV regularization. The optimization problem using trace norm regularization for $A$ is expressed as follows:
\begin{gather}
    \minimize_{A,X} Q_\mathrm{Tr} \subjectto X_{kl}\ge 0, \\
    Q_\mathrm{Tr}\coloneqq \frac{1}{2}\|D-WAX\|_\mathrm{F}^2 +\lambda_A\|A\|_\mathrm{Tr} +\frac{\lambda_X}{2} \det(XX^\top),
    \label{eq:SOU_Trace-VRDet_Q}
\end{gather}
where $\|\cdot\|_\mathrm{Tr}$ denotes trace norm defined in \add{Equation~\eqref{eq:trace_norm}}, and $\lambda_A$ and $\lambda_X$ denote regularization parameters. The subproblem for solving the optimization problem is expressed as follows:
\begin{align}
    \minimize_{A}& q_{A,\mathrm{Tr}}\coloneqq \frac{1}{2}\|D-WAX\|_\mathrm{F}^2 +\lambda_A\|A\|_\mathrm{Tr}, \label{eq:SOU_Trace-VRDet_subproblem_A} \\
    \minimize_{\bm{x}_k}& q_{X,\mathrm{Tr}}\coloneqq \add{q_{X,\ell_1\mathrm{+TSV}}} \nonumber\\
    \subjectto &(\bm{x}_k)_l\ge 0\ (l=1,\ldots,N_l). \label{eq:SOU_Trace-VRDet_subproblem_X}
\end{align}
For trace norm regularization, we do not impose a non-negative condition \add{on $A$} because it is \add{too} difficult to implement. \add{Additionally, we used the determinant type of volume regularization for $X$, same as Section~\ref{sec:SOU_L1TSV-VRDet}. Thus, the subproblem \eqref{eq:SOU_Trace-VRDet_subproblem_X} is the same as the subproblem \eqref{eq:SOU_L1TSV-VRDet_subproblem_X}.} 

We solved the subproblem for $\bm{x}_k$ \eqref{eq:SOU_Trace-VRDet_subproblem_X} using the same scheme as that in Section~\ref{sec:SOU_L1TSV-VRDet}. Let us consider the subproblem of $A$. Given that $f_{\mathrm{Tr}}(A)\coloneqq(1/2)\|D-WAX\|_\mathrm{F}^2$ is \add{differentiable} and $\psi_{\mathrm{Tr}}(A)\coloneqq\lambda_A\|A\|_\mathrm{Tr}$ is \add{non-differentiable}, it can be solved using the proximal gradient method as explained in Appendix~\ref{sec:prox_grad_method}. The derivative of $f_{\mathrm{Tr}}(A) $ and proximal operator of $\psi_{\mathrm{Tr}}(A)$ are as follows:
\begin{align}
    \nabla f_{\mathrm{Tr}}(A) &= -W^\top (D-WAX)X^\top, \\
    \prox \left( W \mid \gamma\psi_{\mathrm{Tr}}\right) &=  U \max\{\Sigma-\gamma\lambda_A I,\bm{O}\} V^\top, \label{eq:prox_Trace}
\end{align}
where $W=U\Sigma V^\top$ denotes the singular value decomposition of $W$ \citep[Lemma 8.4 in][]{tomioka2015}. Hence, the update formula of the proximal gradient method for \eqref{eq:SOU_Trace-VRDet_subproblem_A} can be expressed as follows:  
\begin{align}
    A_{i+1}=U_i \max\{\Sigma_i-\gamma\lambda_A I,\bm{O}\} V_i^\top,
\end{align}
where $Y_i=U_i\Sigma_i V_i^\top$ denotes the singular value decomposition of $Y_i\coloneqq A_i-\gamma \nabla f_{\mathrm{Tr}}(A_i)=A_i+\gamma W^\top (D-WA_iX)X^\top$. Hence, the optimization algorithm for solving \eqref{eq:SOU_Trace-VRDet_Q} can be summarized as follows:
\begin{algorithm}[H]
    \caption{Spin-orbit unmixing with trace norm regularization}
    \label{alg:SOU_using_Trace}
    \begin{algorithmic}
    \STATE Initialization: non-negative matrices $A_0$ and $X_0$
    \FOR{$n$ in $(1,N_\mathrm{try})$}
    \FOR{$k$ in $(1,N_k)$}
    \STATE Update $\bm{x}_k$ using FISTA with the restart method
    \ENDFOR
    \STATE Update $A$ using FISTA with the restart method
    \ENDFOR
    \end{algorithmic}
\end{algorithm}
The code for optimization is publicly available\footnote{https://github.com/atsuki-kuwata/exomap}.

%%%%%%%%%%%%%%%%%%%%%%%%%%%%%%%%%%%
\section{Testing using a cloudless model} \label{chap:experiment}

In this section, we describe the testing of our method using a cloudless Earth model. To compare with the results obtained using sparse modeling, we also consider spin-orbit unmixing with Tikhonov regularization and the determinant type of volume regularization \citep{kawahara2020global}. The optimization algorithm for the latter is provided in Appendix~\ref{sec:SOU_using_L2-VRDet}.

%%%%%%%%%%%%%%%%%%%%%%%%%
\subsection{Cloudless Earth}

\begin{figure}
  \centering
  \subfigure[]{% 
    \includegraphics[width=.45\textwidth]{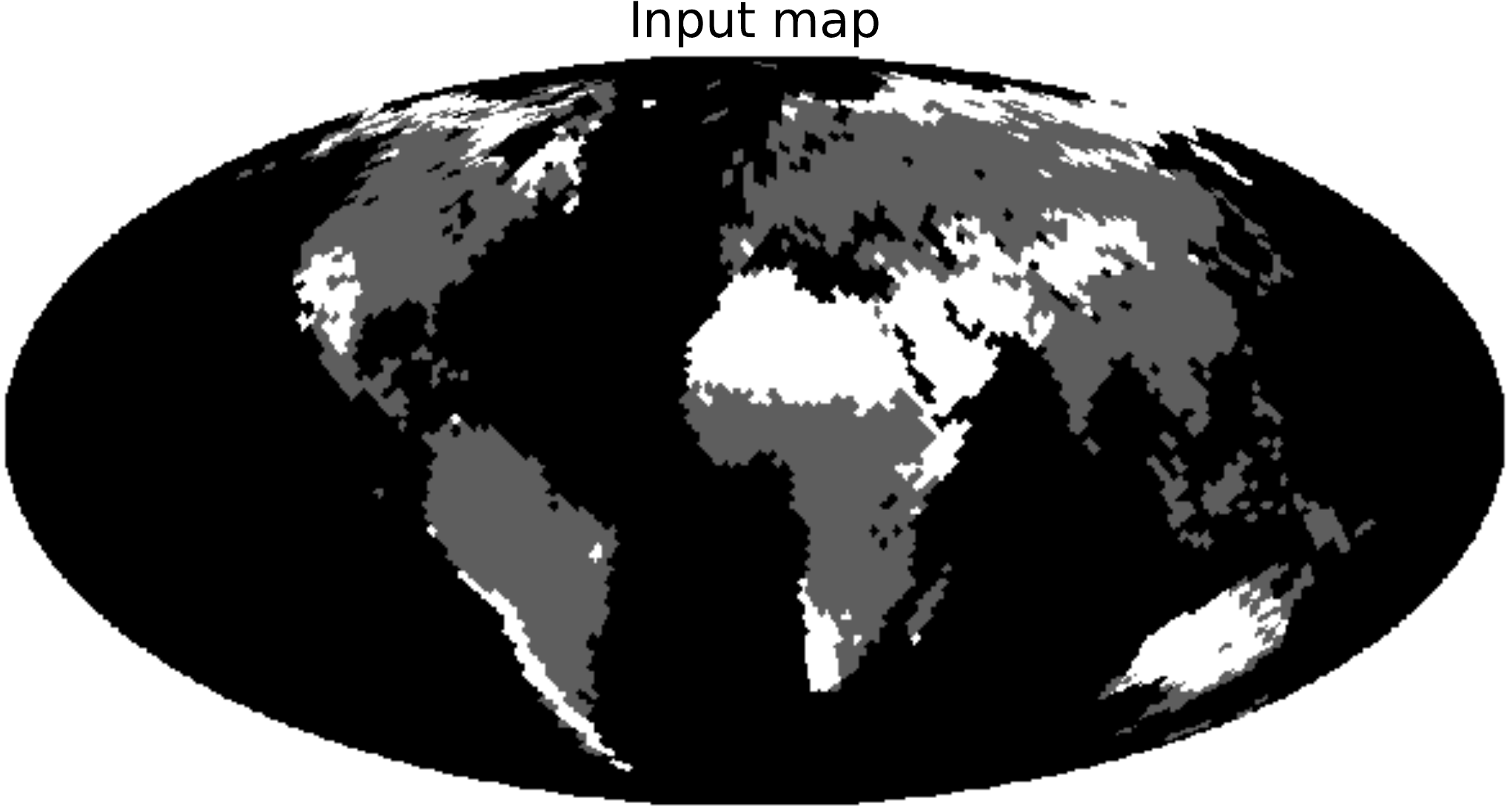}%
    \label{fig:initial_A}%
  }%
  \hspace{.05\textwidth}
  \subfigure[]{%
    \includegraphics[width=.45\textwidth]{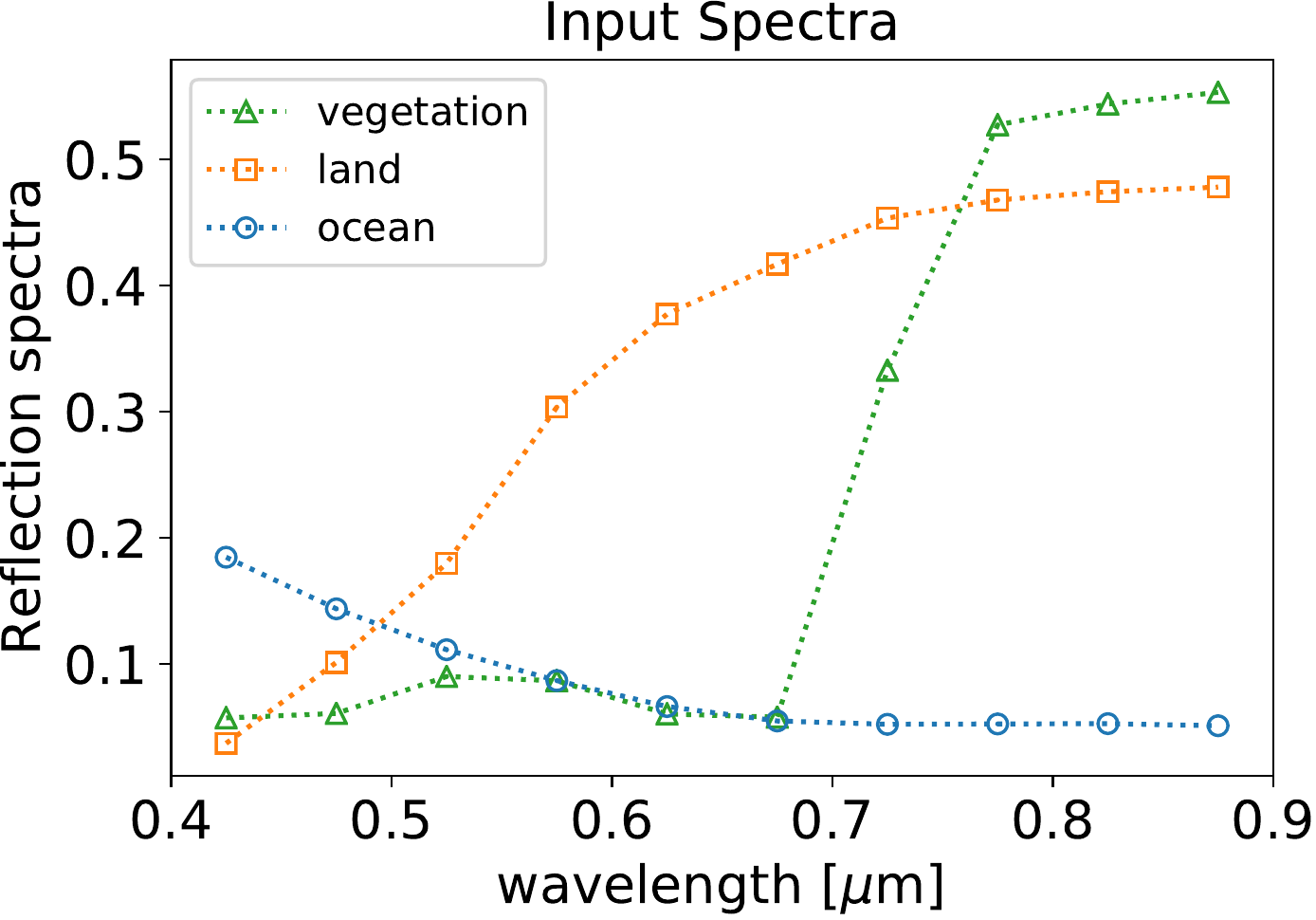}
    \label{fig:initial_X}%
  }
  \caption[]{Input data for the model of a cloudless Earth. (a) Input surface distribution $A_\mathrm{true}$: Gray, white, and black correspond to vegetation, land, and ocean, respectively. (b) Input spectrum $X_\mathrm{true}$.}
\label{fig:initial_A_X}
\end{figure}

First, we generate mock multi-color light curves as follows:
\begin{align}
    D=W A_\mathrm{true} X_\mathrm{true}+E,
\end{align}
where $ A_\mathrm{true}\add{\in\mathbb{R}^{N_{j,\mathrm{true}}\times N_{k,\mathrm{true}}}}$ and $X_\mathrm{true}\in\mathbb{R}^{N_{k,\mathrm{true}}\times N_l}$ denote the input geography and surface spectra, respectively. The time interval is one year divided by $N_i=512$. Furthermore, $N_{j,\mathrm{true}}=12288$, $N_{k,\mathrm{true}}=3$, and $N_l=10$ denote the number of pixels on the planet surface, surface components, and observing bands, respectively. We considered vegetation, land, and ocean as the endmembers of the surface components. Error matrix $E\add{\in\mathbb{R}^{N_i\times N_l}}$ is randomly generated as follows:
\begin{align}
    \add{E_{il}}&\sim 0.01 \overline{D_\mathrm{true}}\ \mathcal{N}(0,1), \\
    \overline{D_\mathrm{true}}&\coloneqq \frac{1}{N_i N_l}\sum_i \sum_l \left(W A_\mathrm{true} X_\mathrm{true}\right)_{il},
\end{align}
where $\mathcal{N}(0,1)$ is a normal distribution with a mean $\mu=0$ and standard deviation $\sigma=1$. We used the classification map provided by the Moderate Resolution Imaging Spectroradiometer (MODIS) as the input surface distribution $A_\mathrm{true}$, ASTER spectral library \citep{baldridge2009aster} for the spectra of vegetation and land, and those produced by \citet{mclinden1997estimating} for the ocean. We set the observation wavelength to $0.425+0.05(l-1)$ \textmu m $ (l=1,\ldots,N_l=10)$. $A_\mathrm{true}$ and $X_\mathrm{true}$ are shown in Figure~\ref{fig:initial_A_X}. We also generated $W$ using the method described in Section~\ref{sec:SOT} with orbital inclination $i=45^\circ$, orbital phase angle at the vernal equinox $\Theta_\mathrm{eq}=90^\circ$, obliquity $\zeta=23.4^\circ$, orbital period $P_\mathrm{orb}=365$~days, and rotation period $P_\mathrm{spin}=23.9344699/24.0$~days. 

Based on $D$ and $W$, we infer $A\in\mathbb{R}^{N_j\times N_k}$ and $X\in\mathbb{R}^{N_k\times N_l}$ by solving the following optimization problem:
\begin{align}
    \minimize_{A,X} \frac{1}{2}\|D-WAX\|_\mathrm{F}^2 +R(A,X) \nonumber \\
    \subjectto \add{A_{jk}}\ge 0, X_{kl}\ge 0, \label{eq:SOU_experiment}
\end{align}
\add{where $R(A,X)$ takes the form in Equation~\eqref{eq:Q_L1TSV} ($\ell_1$-norm and TSV regularization), \eqref{eq:SOU_Trace-VRDet_Q} (trace norm regularization), or \eqref{eq:SOU_L2-VRDet} (Tikhonov regularization). }We set the number of pixels in the inferred map to $N_j=3024$ and number of endmembers to $N_{k}=3$.

%%%%%%%%%%%%%%%%%%%%%%%
\subsection{Results of Spin-Orbit Unmixing with \texorpdfstring{$\ell_1$}{TEXT}-Norm and TSV Regularization} \label{sec:SOU_L1TSV-VRDet_experiment}

We solved Equation~\eqref{eq:Q_L1TSV} with $\lambda_{\ell_1}=10^{-3.5}$, $\lambda_\mathrm{TSV}=10^{-4}$, $ \lambda_X=10^2$, and iteration number $N_\mathrm{try}=10^3$. These parameters were selected based on the Mean-Removed Spectral Angle (MRSA) and the Correct Pixel Rate (CPR) as in \cite{kawahara2020global}. The detailed procedure is presented in Appendix~\ref{sec:L1TSV_evaluate}.

There is a known indefiniteness of matrix factorization, as explained below. The surface distribution at each wavelength $M\in\mathbb{R}^{N_j\times N_l}$ can be written as:
\begin{align}
    M=AX=\sum_{k=1}^{N_k} \bm{a}_k \bm{x}_k,
\end{align}
where $\bm{a}_k$ and $\bm{x}_k$ $(k=1,\ldots,N_k)$ denote the $k$-th column vectors of $A$ and $X^\top$, respectively. By using the constant, $c_k$, we obtain the following.
\begin{align}
    \bm{a}_k \bm{x}_k = \left( c_k^{-1} \bm{a}_k\right) \left( c_k \bm{x}_k\right). \ (k=1,\ldots,N_k)
\end{align}
This implies that the inferred surface distribution and spectrum have an indefiniteness of constant multiples. Hence, we normalize the inferred surface distribution and spectrum for $X$ as follows:
\begin{align}
    \hat{\bm{a}}_k &= \left(\frac{\overline{x_{k,\mathrm{true}}}}{\overline{x_k}}\right)^{-1} \bm{a}_k,\\
    \hat{\bm{x}}_k &= \left(\frac{\overline{x_{k,\mathrm{true}}}}{\overline{x_k}}\right) \bm{x}_k,
\end{align}
where $\bm{x}_{k,\mathrm{true}}$ $(k=1,\ldots,N_{k,\mathrm{true}}=N_k)$ is the $k$-th column vector of $X_\mathrm{true}^\top$. Furthermore, $\overline{x_k}$ and $\overline{x_{k,\mathrm{true}}}$ denote the means of $\bm{x}_k$ and $\bm{x}_{k,\mathrm{true}}$, respectively.
\begin{align}
    \overline{x_k}&=\frac{1}{N_l}\sum_{l=1}^{N_l} \left(\bm{x}_k\right)_l,\\
    \overline{x_{k,\mathrm{true}}}&=\frac{1}{N_{l,\mathrm{true}}}\sum_{l=1}^{N_{l,\mathrm{true}}} \left(\bm{x}_{k,\mathrm{true}}\right)_l.
\end{align}

%Figure~\ref{fig:cost_terms} shows the evolution curve of each term in \eqref{eq:Q_L1TSV} as a function of the iteration number. When the number of iterations exceeded $\sim 10^3$, we found an increase in the objective function $Q_{\ell_1\mathrm{+TSV}}$, which is an indication of overfitting. Therefore, we decided to adopt a method similar to early stopping in the machine learning field. We selected the solution at $N_\mathrm{try}=10^3$ iterations as the best inferred solution. 

%\begin{figure*}[t]
%\centering
%\includegraphics[width=0.8\linewidth]{cost_terms.pdf}
%\caption[]{Evolution of the objective function and regularization terms for SOU with the $\ell_1$-norm and TSV regularization. We selected $N_\mathrm{try}=10^3$ as the optimal number of iterations.}
%\label{fig:cost_terms}%
%\end{figure*}

\begin{figure}
  \centering
  \subfigure[]{% 
    \includegraphics[width=.3\textwidth]{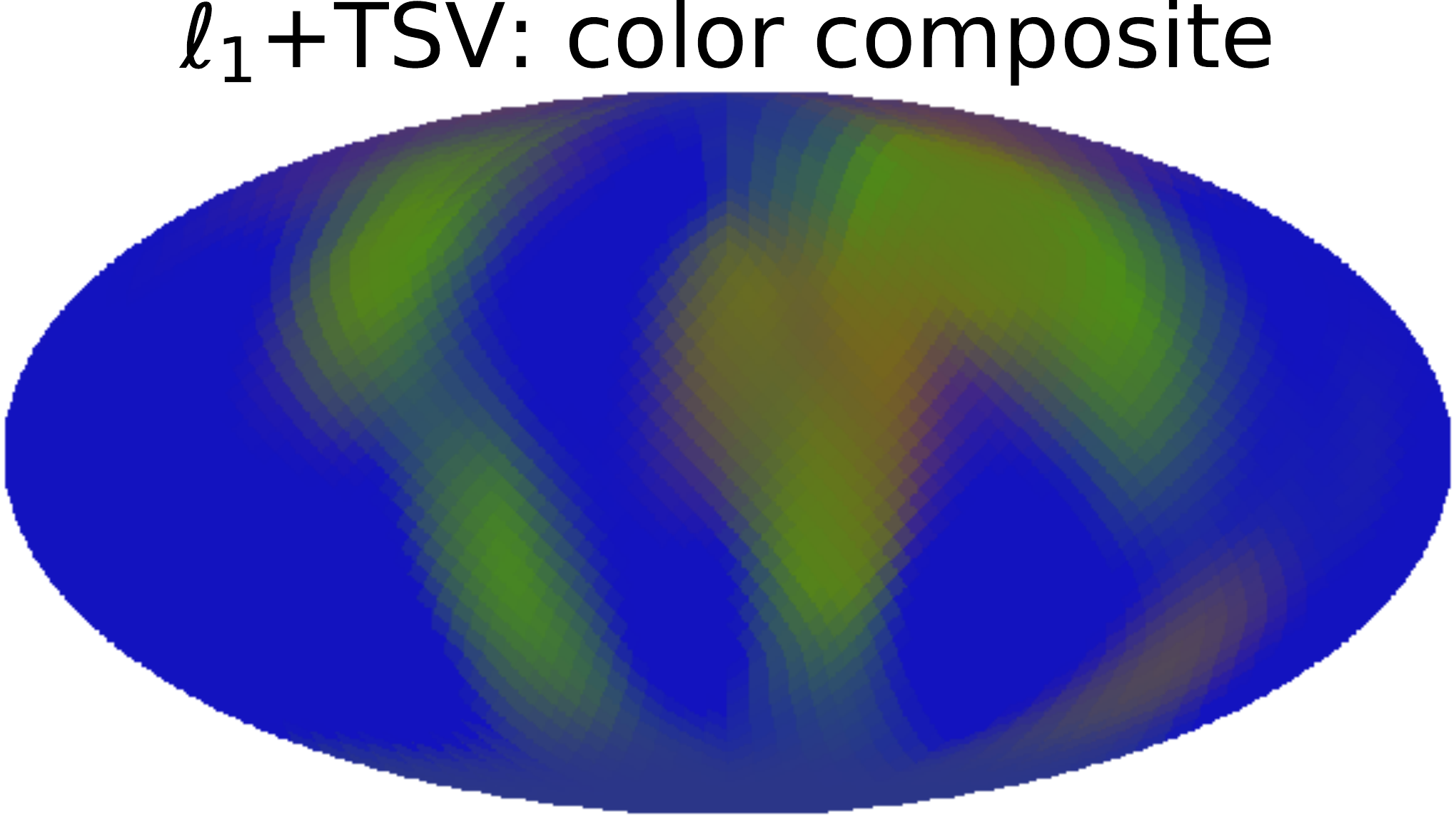}%
    \label{fig:est_A_colormap}%
  }%
  \vspace{-.02\textwidth}
  \hspace{.05\textwidth}
  \subfigure[]{%
    \includegraphics[width=.3\textwidth]{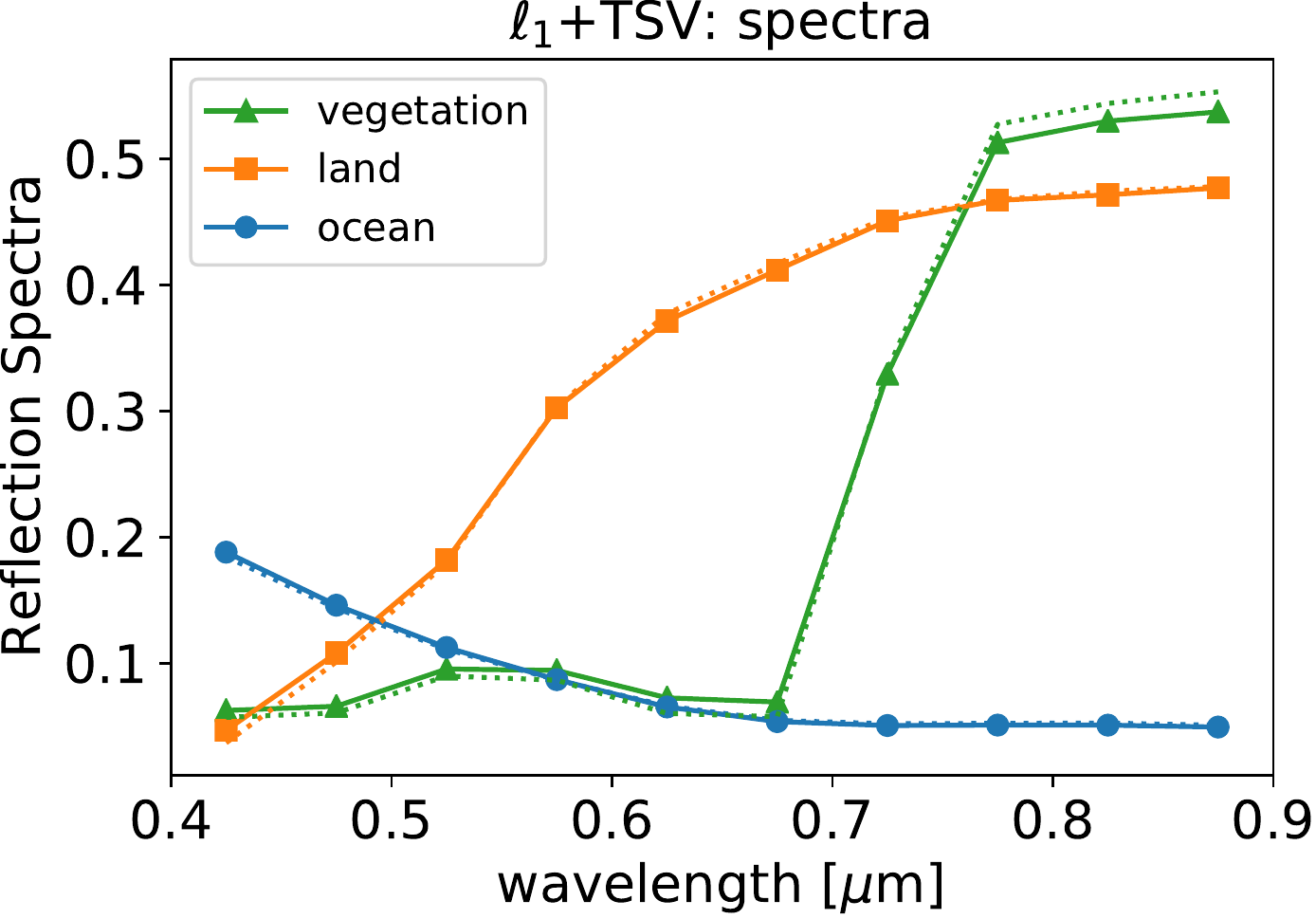}%
    \label{fig:est_X}%
  }
  %%% component %%%
  \subfigure[]{% 
    \includegraphics[width=.25\textwidth]{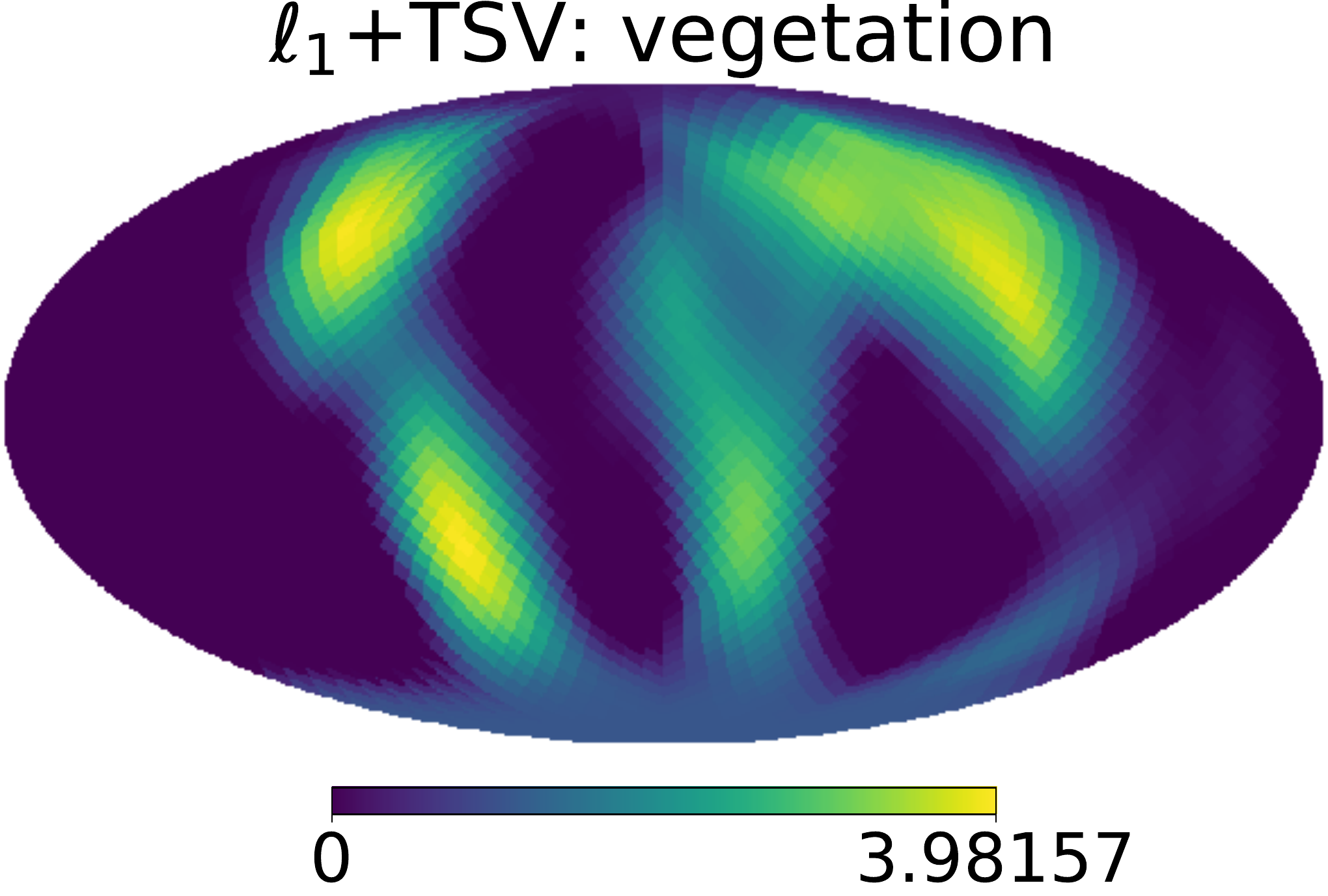}%
    \label{fig:est_A_vegetation}%
  }%
  \hspace{.02\textwidth}
  \subfigure[]{% 
    \includegraphics[width=.25\textwidth]{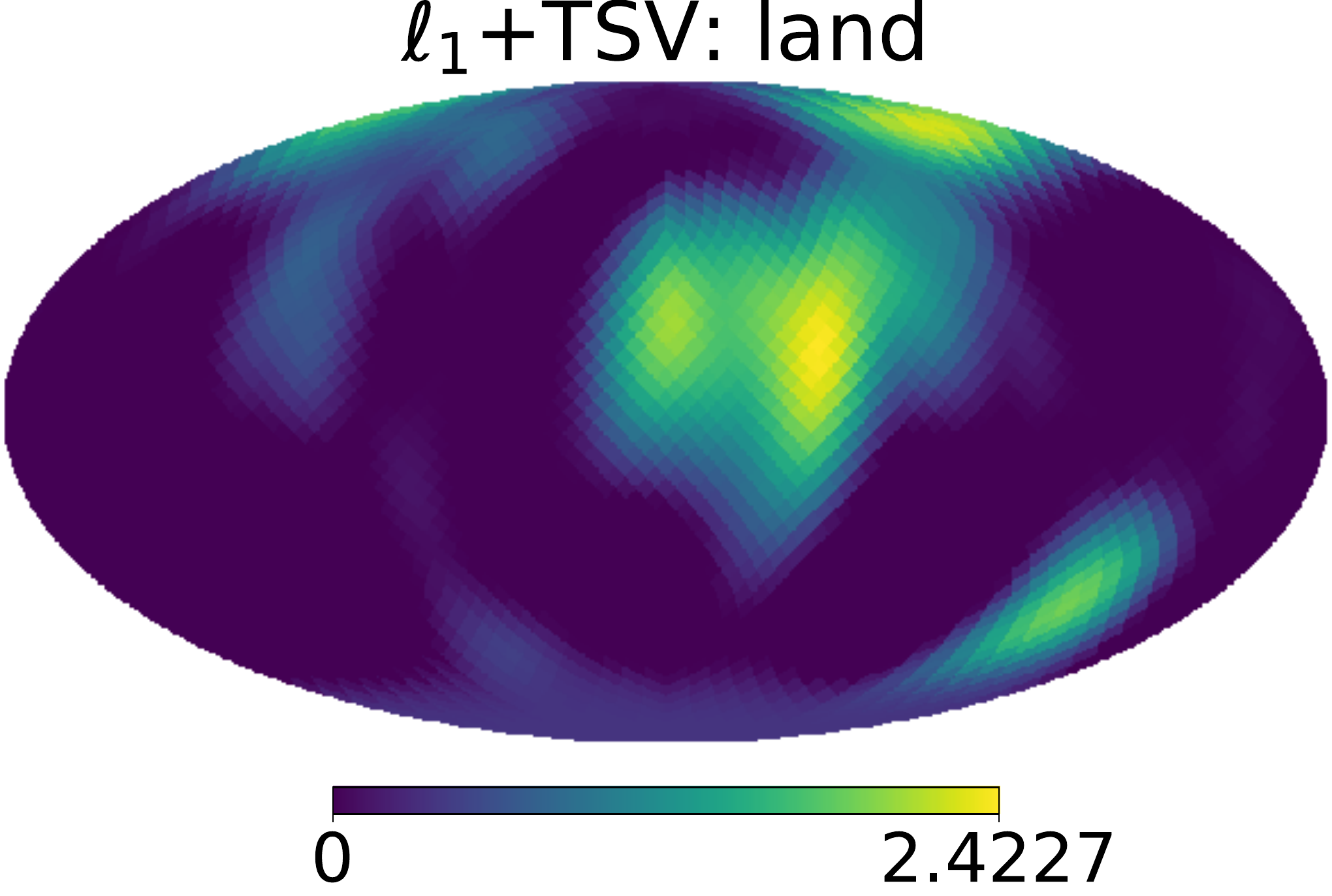}%
    \label{fig:est_A_land}%
  }%
  \hspace{.02\textwidth}
  \subfigure[]{% 
    \includegraphics[width=.25\textwidth]{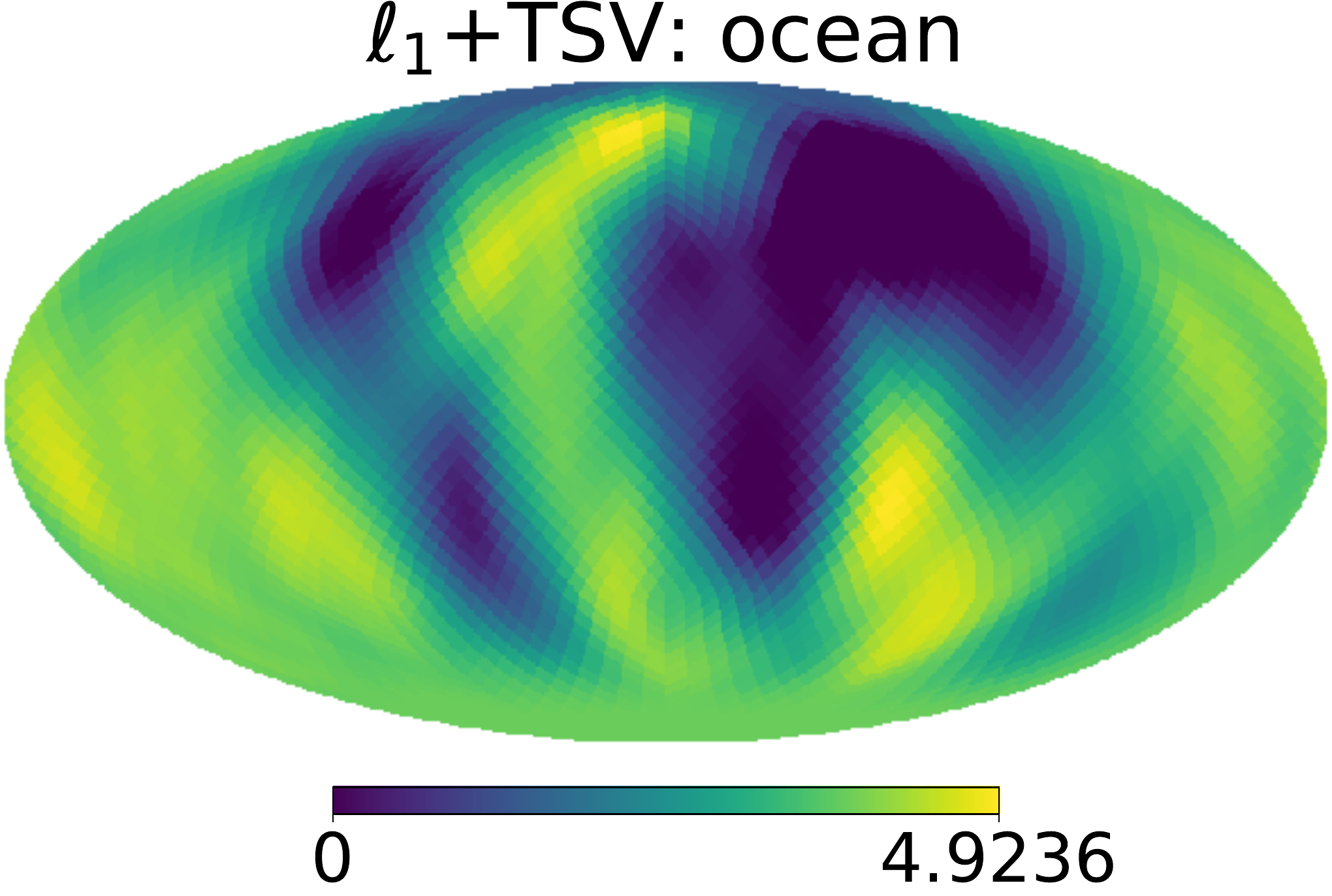}%
    \label{fig:est_A_ocean}%
  }%
  \caption[]{Surface map and spectra inferred by spin-orbit unmixing with the $\ell_1$-norm and TSV regularization \add{using a cloudless Earth model}. (a) Color composite of inferred surface distribution $\hat{A}$. (b) Inferred spectrum $\hat{X}$. Solid and dotted lines denote the inferred and input spectra, respectively. The inferred map for each endmember is displayed in (c) for vegetation, (d) for land, and (e) for the ocean.}
\label{fig:L1TSV_est}
\end{figure}

The normalized surface distribution $\hat{A}$ and spectra $\hat{X}$ are shown in Figure~\ref{fig:L1TSV_est}. The inferred map accurately reproduces the structure of the input. Notably, the inferred spectra are in excellent agreement with the inputs, which is much better than the case of spin-orbit unmixing with Tikhonov regularization \citep{kawahara2020global}.

\begin{figure}
  \centering
  \subfigure[]{% 
    \includegraphics[width=.3\textwidth]{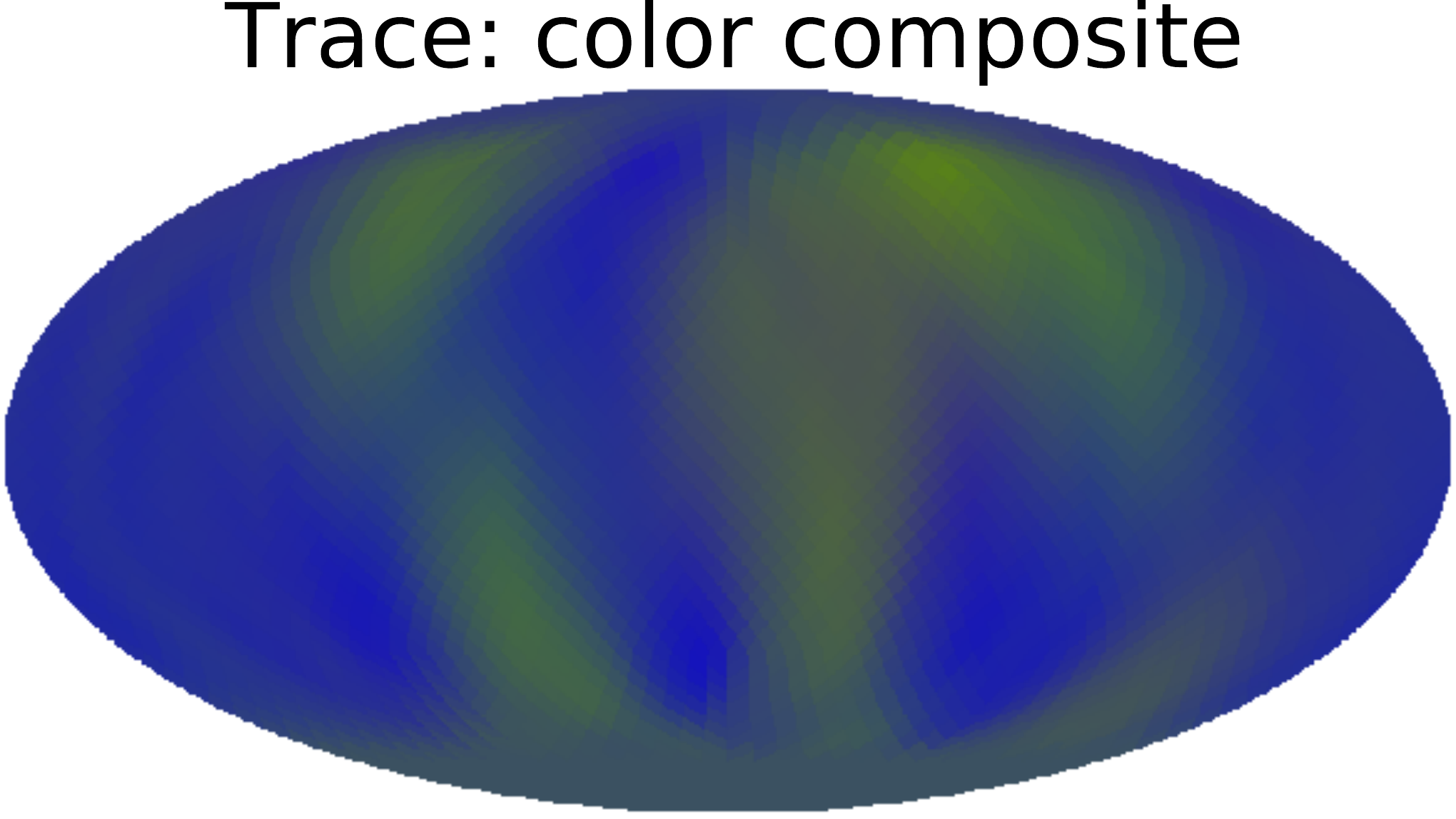}%
    \label{fig:est_Trace_A_colormap}%
  }%
  \vspace{-.02\textwidth}
  \hspace{.05\textwidth}
  \subfigure[]{%
    \includegraphics[width=.3\textwidth]{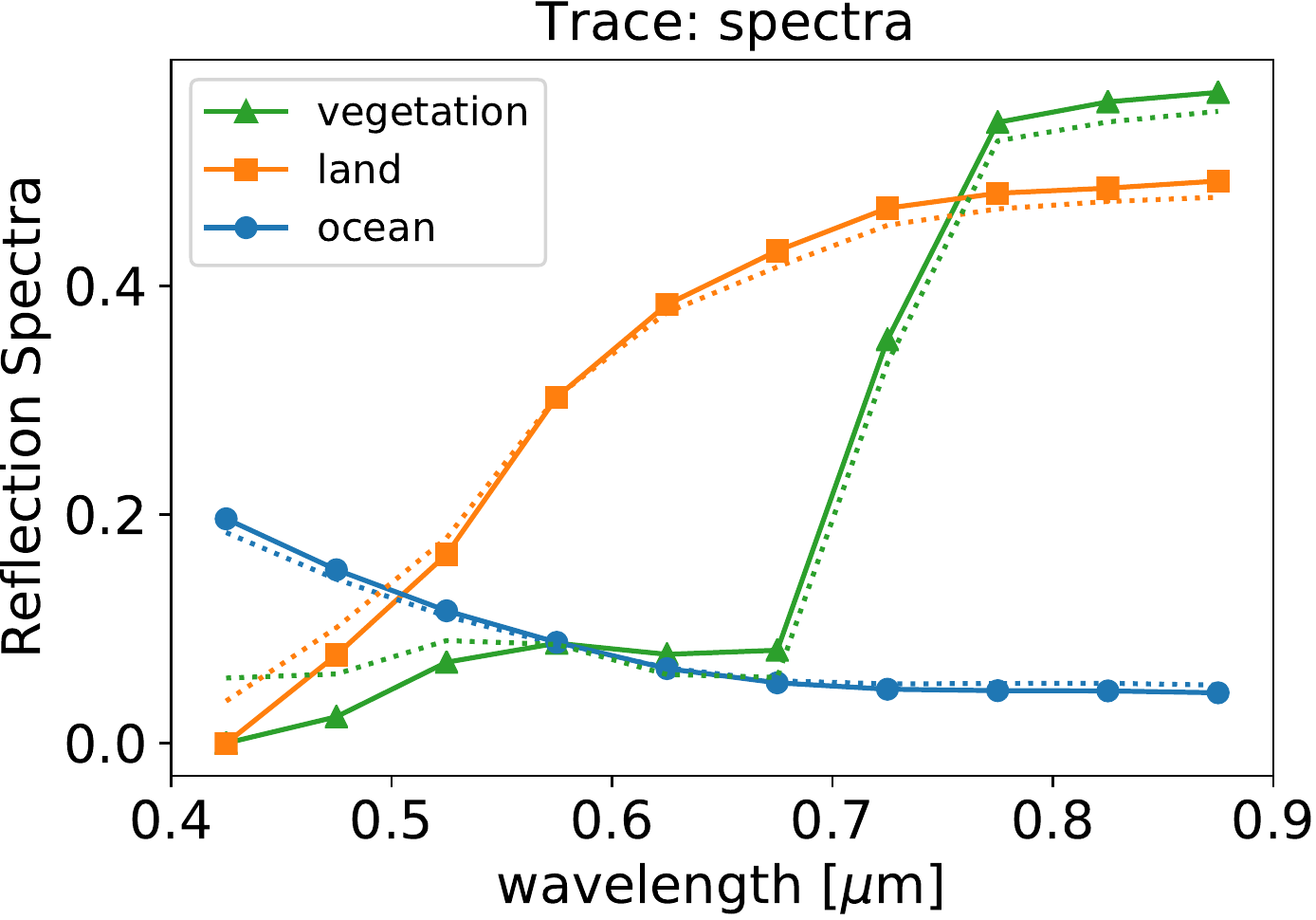}%
    \label{fig:est_Trace_X}%
  }
  \subfigure[]{% 
    \includegraphics[width=.25\textwidth]{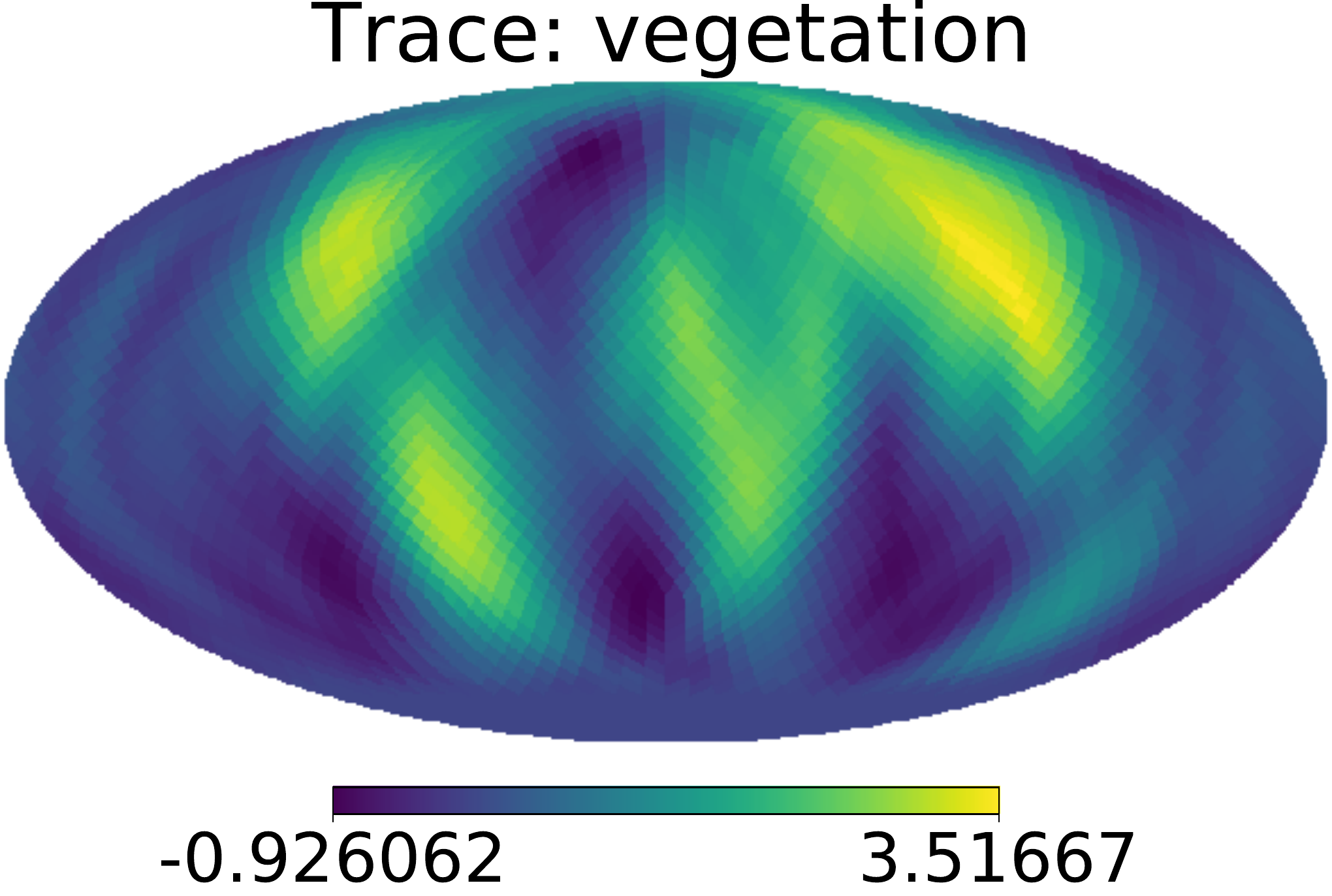}%
    \label{fig:est_Trace_A_vegetation}%
  }%
  \hspace{.02\textwidth}
  \subfigure[]{% 
    \includegraphics[width=.25\textwidth]{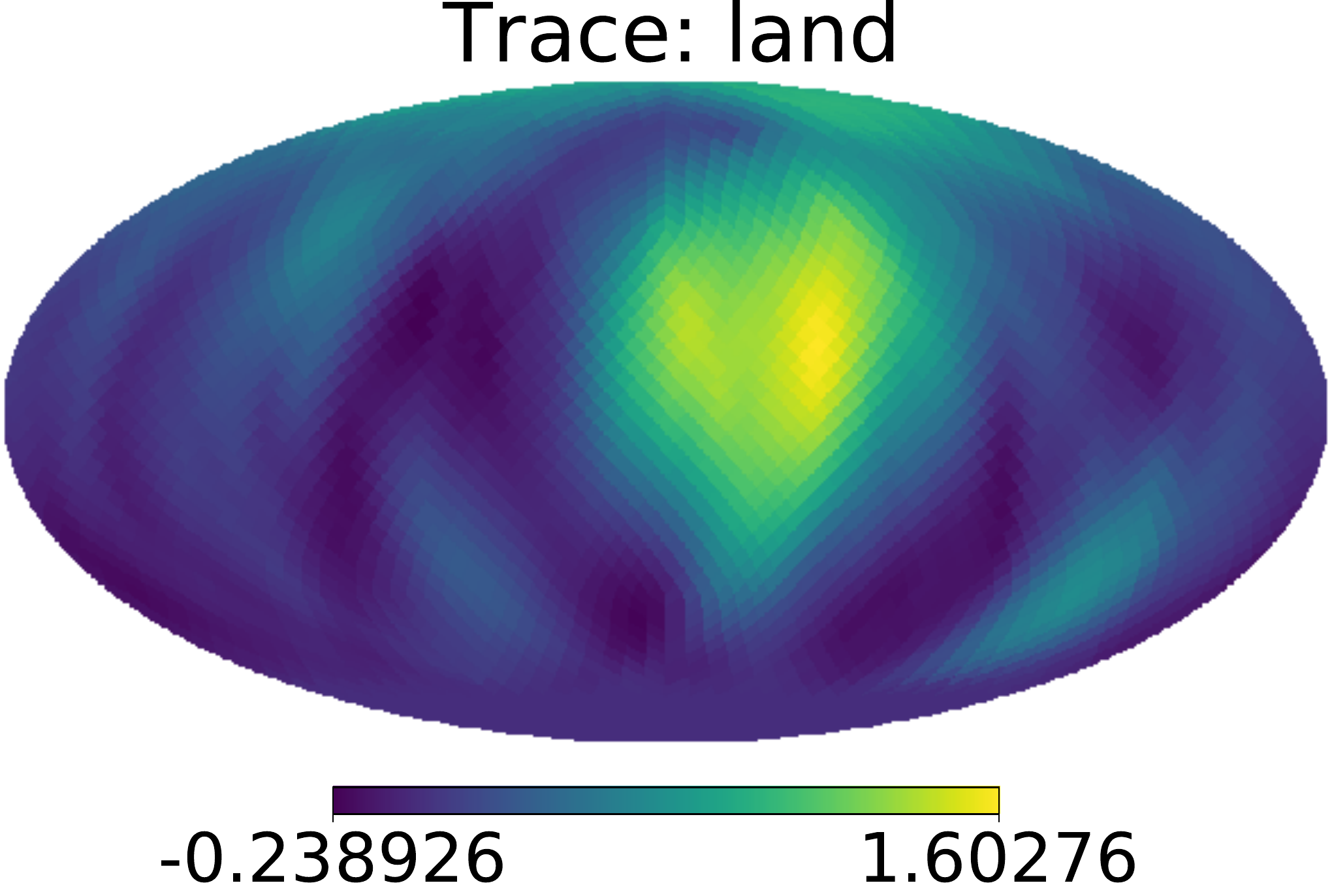}%
    \label{fig:est_Trace_A_land}%
  }%
  \hspace{.02\textwidth}
  \subfigure[]{% 
    \includegraphics[width=.25\textwidth]{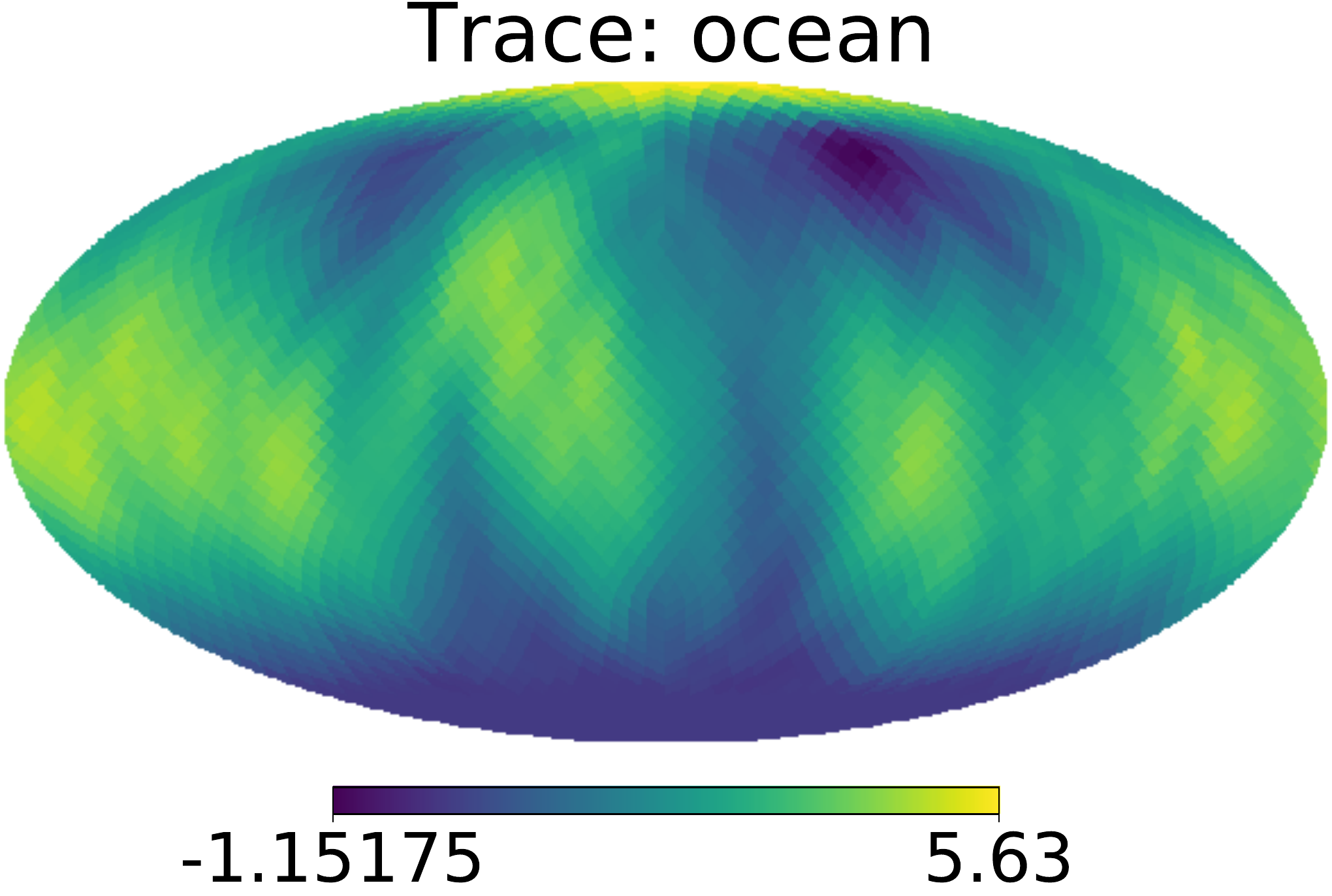}%
    \label{fig:est_Trace_A_ocean}%
  }%
  \caption[]{Same as Figure~\ref{fig:L1TSV_est} but for with trace norm regularization.\\ \\ \\ \\ }
\label{fig:est_Trace}
\end{figure}

%%%%%%%%%%%%%%%%%%%%%%%
\subsection{Results of Spin-Orbit Unmixing with Trace Norm Regularization}\label{sec:SOU_Trace-VRDet_experiment}

Next, we tested spin-orbit unmixing with trace norm regularization by solving Equation~\eqref{eq:SOU_Trace-VRDet_Q} with $\lambda_A=10^{2.1}$, $\lambda_X=10^{3.3}$, and iteration number $N_\mathrm{try}=5\times10^2$.

%Figure~\ref{fig:cost_terms_Trace} shows the values of each term in \eqref{eq:SOU_Trace-VRDet_Q} at iterations, and we can see that $Q_{\mathrm{Tr}}$ converges because the rate of change is short when the number of iterations exceeds $10^2$. The value of the simplex volume minimization term for $X$ increased when the number of iterations exceeded $5\times10^2$. Thus, we choose the solution at $N_\mathrm{try}=5\times10^2$ iterations as the inferred solution.

%\begin{figure}
%\centering
%\includegraphics[width=0.7\textwidth]{cost_terms_Trace.pdf}
%\caption[]{Variation of the values of the objective function and regularization term in iterations of SOU with the trace norm regularization. We selected $N_\mathrm{try}=5\times10^2$ as the number of iterations.}
%\label{fig:cost_terms_Trace}%
%\end{figure}

We used the same evaluation measures as those in \add{Appendix}~\ref{sec:L1TSV_evaluate} to select the regularization parameters $\lambda_A$ and $\lambda_X$. %Figure~\ref{fig:model_Trace_evaluate} shows the values of each evaluation measure calculated by varying these parameters. 
Furthermore, we adopted $\lambda_A$ and $\lambda_X$ with a local minimum value of $\overline{\mathrm{MRSA}}$ similar to that in \add{Appendix}~\ref{sec:L1TSV_evaluate}.

In the same manner as in Section~\ref{sec:SOU_L1TSV-VRDet_experiment}, the normalized surface distribution $\hat{A}$ and spectra $\hat{X}$ are shown in Figure~\ref{fig:est_Trace}.

\subsection{Comparison by Varying The Regularization Term}

\begin{figure*}
  \centering
  %%% true %%%
  \subfigure[]{% 
    \includegraphics[width=.3\textwidth]{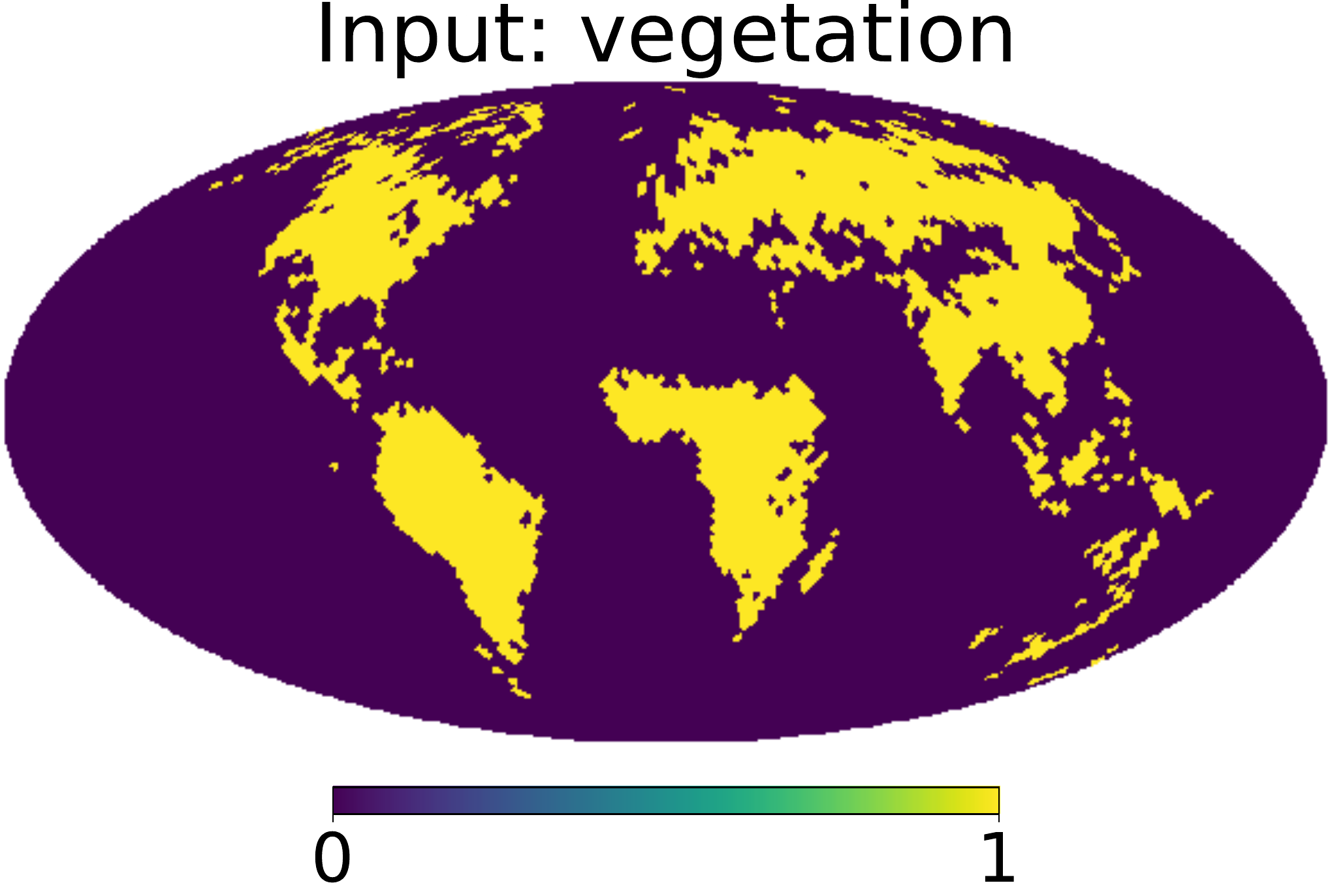}%
    \label{fig:input_map_vegetation_2}%
  }%
  \hspace{.02\textwidth}
  \subfigure[]{% 
    \includegraphics[width=.3\textwidth]{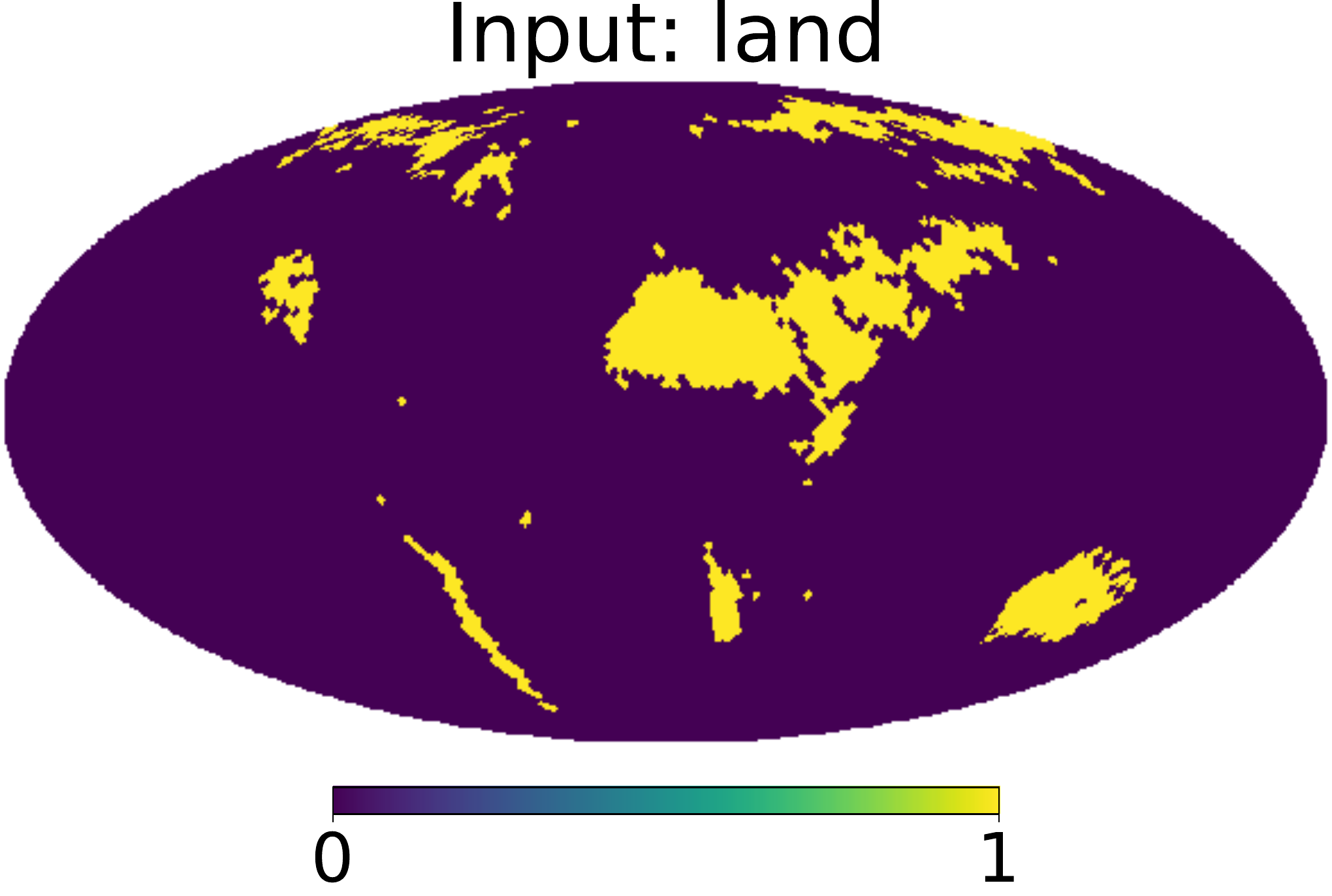}%
    \label{fig:input_map_land_2}%
  }%
  \hspace{.02\textwidth}
  \subfigure[]{% 
    \includegraphics[width=.3\textwidth]{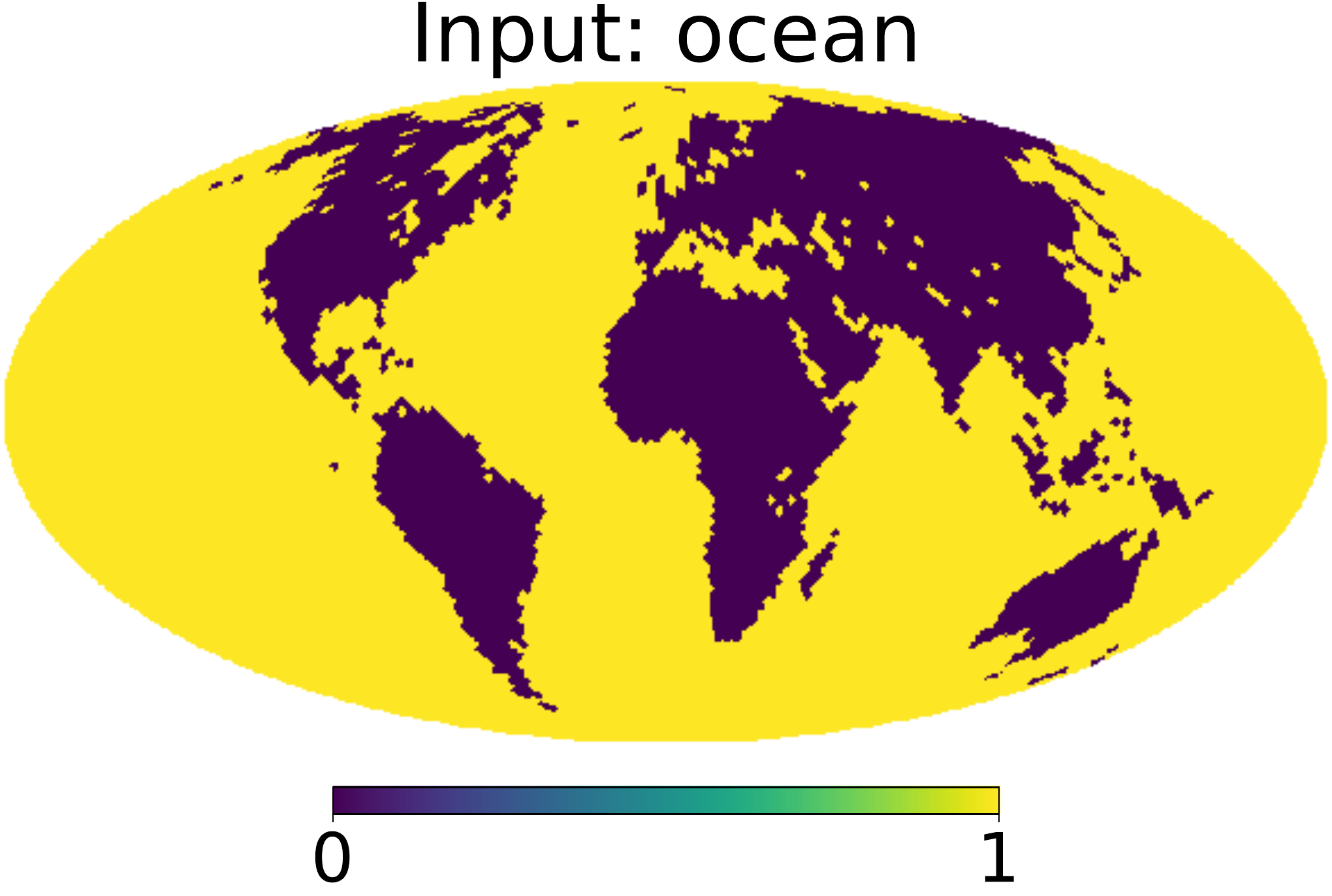}%
    \label{fig:input_map_ocean_2}%
  }%
  \vspace{-0.02\textwidth}
  %%% L1TSV %%%
  \subfigure[]{% 
    \includegraphics[width=.3\textwidth]{L1TSV_est_A_l-35tsv-4x2_vegetation.pdf}%
    \label{fig:est_A_vegetation_compare2}%
  }%
  \hspace{.02\textwidth}
  \subfigure[]{% 
    \includegraphics[width=.3\textwidth]{L1TSV_est_A_l-35tsv-4x2_land.pdf}%
    \label{fig:est_A_land_compare2}%
  }%
  \hspace{.02\textwidth}
  \subfigure[]{% 
    \includegraphics[width=.3\textwidth]{L1TSV_est_A_l-35tsv-4x2_ocean.pdf}%
    \label{fig:est_A_ocean_compare2}%
  }%
  \vspace{-0.02\textwidth}
  %%% Trace %%%
  \subfigure[]{% 
    \includegraphics[width=.3\textwidth]{Trace_est_A21X33_vegetation.pdf}%
    \label{fig:Trace_compare_vegetation}%
  }%
  \hspace{.02\textwidth}
  \subfigure[]{% 
    \includegraphics[width=.3\textwidth]{Trace_est_A21X33_land.pdf}%
    \label{fig:Trace_compare_land}%
  }%
  \hspace{.02\textwidth}
  \subfigure[]{% 
    \includegraphics[width=.3\textwidth]{Trace_est_A21X33_ocean.pdf}%
    \label{fig:Trace_compare_ocean}%
  }%
  \vspace{-0.02\textwidth}
  %%% Trace %%%
  \subfigure[]{% 
    \includegraphics[width=.3\textwidth]{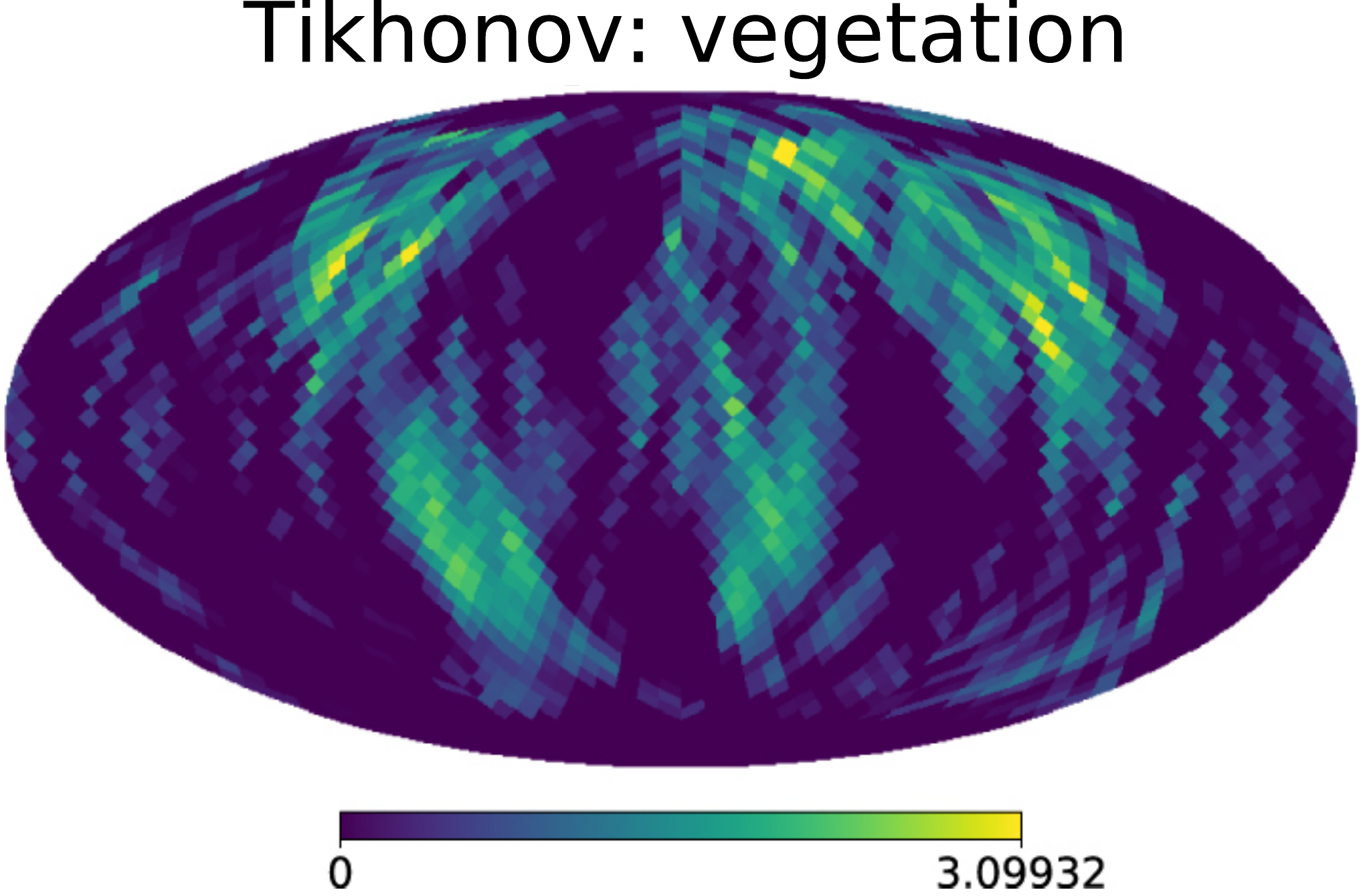}%
    \label{fig:Kawahara2020_vegetation}%
  }%
  \hspace{.02\textwidth}
  \subfigure[]{% 
    \includegraphics[width=.3\textwidth]{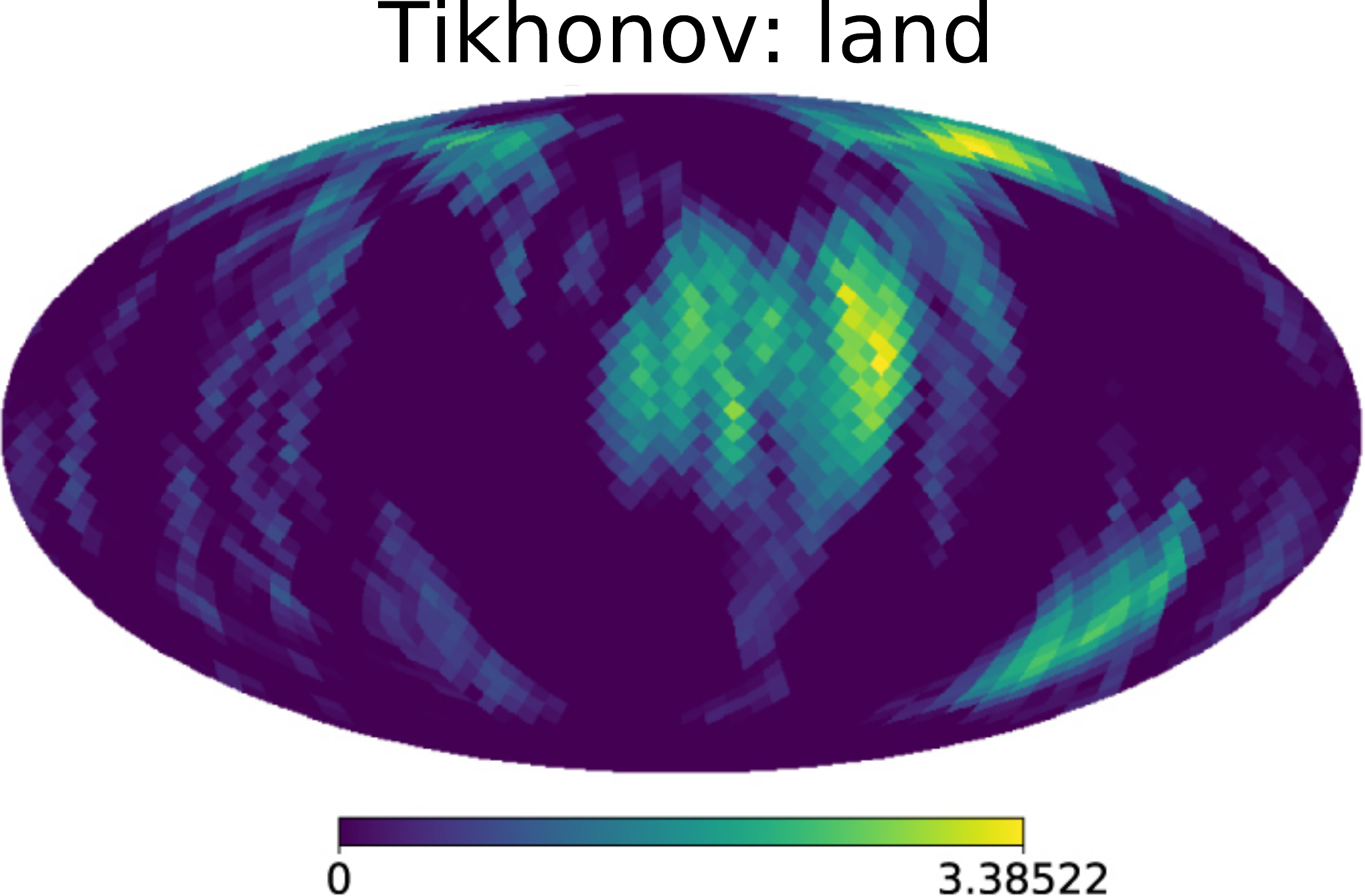}%
    \label{fig:Kawahara2020_land}%
  }%
  \hspace{.02\textwidth}
  \subfigure[]{% 
    \includegraphics[width=.3\textwidth]{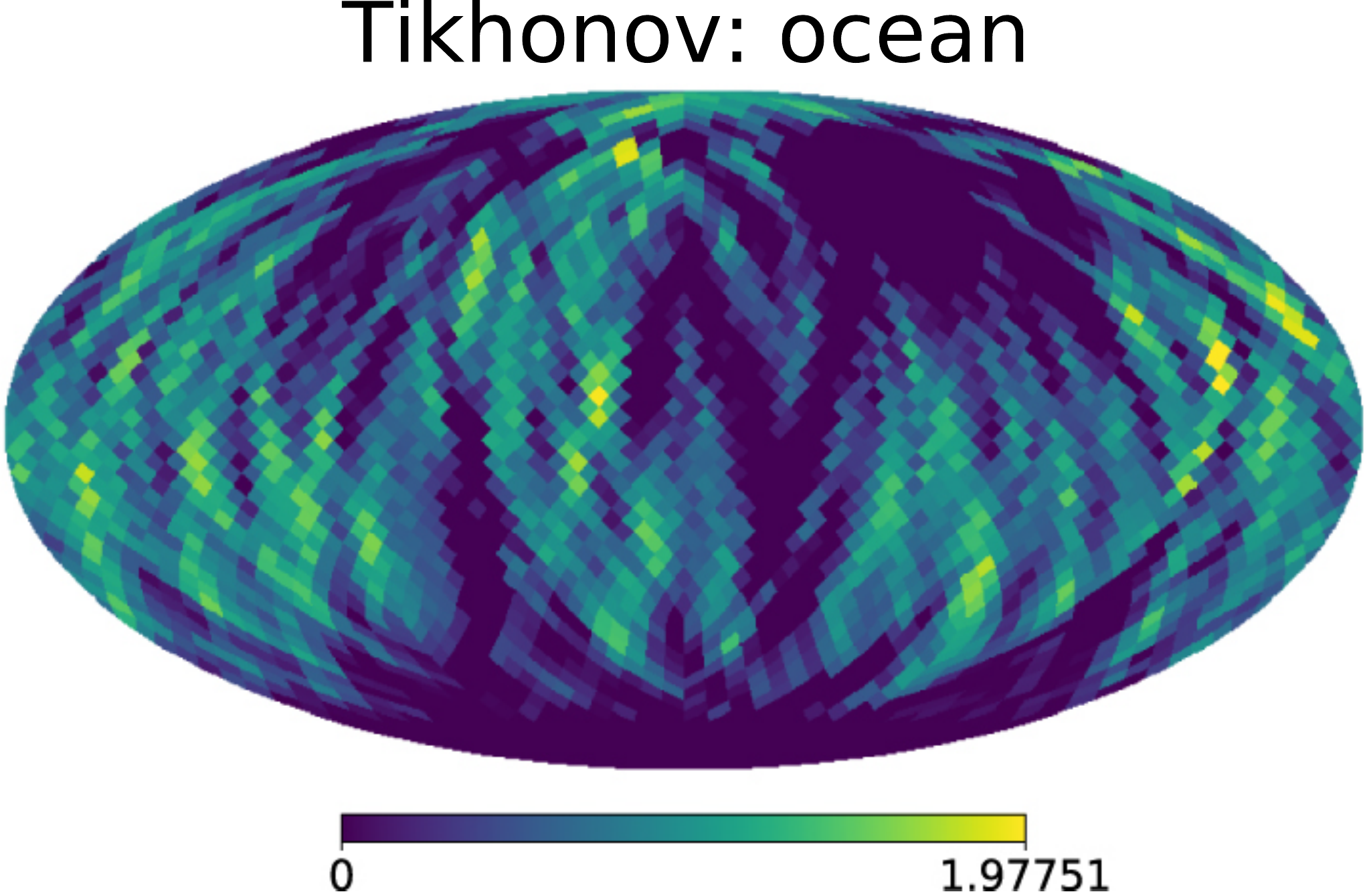}%
    \label{fig:Kawahara2020_ocean}%
  }%
  \vspace{-0.02\textwidth}
  %%% spectra %%%
  \subfigure[]{% 
    \includegraphics[width=.3\textwidth]{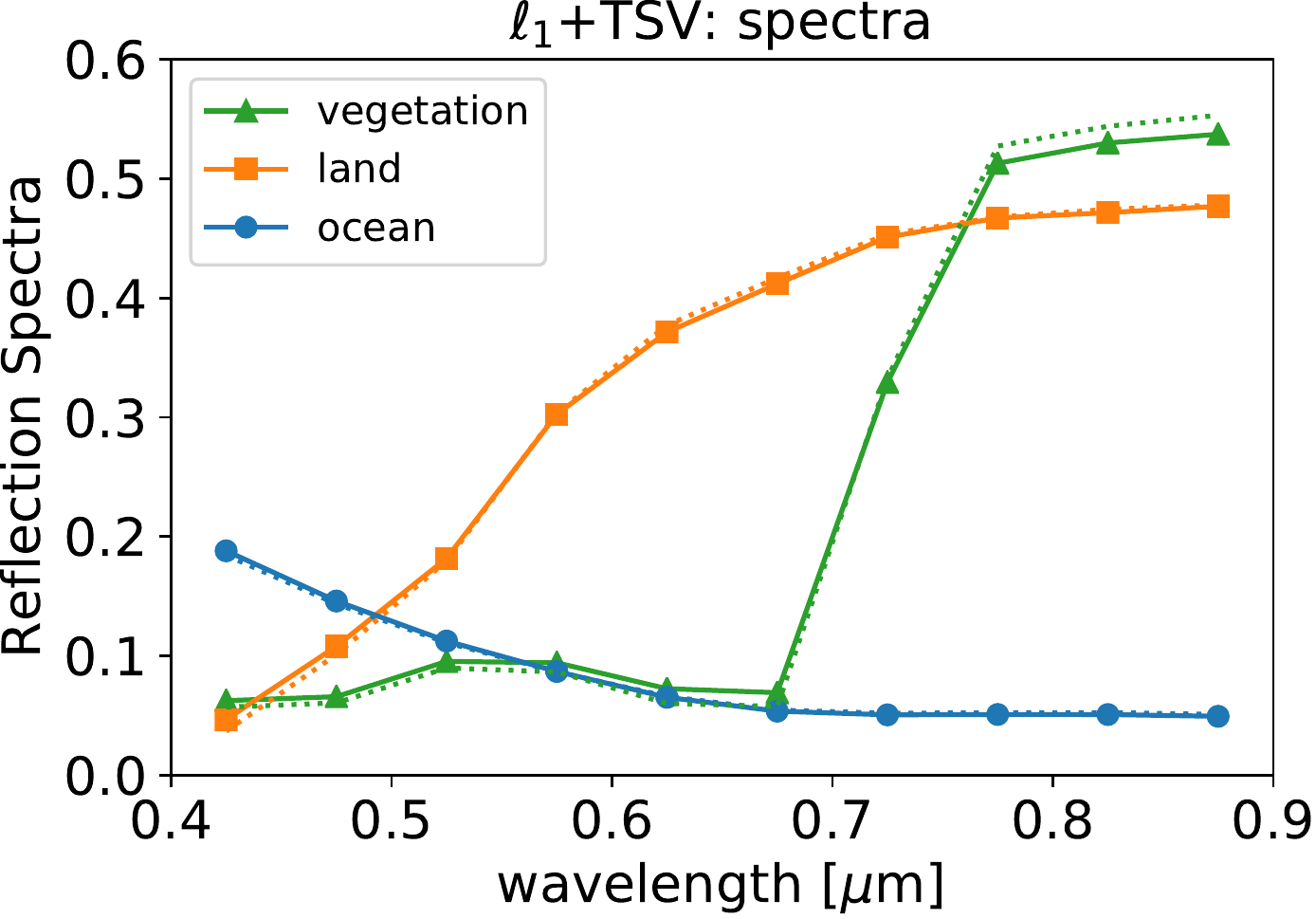}%
    \label{fig:L1TSV_X_compare2}%
  }%
  \hspace{.02\textwidth}
  \subfigure[]{% 
    \includegraphics[width=.3\textwidth]{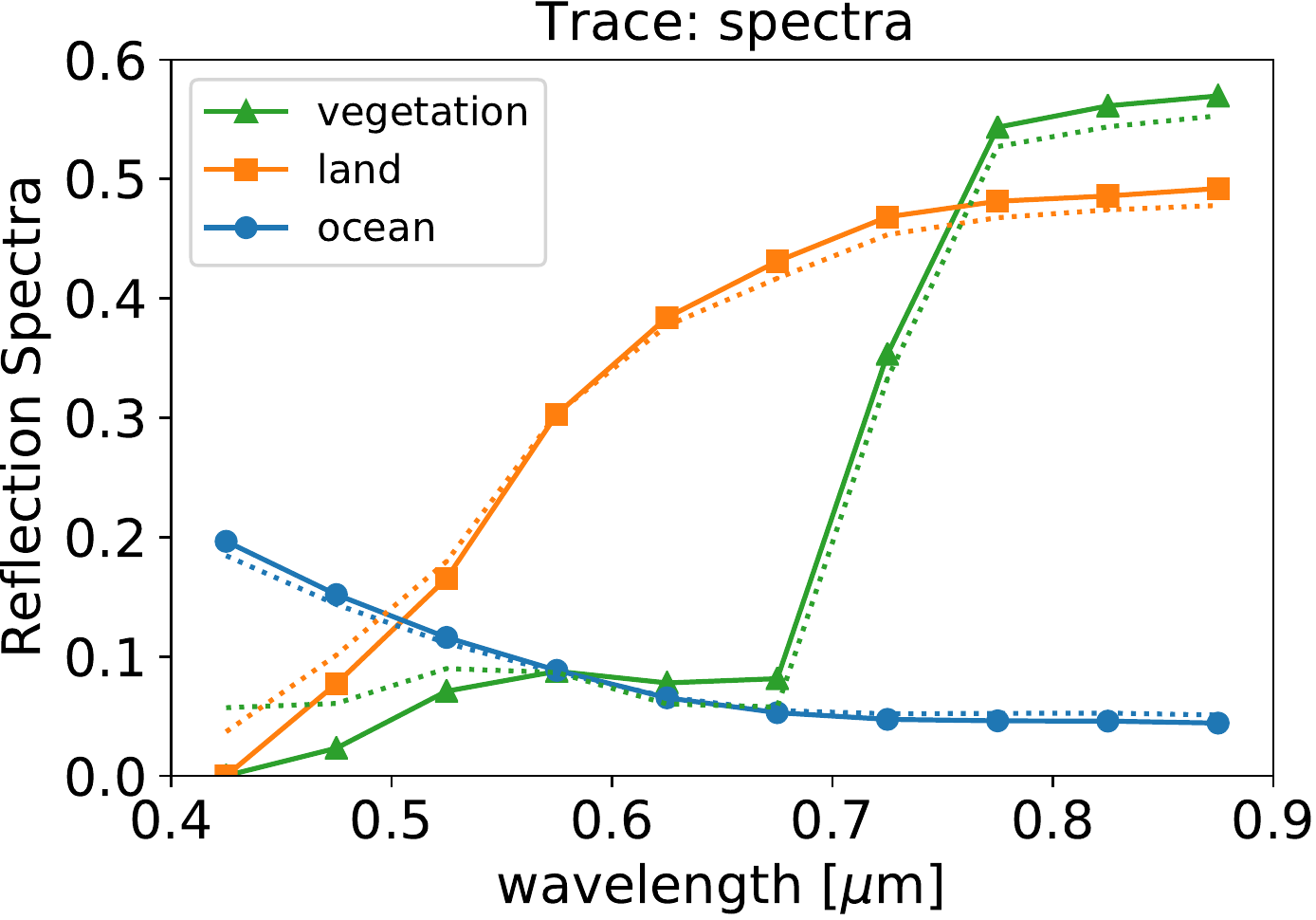}%
    \label{fig:Trace_compare_x}%
  }%
  \hspace{.02\textwidth}
  \subfigure[]{% 
    \includegraphics[width=.3\textwidth]{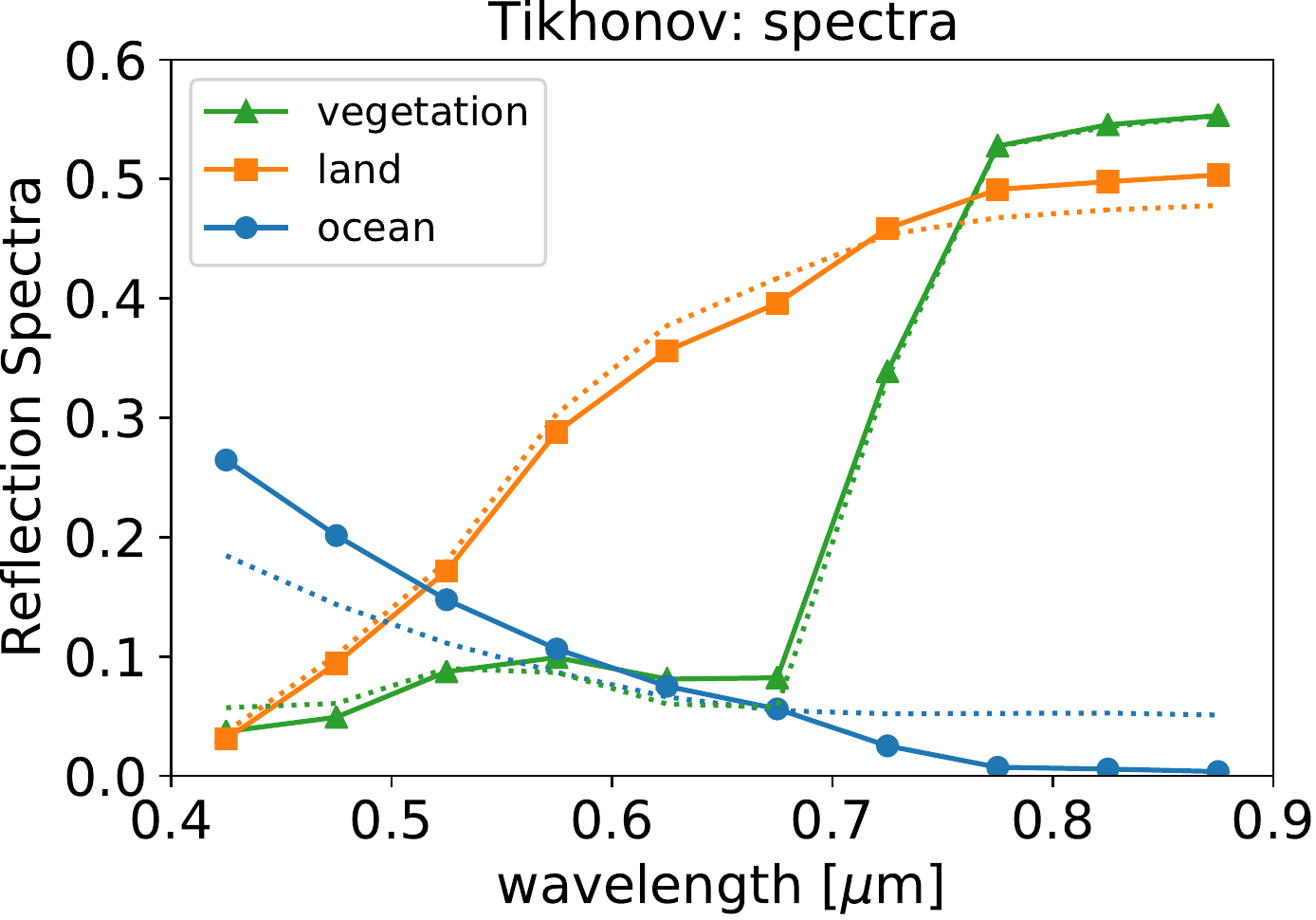}%
    \label{fig:Kawahara2020_X}%
  }%
  \caption[]{(a)--(c) Input surface distributions and (d)(e)(f)(m) inferred solutions by spin-orbit unmixing with $\ell_1$-norm and TSV regularization \add{using a cloudless Earth model}, (g)(h)(i)(n) trace norm regularization, and (j)(k)(l)(o) Tikhonov regularization. The first, second, and third columns of (a)--(l) are vegetation, land, and ocean distributions, respectively. In spectra (m)--(o), the solid lines is the inferred spectra and the dotted lines represents the input spectra. (j), (k), (l), and (o) are based on \citet{kawahara2020global}.}
\label{fig:compare_other_regularization}
\end{figure*}

We compare the inferred solutions with $\ell_1$-norm and TSV regularization, trace norm regularization, and Tikhonov regularization (Figure~\ref{fig:compare_other_regularization}). Figure~\ref{fig:est_A_vegetation_compare2}\subref{fig:est_A_land_compare2}\subref{fig:est_A_ocean_compare2} are the inferred surface distributions by spin-orbit unmixing with $\ell_1$-norm and TSV regularization. We can see less noise in the area with zero values and a more continuous surface with values in each endmember than the case with other regularizations. The sparseness and continuousness of \add{the} inferred map are induced by $\ell_1$-norm and TSV norm regularization, respectively. In general, it is difficult to retrieve the geography that contributes less to the data such as vegetation in Oceania and land in South Africa whichever regularization we use. However, the land distribution in Chile inferred with the $\ell_1$-norm and TSV regularization seems consistent with the input map while one with Tikhonov norm regularization is equivalent to the noise. \add{Additionally, the inferred maps obtained using spin-orbit unmixing with $\ell_1$-norm and TSV regularization (Figure~\ref{fig:est_A_vegetation_compare2}\subref{fig:est_A_land_compare2}\subref{fig:est_A_ocean_compare2}) are smoother than that with Tikhonov regularization (Figure~\ref{fig:Kawahara2020_vegetation}\subref{fig:Kawahara2020_land}\subref{fig:Kawahara2020_ocean}).} 
%This could be because of the differences in properties between TSV and Tikhonov regularizations, both of which tend to exhibit smooth solutions. Furthermore, the non-negative condition can affect the smoothness or noise.
\add{This could be because of the differences in properties between TSV and Tikhonov regularizations, both of which tend to exhibit smooth solutions. TSV regularization induces a smooth map by minimizing the square of the difference between the values of neighboring pixels (described in Appendix~\ref{sec:TV_and_TSV}), while Tikhonov regularization induces a smooth solution by preventing overfitting. Furthermore, the choice of regularization parameters and the non-negative condition can affect the smoothness and noise of the inferred map. The results can be compared using various regularizations or constraints in future studies.}

%On the other hand, concerning the land distribution (Figure~\ref{fig:est_A_land_compare2}), we can see that the values in the land areas of South America and southern Africa, which are narrower, are smaller than those obtained by applying other regularizations. This may be due to the nature of inducing sparsity of the $\ell_1$-norm regularization. Additionally, concerning the vegetation distribution (Figure ~\ref{fig:est_A_vegetation_compare2}), the details around Oceania are hardly recovered. Therefore, while spin-orbit unmixing with $\ell_1$-norm and TSV regularization can induce sparsity across the entire surface distribution, there are still issues with the accuracy of the recovery in the estimation of the details of the surface distribution.

In the \add{inference} of the spectra (Figure~\ref{fig:L1TSV_X_compare2}\subref{fig:Trace_compare_x}\subref{fig:Kawahara2020_X}), we used \add{simplex} volume regularization in all cases. Although regularization terms for the spectra are the same, the inferred spectra are affected by the change in the regularization term for the surface distribution. Specifically, focusing on the ranges of $0.425$--$0.575$ \textmu m and $0.725$--$0.875$ \textmu m, the one using $\ell_1$-norm and TSV regularization (Figure ~\ref{fig:L1TSV_X_compare2}) is the closest to the input spectrum. These results indicate that the surface distribution and spectrum inferred by spin-orbit unmixing with $\ell_1$-norm and TSV regularization are superior to other regularizations.

Let us also note the solution inferred by spin-orbit unmixing with trace norm. The inferred surface distributions (Figure ~\ref{fig:Trace_compare_vegetation}\subref{fig:Trace_compare_land}\subref{fig:Trace_compare_ocean}) capture the features of continuous surfaces such as continents, but \add{pixels} with small values are noisier than those inferred with $\ell_1$-norm and TSV regularization. This may be due to the similarity of the vectors corresponding to the surface distributions of the endmembers as a result of the low-rank matrix induced by using trace norm regularization. In this study, we \add{employed} only trace norm regularization in the \add{inference}. \add{We can consider adding other regularizations or constraints, especially the non-negative condition. The non-negative condition leads to a map in which large parts are zero, as shown in Section~\ref{sec:SOU_L1TSV-VRDet_experiment} (spin-orbit unmixing with $\ell_1$-norm and TSV regularization), \citet{kawahara2020global} (spin-orbit unmixing with Tikhonov regularization), and \citet{kawahara2010global} (spin-orbit tomography with Bounded Variable Least-Squares Solver including the non-negative condition). Therefore, by including the non-negative condition in spin-orbit unmixing with trace norm, we expect improvement in inferences of geography and spectrum.} The application of trace norm regularization is subject to future investigation.

\section{Application to real observed data} \label{chap:dscovr}

In this section, we apply our method with $\ell_1$-norm and TSV regularization to real long-monitoring data of Earth as observed by DSCOVR/Earth Polychromatic Imaging Camera (EPIC) \citep{jiang2018using}. Since 2015, DSCOVR has been continuously observing the dayside of the Earth from the first Sun--Earth Lagrangian point (L1). Given that Earth's rotation axis is tilted relative to its orbit, although it is not the same as that in the case of direct imaging, the observed data contain two-dimensional information about the planet surface. This allowed us to perform two-dimensional mapping \citep{2019ApJ...882L...1F}. \citet{kawahara2020global} inferred the surface distribution and unmixed spectra by applying spin-orbit unmixing with Tikhonov regularization to the DSCOVR data. In our experiment, we applied spin-orbit unmixing with the $\ell_1$-norm and TSV norm described in \add{Section~\ref{sec:SOU_L1TSV-VRDet}} to the DSCOVR data. Following the same setup as that in \citet{kawahara2020global}, we used a quarter of the two-year data (i.e., one in each of the four bins) used in \citet{2019ApJ...882L...1F}. The observed wavelengths are the seven optical bands used in the EPIC instrument (0.388, 0.443, 0.552, 0.680, 0.688, 0.764, and 0.779 \textmu m). There are strong oxygen B and A absorptions at 0.688 and 0.764 \textmu m, respectively.

\begin{figure}
\centering
\includegraphics[width=.45\textwidth]{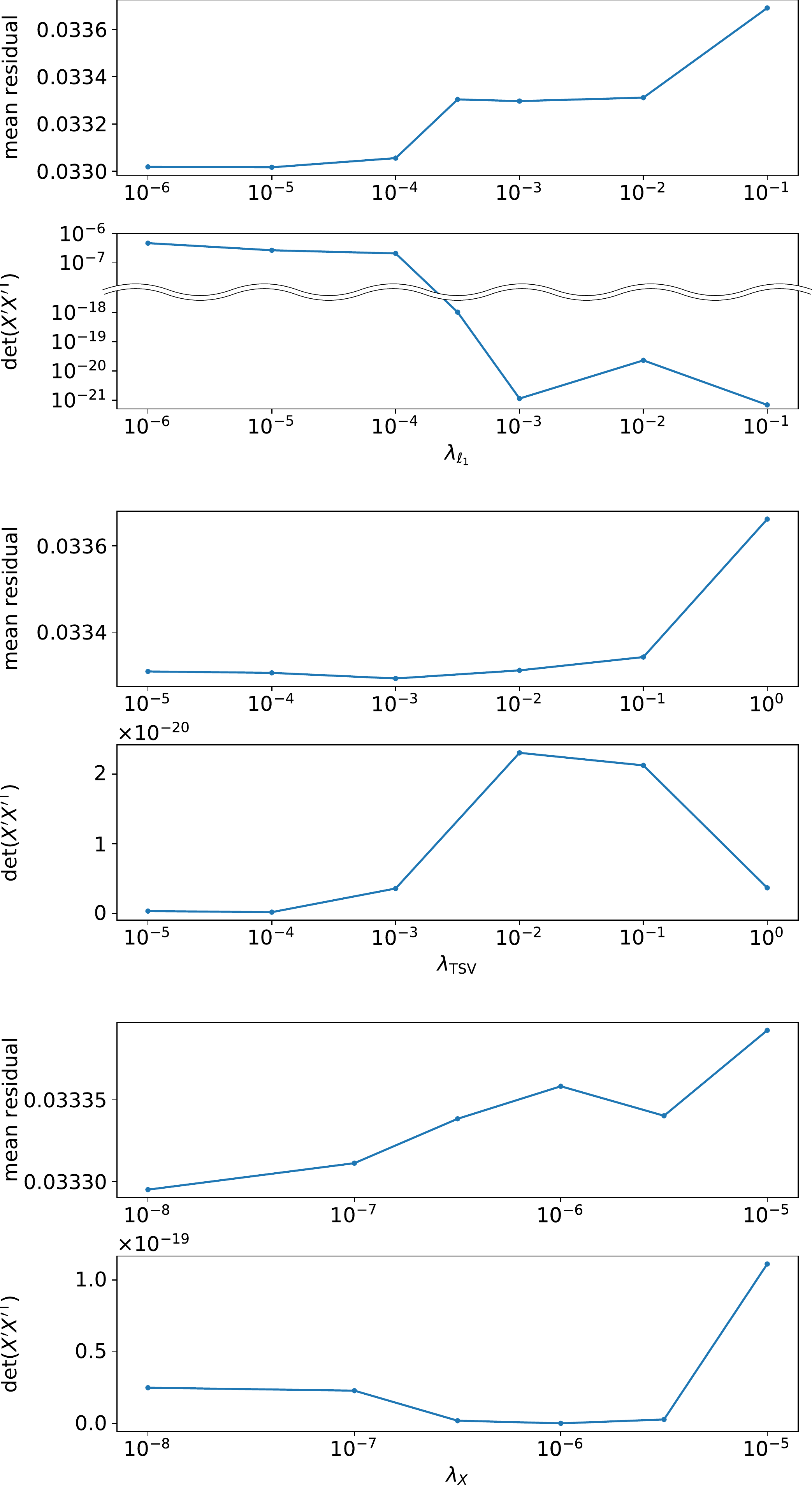}%
\label{fig:DSCOVR_lambda_choice_sub}%
\caption[]{The mean residual and the normalized volume of the spectrum \add{by spin-orbit unmixing with the $\ell_1$-norm and TSV regularization using DSCOVR data}. \add{We select} $\lambda_{\ell_1}=10^{-2}, \lambda_\mathrm{TSV}=10^{-2}$, and $\lambda_X=10^{-7}$.}
\label{fig:DSCOVR_lambda_choice}
\end{figure}

\begin{figure}
\centering
\includegraphics[width=.45\textwidth]{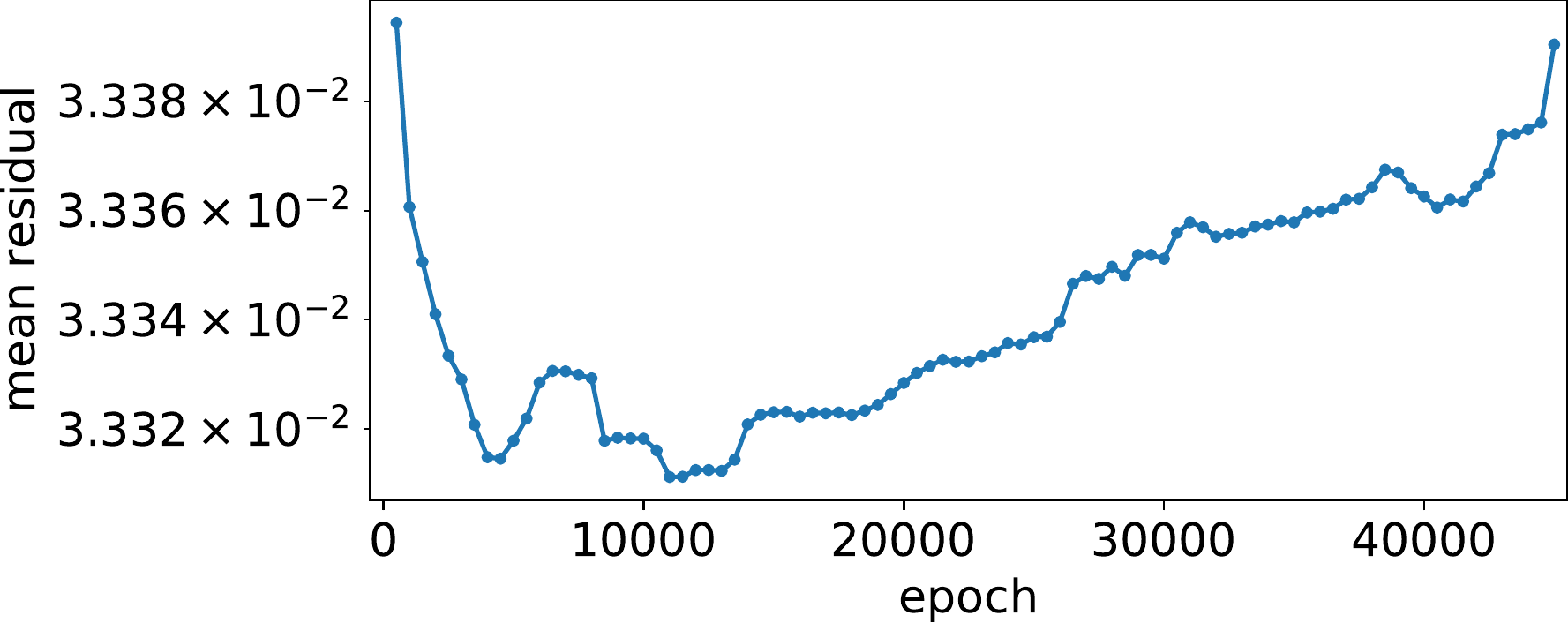}%
\caption[]{Evolution of the mean residual \add{by spin-orbit unmixing with the $\ell_1$-norm and TSV regularization using DSCOVR data} at $\lambda_{\ell_1}=10^{-2}, \lambda_\mathrm{TSV}=10^{-2}$, and $\lambda_X=10^{-7}$}
\label{fig:DSCOVR_mean_residual}
\end{figure}

We selected the regularization parameters using the procedure described in \add{Appendix~\ref{sec:L1TSV_evaluate}, same as Section~\ref{sec:SOU_L1TSV-VRDet_experiment}}. However, it is not possible to calculate $\overline{\mathrm{MRSA}}$ and CPR for actual exoplanet observations because the true surface distribution and reflection spectra are unknown. Therefore, only the mean residual and the normalized volume of the spectrum were used to determine the optimal parameters. Figure~\ref{fig:DSCOVR_lambda_choice} shows the mean residuals and normalized spectral volumes calculated by varying one of $\lambda_{\ell_1}, \lambda_\mathrm{TSV}$, and $\lambda_X$. We selected $\lambda_{\ell_1}=10^{-2}, \lambda_\mathrm{TSV}=10^{-2}$, and $\lambda_X=10^{-7}$ as optimal values because \add{the mean residual significantly increased at a range higher than these values, and the value of normalized spectral volume was sufficiently small ($\sim10^{-20}$) at these values. As shown in Figure~\ref{fig:DSCOVR_mean_residual}, as the calculation proceeds, the mean residual increases at a point. Therefore, we used the inferred solution at the epoch where the mean residual is minimal.}

\begin{figure}
  \centering
  \subfigure[]{% 
    \includegraphics[width=.37\textwidth]{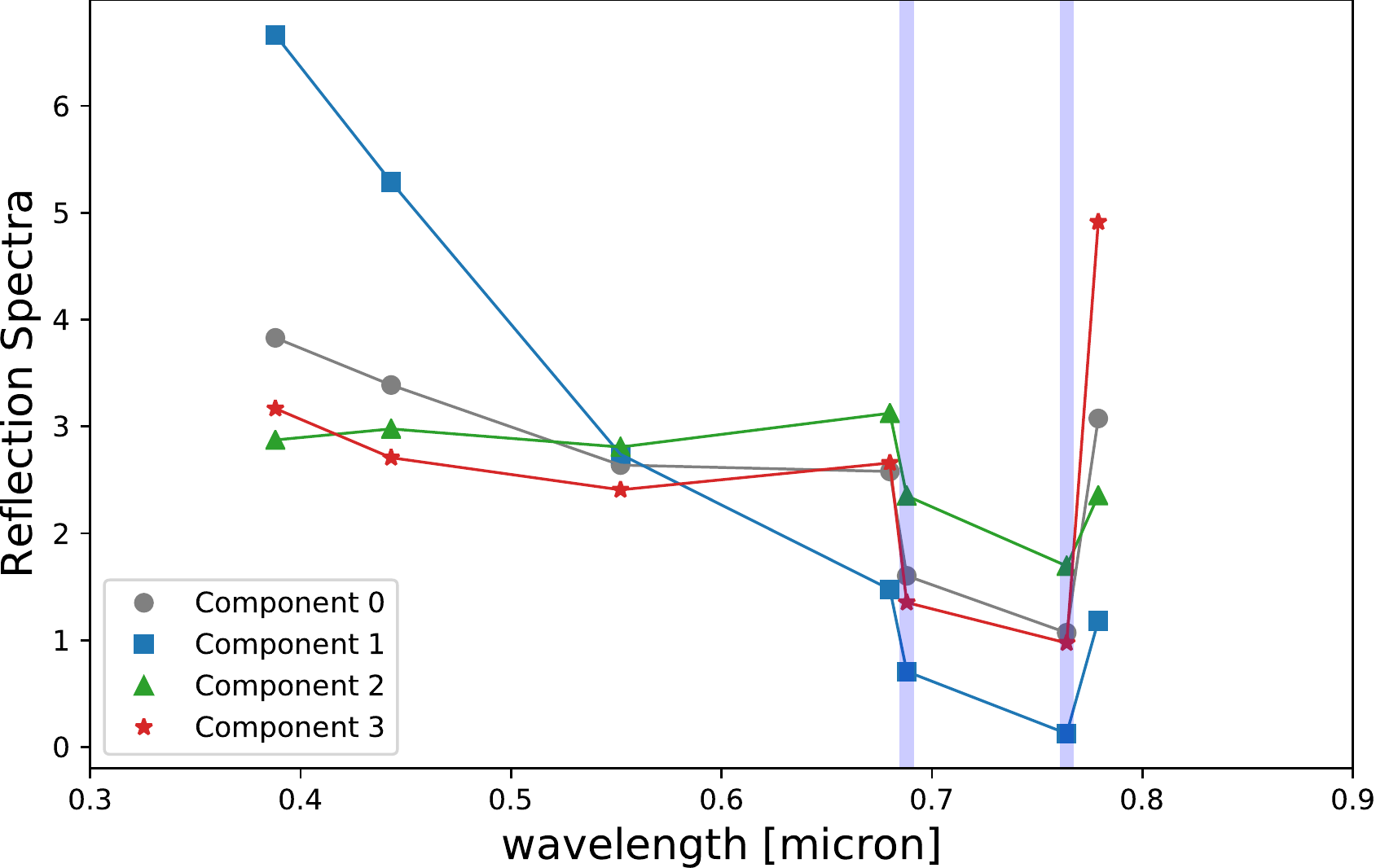}%
    \label{fig:descovr_ref}%
  }%
  \vspace{-.02\textwidth}
  \hspace{.05\textwidth}
  \subfigure[]{%
    \includegraphics[width=.37\textwidth]{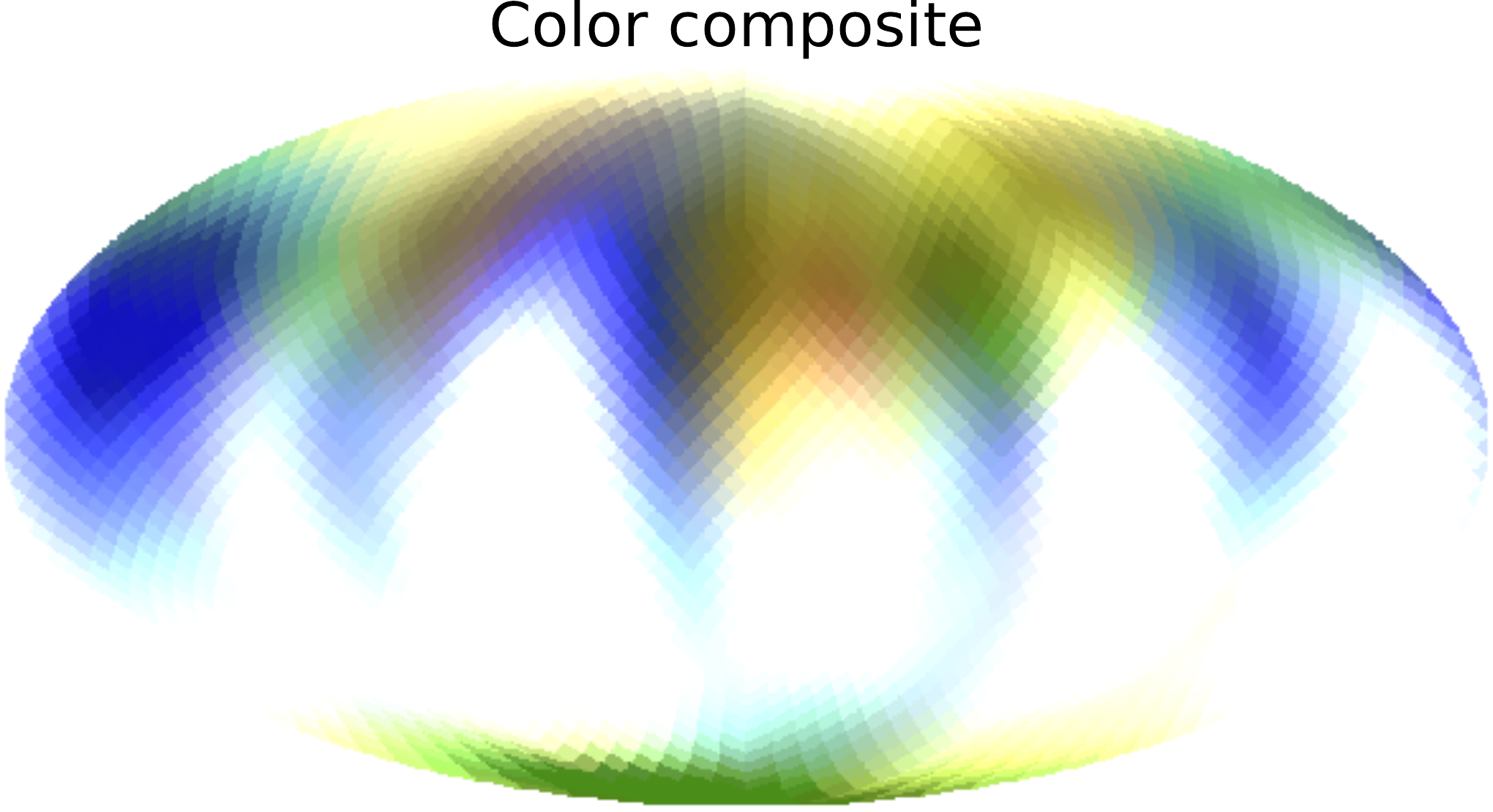}%
    \label{fig:dscovr_classmap4c}%
  }
  %\vspace{-.01\textwidth}
  \hspace{.05\textwidth}
  \subfigure[]{% 
    \includegraphics[width=.37\textwidth]{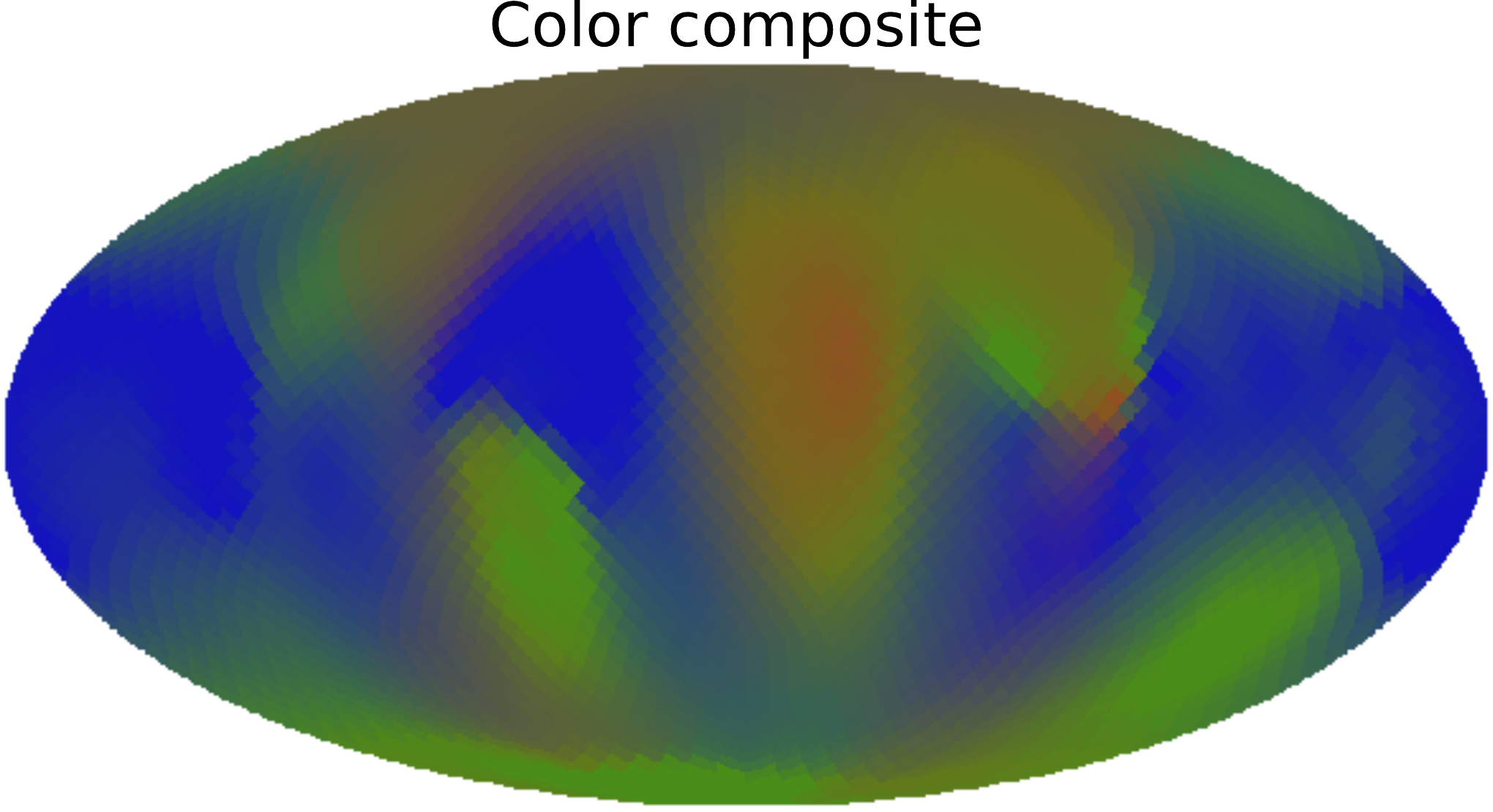}%
    \label{fig:dscovr_classmap3c}%
  }%
  \hspace{.02\textwidth}
  \caption[]{Solutions inferred and normalized by spin-orbit unmixing using DSCOVR data. (a) Spectra, (b) color map, and (c) color map that excludes Component 0.}
\label{fig:dscovr_result}
\end{figure}

Figure~\ref{fig:dscovr_result} shows the inferred spectra, color composite maps, and the map that excludes Component 0 using the aforementioned procedure (we assume that $N_k = 4$). In the unmixed spectra, the strong oxygen B and A absorption features were observed at 0.688 and 0.764 \textmu m, respectively. 
%\add{For Component 0, \citet{kawahara2020global} using Tikhonov regularization resulted in patchy distributions (Figure~\ref{fig:dscovr_L2})\footnote{Compared to \add{Figure~11} in \citet{kawahara2020global}, the red and green in the color map are swapped. This is due to a numerical error in \citet{kawahara2020global}.} that was inconsistent with the real distribution. In contrast, we were able to obtain more continuous maps of cloud or ice components than Tikhonov regularization.}\add{For Component 0, Component 0 is distributed more in the mid-latitude zone than in the high-latitude zone. This is may be because of the low observational weights of the high latitude zone.}

\begin{figure}
  \centering
  \subfigure[]{% 
    \includegraphics[width=.37\textwidth]{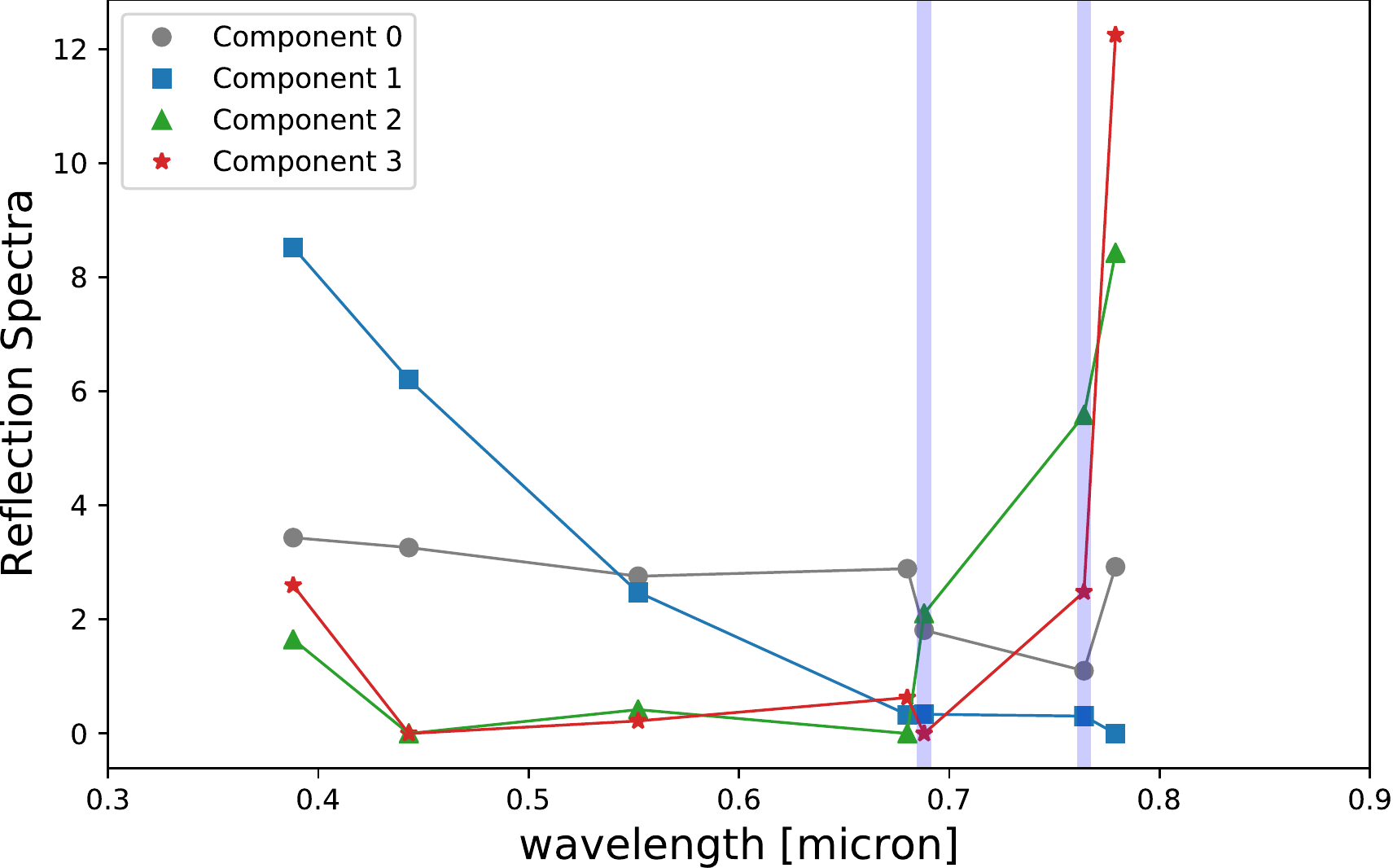}%
    \label{fig:descovr_L2_ref}%
  }%
  \vspace{-.02\textwidth}
  \hspace{.05\textwidth}
  \subfigure[]{%
    \includegraphics[width=.37\textwidth]{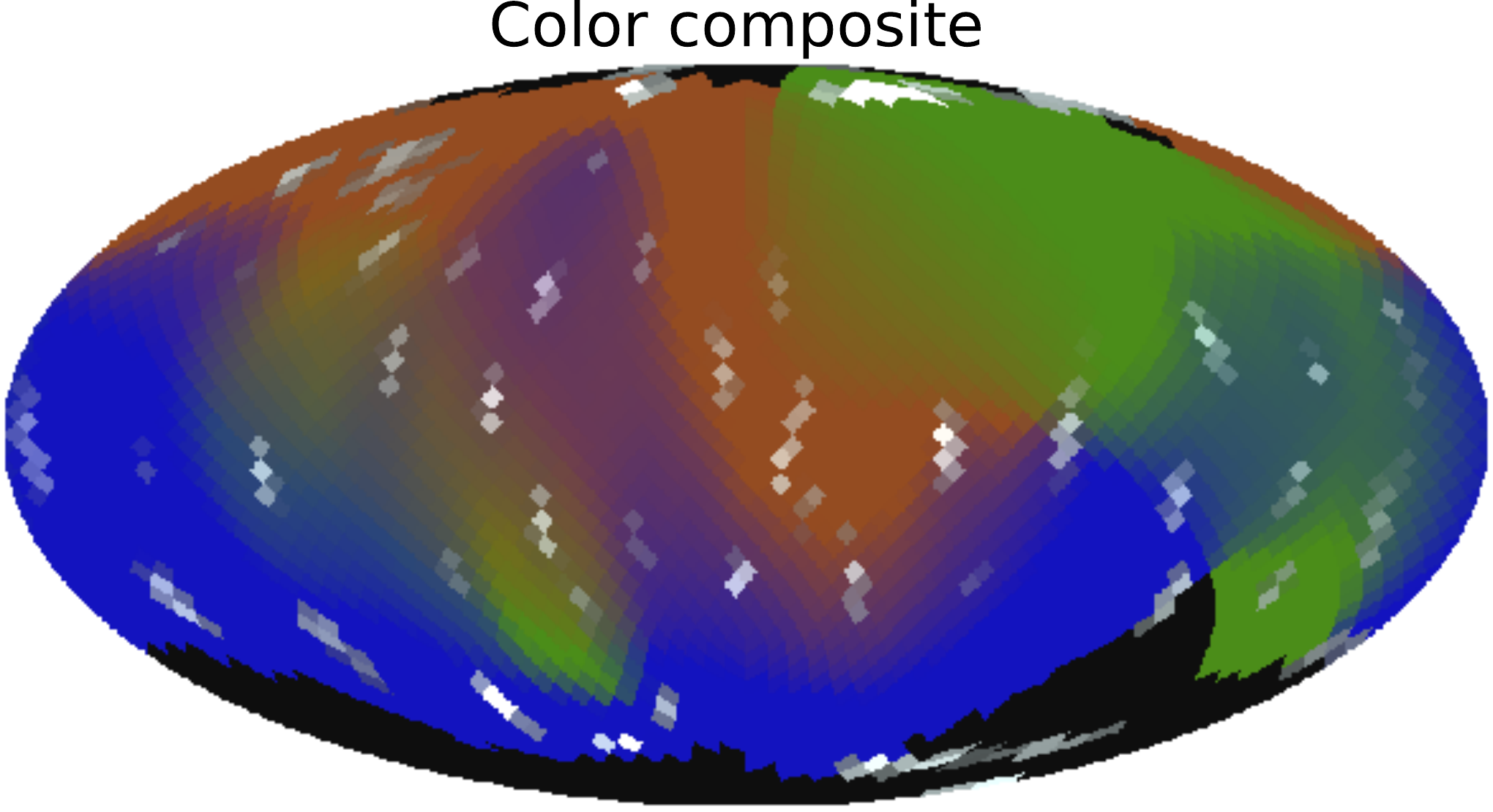}%
    \label{fig:dscovr_L2_classmap4c}%
  }
  \caption[]{Reproduction of the retrieval using Tikhonov regularization, corresponding to \add{Figure~11} in \citet{kawahara2020global}. (a) Spectra and (b) color map.}
\label{fig:dscovr_L2}
\end{figure}

\add{As shown in Figure~\ref{fig:dscovr_classmap4c}, we obtained the distribution of real clouds in the mid-latitudes of the Southern Hemisphere, depicted by Component 0. In addition, there are few in the vicinity of the Sahara Desert. Therefore, we can interpret that Component 0 corresponds to the cloud. \citet{kawahara2020global} using Tikhonov regularization resulted in patchy \add{cloud} distributions (Figure~\ref{fig:dscovr_L2})\footnote{Compared to \add{Figure~11} in \citet{kawahara2020global}, the red and green in the color map are swapped. This is due to a numerical error in \citet{kawahara2020global}.} that were inconsistent with the real distribution. In contrast, we were able to obtain more continuous maps of cloud than that from Tikhonov regularization. Note that real clouds are also distributed in the mid-latitude zone of the Northern Hemisphere (described in Figure 6 in \citet{kawahara2020bayesian}), but we were unable to retrieve the distribution. This may be due to the degeneracy with the continents.}

The map that excludes Component 0 (Figure~\ref{fig:dscovr_classmap3c}) accurately resembles the real continental distribution. \add{The structure of South America and Australia, depicted by Components 2 and 3, is consistent with \citet{aizawa2020global} (single band mapping using sparse modeling).} Component 1 depicts the geography of the ocean\add{. The spectrum of Component 1 also reasonably reproduces that of the ocean. Hence, we can interpret that Component 1 corresponds to the ocean.} 

\add{On the other hand, Components 2 and 3 seem to be degenerate. However, we can see that the geographical features of North America and Australia are depicted by Component 2, and that of the Sahara Desert and Chile by Component 3. Furthermore, South America and Eurasia are depicted by both Components 2 and 3. This is probably because these continents contain both soil and vegetation on the surface. For unmixed spectra,} \add{Component} 2 exhibits larger values at 0.688 and 0.764 \textmu m than Component 3, but smaller for 0.779 \textmu m\add{, that is, Component 3 appears redder than Component 2}. \add{Additionally, the} increase at 0.688 and 0.764 \textmu m might be interpreted that the spectrum of Component 2 captures the red edge of vegetation although the strong oxygen absorption bands make the interpretation difficult. \add{Thus, we can interpret that Component 2 corresponds to vegetation and Component 3 to soil or sands.}

\add{Let us also note the South Pole should be depicted as ice, namely Component 0, but it was depicted as Component 2 in Figure~\ref{fig:dscovr_result} and not visible in Figure~\ref{fig:dscovr_L2}. This inconsistency may be due to the low observational weights on the poles of DSCOVR (presented as Figure~3(b) in \citet{aizawa2020global}). When compared with the inferred map obtained using Tikhonov regularization in \citet{kawahara2020global} (Figure~\ref{fig:dscovr_L2}), the continents are better separated from each other in the map obtained using $\ell_1$-norm and TSV regularization, especially for the Arabian Sea and North Atlantic Ocean.}

%This implies that Components 2 and 3 cannot be sufficiently distinguished. Hence, Eurasia is depicted by Component 3, which should be depicted by Component 2. Moreover, Component 0 is distributed more in the mid-latitude zone than in the high-latitude zone. This is potentially due to the low observational weights of the high latitude zone.

\section{Conclusion}\label{chap:conclusion}

In this study, we introduced sparse modeling ($\ell_1$-norm and TSV regularization) to spin-orbit unmixing for the global mapping of planetary surfaces\add{. For this purpose, we combined and improved the methods proposed by \citet{aizawa2020global} and \citet{kawahara2020global}, and modified the method proposed by \citet{2019ApJ...882L...1F}}. Test calculations on a cloudless toy model of the Earth yielded surface distributions with sparsity and continuity. The inferred unmixed spectra were closer to the input model than those inferred \add{by \citet{kawahara2020global}}. Applying our method to real observation data of the Earth obtained by DSCOVR, we also found that the surface distributions and spectra were reasonably recovered by the current method. We concluded that sparse modeling \add{provides} better \add{inferences} of the surface distribution and unmixed spectra than the method based on Tikhonov regularization. 

This study can be extended in several ways. In this study, we focused on the $\ell_1$-norm and TSV regularization, which prefers sparsity and continuity. However, other choices of regularizations for surface distributions and spectrum can be considered. Furthermore, another type of sparse modeling based on matrices was proposed in previous studies \citep[\add{e.g.,}][]{candes2011robust}, and different types of volume regularization in remote sensing can also be used \citep[e.g.,][]{ang2019algorithms}. Additionally, we assumed the surface distribution of the endmember as static, but we should also consider the dynamical motion of surfaces, especially for clouds. Recently, \citet{kawahara2020bayesian} developed \textit{dynamic} spin-orbit tomography to retrieve the geometry and surface maps in a single band using Tikhonov regularization, and we can extend their method based on sparse modeling. Ultimately, we might be able to combine dynamical mapping \citep{kawahara2020bayesian} and spectral unmixing \citep{kawahara2020global} into \textit{dynamic spin-orbit unmixing} to solve the dynamical motions of planetary surfaces. These issues will be addressed in future research.

The authors are indebted to the DSCOVR team for making the data publicly available. We are grateful to Siteng Fan and Yuk L. Yung for providing the processed light curves and their geometric kernel from the DSCOVR dataset. We are also grateful to Kento Masuda and Shiro Ikeda for insightful discussions. \add{We would also like to thank the anonymous reviewer for an attentive reading and fruitful suggestions.} This study was supported by JSPS KAKENHI Grant No.\ JP18H04577, JP18H01247, JP20H00170, JP21H04998 (H.\ K.\ ), JP22000005, JP15H02063, and JP18H05442 (M.\ T.\ ). A.\ K.\ was also supported by JST SPRING, Grant Number JPMJSP2108. This study was supported by the JSPS Core-to-Core Program Planet$^2$ and SATELLITE Research from the Astrobiology Center (H.\ K.\ ). 

\appendix

\section{Spectral Unmixing}\label{sec:spectral_unmixing}
In this section, we review spectral unmixing, which originates from remote sensing techniques. First, we consider hyperspectral images, which are targets of spectral unmixing. A hyperspectral image has dimensions in the spatial direction and also in the wavelength direction \citep{bioucas2013hyperspectral}. Each pixel in the image contains multiple components (e.g., vegetation, land, and ocean). While we term it a pure pixel that a pixel \add{contains} only a single component, we term it a mixel that contains multiple components due to the observational resolution, and each component is termed an endmember. Decomposition of the observed image into the spectra of the endmembers and their abundance is termed as spectral unmixing. 

We now consider spectral unmixing with non-negative matrix factorization (NMF) \citep{paatero1994positive}. Let $N_j$, $N_k$, and $N_l$ denote the number of pixels of the image, endmembers, and wavelengths of observation, respectively. The hyperspectral image obtained from the observation is $M\in\mathbb{R}^{N_j\times N_l}$, where $M_{jl}\coloneqq m(r_j,\lambda_l) $ represents the observational data at position $r_j$ and wavelength $\lambda_l$. The linear mixing model is expressed as follows:
\begin{align}
    M=AX, \label{eq:MF}
\end{align}
where $A\in\mathbb{R}^{N_j\times N_k}$ denotes the surface distribution matrix, and $X\in\mathbb{R}^{N_k\times N_l}$ denotes the endmember matrix. Here, $\bm{a}_k$, which denotes the $k$-th column vector of $A$, denotes the surface distribution of the $k$-th endmember; namely, an abundance of $A_{jk}=a_k(r_j)$. $\bm{x}_k$, which denotes the $k$-th column vector of $X^\top$, \add{denotes} the reflection spectrum of the $k$-th endmember $X_{kl}=x_k(\lambda_l)$. The problem that is described by matrices, such as spectral unmixing, is expressed as optimization problem:
\begin{align}
    \minimize_{A,X} \frac{1}{2}\|M-AX\|_\mathrm{F}^2,
\end{align}
where $\|\cdot\|_\mathrm{F}$ denotes the Frobenius norm defined as  
\begin{align}
    \|Y\|_\mathrm{F}\coloneqq \sqrt{\sum_i\sum_j Y_{ij}^2}.
\end{align}
The optimization problem of adding the constraint that the entries of each matrix are non-negative is NMF.
\begin{align}
    \minimize_{A,X} \frac{1}{2}\|M-AX\|_\mathrm{F}^2 \subjectto A_{jk}\ge 0, X_{kl}\ge 0. \label{eq:unmixing_NMF}
\end{align}

On the other hand, we can generate matrices $A'$ and $X'$ using the regular matrix $P\in\mathbb{R}^{N_k\times N_k}$ in \eqref{eq:MF}:
\begin{align}
    M&=AX=A'X', \\
    A'&=AP, \\
    X'&=P^{-1}X.
\end{align}
Thus, in general, $A$ and $X$ that satisfy \eqref{eq:MF} are not unique. Additionally, NMF is known to be NP-hard, and thus, it is difficult to determine the optimal solution. To address the above problems, it is necessary to add appropriate constraints or regularization terms.
\begin{align}
    \minimize_{A,X} \frac{1}{2}\|M-AX\|_\mathrm{F}^2 +R(A,X) \subjectto A_{jk}\ge 0, X_{kl}\ge 0,
\end{align}
where $R(A,X)$ denotes the regularization term. With respect to NMF, using simplex volume minimization as a regularization term can reproduce the high-resolution spectrum components \citep{craig1994minimum,lin2015identifiability,fu2019nonnegative,ang2019algorithms}. In the simplex volume regularization, $\det(XX^\top)/(N_l!)$, the volume of an $(N_l-1)$-simplex in $(N_l-1)$-dimensional space with vertices$\{\bm{x}_1,\bm{x}_2,\ldots,\bm{x}_{N_l}\}$, is used as the regularization term.
\begin{align}
    \minimize_{A,X} \frac{1}{2}\|M-AX\|_\mathrm{F}^2 +\frac{\lambda_X}{2} \det(XX^\top) \subjectto \add{A_{jk}}\ge 0, X_{kl}\ge 0,
\end{align}
where $\lambda_X$ denotes the regularization parameter.

%%%%%%%%%%%%%%%%%%%%%%%%%
\section{Optimization of A Non-differentiable Function} \label{sec:optimization_algorithm}

Let $f: \mathbb{R}^n \rightarrow \mathbb{R}$ be a function. The optimization problem of obtaining a solution $\bm{x}\in\mathbb{R}^n$ that minimizes $f$ under constraint $\bm{x}\in S \subset \mathbb{R}^n$ is expressed as follows:  
\begin{align}
    \minimize_{\bm{x}\in\mathbb{R}^n} f(\bm{x}) \subjectto \bm{x}\in S \mbox{\,\,\,\,(constrained problem)}.
\end{align}
Function $f$ is termed the objective function. Specifically, when $S=\mathbb{R}^n$, the optimization problem can be re-expressed as follows:  
\begin{align}
    \minimize_{\bm{x}\in\mathbb{R}^n} f(\bm{x}) \mbox{\,\,\,\,(unconstrained problem)}. 
\end{align}
In the following, we consider the objective function $f$ to be a convex function (Definition~\ref{dfn:convex}). Optimization problems for convex functions are extensively studied due to their tractable properties \citep[for example, ][]{rockafellar1970convex}.

If $f$ is a differentiable function, then the update formula for the gradient descent method, which is the simplest update method, can be provided as follows:  
\begin{align}
    \bm{x}_{i+1}=\bm{x}_i - \gamma\nabla f(\bm{x}_i), \label{eq:gradient_descent}
\end{align}
where $\gamma>0$ denotes the parameter indicating the step size. 

%%%%%%%%%%%%%
\subsection{Proximal Point Algorithm}

If $f$ is a non-differentiable function, then we cannot use the gradient descent method because $\nabla f$ does not exist. An algorithm used to solve this problem is the proximal point algorithm \citep{kanamori2016}. Let $\psi$ be a closed proper convex function (Definition~\ref{dfn:proper_convex},~\ref{dfn:closed_func}) that is not necessarily differentiable. We consider the following optimization problem:
\begin{align}
    \minimize_{\bm{x}\in\mathbb{R}^n} \psi(\bm{x}).
\end{align}
The update formula for the proximal point algorithm is expressed as  
\begin{align}
    \bm{x}_{i+1}= \prox (\bm{x}_i \mid \gamma\psi) \coloneqq \argmin_{\bm{y}\in\mathbb{R}^n} \left( \psi(\bm{y})+\frac{1}{2\gamma}\|\bm{y}-\bm{x}_i\|_2^2 \right),
\end{align}
where $\prox(\bm{x}\mid\gamma\psi)$ denotes the proximal operator of $\psi$ (Definition~\ref{dfn:prox_ope}), and its value is unique for any $\bm{x}$ (Proposition~\ref{prp:prox_unique}).

Now, we consider the following function for $\psi$:
\begin{align}
    \psi_\gamma(\bm{x})\coloneqq \min_{\bm{y}\in\mathbb{R}^n} \left( \psi(\bm{y})+\frac{1}{2\gamma}\|\bm{y}-\bm{x}\|_2^2 \right).
\end{align}
Subsequently, the function is expressed as 
\begin{align}
    \psi_\gamma(\bm{x})&=\frac{1}{2\gamma}\|\bm{x}\|_2^2 - \frac{1}{\gamma} \max_{\bm{y}\in\mathbb{R}^n} \left(\bm{x}^\top\bm{y}-\gamma\psi(\bm{y})-\frac{1}{2}\|\bm{y}\|_2^2\right) \\
    &=\frac{1}{2\gamma}\|\bm{x}\|_2^2 - \frac{1}{\gamma} \left( \gamma\psi(\bm{y})+\frac{1}{2}\|\bm{y}\|_2^2\right)^*,
\end{align}
where $ ^*$ is the conjugate function (Definition~\ref{dfn:conjugate_func}). Given that $\gamma\psi(\bm{y})+(1/2)\|\bm{y}\|_2^2$ is a $1$-strongly convex function (Theorem~\ref{thm:strongly_convex_equiv}), its conjugate function is a $1$-smooth function (Theorem~\ref{thm:strong_smooth}). Hence, $\left( \gamma\psi(\bm{y})+(1/2)\|\bm{y}\|_2^2 \right)^*$ is differentiable and $\psi_\gamma(\bm{x})$ is differentiable. We term $\psi_\gamma(\bm{x})$ the Moreau envelope of $\psi$, which smoothens $\psi$. The gradient of $\psi_\gamma(\bm{x})$ is
\begin{align}
    \nabla\psi_\gamma(\bm{x})&=\frac{1}{\gamma}\bm{x}-\frac{1}{\gamma} \argmax_{\bm{y}\in\mathbb{R}^n} \left( \bm{x}^\top\bm{y}-\gamma\psi(\bm{y})-\frac{1}{2}\|\bm{y}\|_2^2 \right) \ \ \mbox{(Corollary~\ref{cor:grad_conjugate_func})}\\
    &=\frac{1}{\gamma}\bm{x}-\frac{1}{\gamma} \argmin_{\bm{y}\in\mathbb{R}^n} \left( \psi(\bm{y})+\frac{1}{2\gamma}\|\bm{y}-\bm{x}\|_2^2 \right) \\
    &=\frac{1}{\gamma}\bm{x}-\frac{1}{\gamma}\prox(\bm{x}\mid\gamma\psi),
\end{align}
then we have
\begin{align}
    \prox(\bm{x}\mid\gamma\psi)=\bm{x}-\gamma\nabla\psi_\gamma(\bm{x}).
\end{align}
Hence, the proximal point algorithm smoothens the objective function prior to applying the gradient descent method.

%%%%%%%%%%%%%%
\subsection{Proximal Gradient Method} \label{sec:prox_grad_method}

Let $f$ be a proper convex function that is differentiable and let $\psi$ be a proper convex function that is not necessarily differentiable. We consider the following optimization problem:
\begin{align}
    \minimize_{\bm{x}\in\mathbb{R}^n} f(\bm{x})+\psi(\bm{x}). \label{eq:prox_grad_f+psi_problem}
\end{align}
An algorithm used to solve this problem is the proximal gradient method \citep{kanamori2016}. It first updates
\begin{align}
    \bm{w}_{i}=\bm{x}_i - \gamma\nabla f(\bm{x}_i)
\end{align}
using the gradient descent method for $f$ and then updates
\begin{align}
    \bm{x}_{i+1}= \prox (\bm{w}_i \mid \gamma\psi)
\end{align}
using the proximal point algorithm for $\psi$. These are summarized as 
\begin{align}
    \bm{x}_{i+1}= \prox \left(\bm{x}_i - \gamma\nabla f(\bm{x}_i) \mid \gamma\psi\right). \label{eq:prox_grad_update}
\end{align}

Next, we consider the optimization problem for a differentiable proper convex function $f$ with non-negative constraints as follows:
\begin{align}
     \minimize_{\bm{x}\in\mathbb{R}^n} f(\bm{x}) \subjectto x_j\ge 0.\ (j=1,\ldots,n) \label{eq:nonneg_opt_prob}
\end{align}
The problem is equivalent to
\begin{align}
     \minimize_{\bm{x}\in\mathbb{R}^n} f(\bm{x})+\delta_+ (\bm{x}),
\end{align}
where $\delta_+$ \add{denotes the indicator function of a non-negative set} defined as 
\begin{align}
     \delta_+(\bm{x}) \coloneqq 
        \begin{cases}  
            0 & ( \bm{x}\ge\bm{0} ) \\
            \infty & ( \mathrm{otherwise} ),
        \end{cases} \label{eq:nonnegative_indicator}
\end{align}
where $\bm{x}\ge\bm{0}$ denotes $\bm{x}\in \{\bm{x}\in\mathbb{R}^n\mid x_j\ge 0\ (j=1,\ldots,n)\}$.
Given that $\delta_+$ is a closed proper convex function (Proposition~\ref{prp:indicator_proper}), the proximal operator of $\delta_+$ can be defined.
\begin{align}
    \prox(\bm{x}\mid\gamma\delta_+) &=  \argmin_{\bm{y}\in\mathbb{R}^n} \left( \delta_+(\bm{y})+\frac{1}{2\gamma}\|\bm{y}-\bm{x}\|_2^2 \right) \\
    &= \argmin_{\bm{y}\ge\bm{0}} \|\bm{y}-\bm{x}\|_2^2 \\
    &= \max \{\bm{x},\bm{0}\},
\end{align}
where $\max$ denotes the element-wise maximum. The update fomula of the proximal gradient method for problem \eqref{eq:nonneg_opt_prob} is written as:
\begin{align}
    \bm{x}_{i+1} &= \max \{\bm{x}_i - \gamma\nabla f(\bm{x}_i),\bm{0}\}. \label{eq:prox_grad_update_nonneg}
\end{align}

%%%%%%%%%%%%%%%%%%%%%%%%%%%%%%%%%%%%%%
\section{Sparse Optimization Problem} \label{sec:sparse}

Sparse modeling is a technique that extracts and analyzes low-dimensional information to explain high-dimensional data. We consider a method to infer the sparse solution to optimization problems for a differentiable proper convex function $f$:
\begin{align}
    \minimize_{\bm{x}\in\mathbb{R}^n} f(\bm{x}).
\end{align}

%%%%%%%%%%%%%%%%%%

\subsection{\texorpdfstring{$\ell_1$}{TEXT}-Norm Regularization}

The sparsity of a solution implies that most of its elements are zero. We then infer with the constraint to reduce the number of non-zero elements. A straightforward method to infer a sparse solution involves solving the following optimization problem:
\begin{align}
    \minimize_{\bm{x}\in\mathbb{R}^n} f(\bm{x}) \subjectto \|\bm{x}\|_0\le C, \label{eq:optimization_L0}
\end{align}
where $\|\bm{x}\|_0\coloneqq \#\{j\mid x_j\neq0\}$ denotes the $\ell_0$-norm of $\bm{x}$, and $C$ denotes a parameter. However, solving \eqref{eq:optimization_L0} incurs a huge computational cost because function values must be calculated continuously while changing the value of $\|\bm{x}\|_0$. A method of optimization using the $\ell_1$-norm instead of the $\ell_0$-norm is proposed as follows \citep{tibshirani1996regression}:
\begin{align}
    \minimize_{\bm{x}\in\mathbb{R}^n} f(\bm{x}) \subjectto \|\bm{x}\|_1\le C, \label{eq:optimization_L1} 
\end{align}
where $\|\bm{x}\|_1 \coloneqq \sum_{i=1}^n |x_i|$ denotes the $\ell_1$-norm. A sparse solution can be inferred by using this method. The $\ell_1$-norm is an approximation of the $\ell_0$-norm because $\|\bm{x}\|_1$ is a convex hull of $\|\bm{x}\|_0$ in $[-1,1]^n$. Moreover, \eqref{eq:optimization_L1} is equivalent to
\begin{align}
    \minimize_{\bm{x}\in\mathbb{R}^n} f(\bm{x}) +\lambda \|\bm{x}\|_1, \label{eq:optimization_regularization_L1}
\end{align}
where $\lambda$ denotes a parameter \citep{tomioka2015}. We can solve the problem \eqref{eq:optimization_regularization_L1} using the proximal gradient method because $\|\bm{x}\|_1$ is a non-differentiable proper convex function.

%%%%%%%%%%%%%%%%%%%%%%%
\subsection{Total Variation and Total Squared Variation} \label{sec:TV_and_TSV}

Another example of sparse modeling is the Total Variation (TV) regularization defined as 
\begin{align}
    \minimize_{\bm{x}\in\mathbb{R}^n} f(\bm{x}) +\lambda \|\bm{x}\|_\mathrm{TV}, \label{eq:optimization_regularization_TV}
\end{align}
where  $\lambda$ denotes the regularization parameter. We then define matrix $N$ to represent adjacent pixels:
\begin{align}
    N_{ij} \coloneqq 
        \begin{cases}
            1 & (\mathrm{if\ } i\mathrm{\mathchar`-th\ and\ } j\mathrm{\mathchar`-th\ pixels\ are\ adjacent}) \\
            0 & (\mathrm{otherwise}).
        \end{cases}
\end{align}
The TV regularization term is written using $N$ as:
\begin{align}
    \|\bm{x}\|_\mathrm{TV} &\coloneqq \sum_{i=1}^n \sum_{j=1}^n \frac{1}{2} N_{ij} |x_i-x_j|.
\end{align}
TV regularization minimizes the difference between the values of neighboring pixels. This is expected to smooth the values of the neighboring pixels and reduce noise in the solution.

Moreover, the Total Squared Variation (TSV) \citep{kuramochi2018superresolution} is expressed as an extension of TV regularization:
\begin{gather}
    \minimize_{\bm{x}\in\mathbb{R}^n} f(\bm{x}) +\lambda \|\bm{x}\|_\mathrm{TSV}. \label{eq:optimization_regularization_TSV}\\
    \|\bm{x}\|_\mathrm{TSV} \coloneqq \sum_{i=1}^n \sum_{j=1}^n \frac{1}{2} N_{ij} (x_i-x_j)^2. \label{eq:TSV}
\end{gather}
TSV regularization allows us to infer solutions with smooth boundaries in addition to the effects of TV regularization. To consider the discretized distribution using HEALPix, the TSV regularization term can be re-expressed as: 
\begin{align}
    \|\bm{x}\|_\mathrm{TSV}&=\bm{x}^\top\tilde{N}\bm{x}, \label{eq:TSV_equiv}\\
    \tilde{N}&\coloneqq 8I-N, \label{eq:tilde_N}
\end{align}
because
\begin{align}
    \| \bm{x} \|_\mathrm{TSV} &= \sum_i \sum_j \frac{1}{2} N_{ij} ( x_i - x_j )^2 \\
    &= \sum_i \sum_j \frac{1}{2} N_{ij} x_i^2 + \sum_i \sum_j \frac{1}{2} N_{ij} x_j^2 - \sum_i \sum_j N_{ij} x_i x_j \\
    &= \sum_i \sum_j \frac{1}{2} N_{ij} x_i^2 + \sum_j \sum_i \frac{1}{2} N_{ji} x_i^2 - \sum_i \sum_j N_{ij} x_i x_j \\
    &= \sum_i \sum_j N_{ij} x_i^2 - \sum_i \sum_j N_{ij} x_i x_j \ (N_{ij}=N_{ji})\\
    &= \sum_i 8 x_i^2 - \sum_i \sum_j N_{ij} x_i x_j \ \mbox{(eight neighboring pixels per pixel)} \\
    &= \sum_i 8 x_i \sum_j \delta_{ij} x_j- \sum_i \sum_j N_{ij} x_i x_j \\
    &= \sum_i \sum_j x_i ( 8\delta_{ij} - N_{ij} ) x_j \\
    &=\bm{x}^\top(8I-N)\bm{x}.
\end{align}

%%%%%%%%%%%%%%%%%%%%%%%
\subsection{Trace Norm Regularization} \label{sec:trace_norm}

A type of sparse modeling that uses the structure of matrices is trace norm regularization. First, the singular value decomposition of matrix $X$ is written as $X=U\Sigma V^\top$, where $U\in\mathbb{R}^{m\times r}$ and $V\in\mathbb{R}^{n\times r}$ are orthogonal matrices, $\Sigma\in\mathbb{R}^{r\times r}$ is a diagonal matrix, and $r=\min\{m,n\}$.  When $\Sigma=\diag \left(\sigma_1(X),\ldots,\sigma_r(X)\right) (\sigma_1(X)>\sigma_2(X)>\cdots>\sigma_r(X))$, $\sigma_j(X)$ $(j=1,\ldots,r)$ is termed as the $j$-th singular value. We define the trace norm of the matrix $X$ as follows:
\begin{align}
    \|X\|_\mathrm{Tr} \coloneqq \sum_{j=1}^r \sigma_j(X). \label{eq:trace_norm}
\end{align}
Let $\bm{\sigma}(X)$ be a singular value vector of $X$, defined as $\left(\bm{\sigma}(X)\right)_j\coloneqq \sigma_j(X)$; then, the trace norm of $X$ is re-expressed as: 
\begin{align}
    \|X\|_\mathrm{Tr} = \|\bm{\sigma}(X)\|_1.
\end{align}
Hence, we can infer the matrix with a sparse singular value vector by solving the optimization problem with the trace norm as the regularization term.
\begin{align}
    &\minimize_{X\in\mathbb{R}^{m\times n}} f(X) +\lambda \|X\|_\mathrm{Tr}, \label{eq:optimization_regularization_Trace}
\end{align}
where $\lambda$ denotes the regularization term. The number of non-zero singular values of a matrix X is equal to the rank of X, and thus trace norm regularization allows us to infer a low-rank matrix.

\section{Spin-Orbit Unmixing with Tikhonov Regularization} \label{sec:SOU_using_L2-VRDet}

For comparison with sparce modeling, we also consider spin-orbit unmixing with Tikhonov regularization for geography $A$ and volume regularization for $X$,
\begin{gather}
    \minimize_{A,X} Q_\mathrm{Tik} \subjectto \add{A_{jk}}\ge 0, X_{kl}\ge 0, \label{eq:SOU_L2-VRDet_Q}\\
    Q_\mathrm{Tik}\coloneqq\frac{1}{2}\|D-WAX\|_\mathrm{F}^2 +\frac{\lambda_A}{2}\|A\|_\mathrm{F}^2     +\frac{\lambda_X}{2} \det(XX^\top). \label{eq:Q_Tik}
\end{gather}
By rewriting $Q_\mathrm{Tik}$ as the quadratic form of $\bm{a}_k$, the $k$-th column vector of $A$, we obtain
\begin{align}
    Q_\mathrm{Tik} &= \frac{1}{2} \bm{a}_k^\top ( \mathcal{L}_A + \mathcal{T}_A ) \bm{a}_k - \bm{l}_A^\top \bm{a}_k +(\mathrm{constant\ for\ }\bm{a}_k ), \label{eq:Q_Tik_A_quad}
\end{align}
where $\mathcal{L}_A \coloneqq \bm{x}_k^\top \bm{x}_k W^\top W$, $\mathcal{T}_A \coloneqq \lambda_A I$, $\bm{l}_A \coloneqq W^\top \Delta \bm{x}_k$, and $\Delta$ denotes a matrix defined as $\Delta_{il} \coloneqq D_{il} - \sum_j \sum_{s \neq k} W_{ij} A_{js} X_{sl}$. Furthermore, from $Q_\mathrm{Tik}$ to the quadratic form of $\bm{x}_k$, the $k$-th column vector of $X^\top$, we obtain
\begin{align}
    Q_\mathrm{Tik} &= \frac{1}{2} \bm{x}_k^\top ( \mathcal{L}_X + \mathcal{D}_X ) \bm{x}_k - \bm{l}_X^\top \bm{x}_k +(\mathrm{constant\ for\ }\bm{x}_k),\label{eq:Q_Tik_X_quad}
\end{align}
where $\mathcal{L}_X \coloneqq \| W \bm{a}_k \| _2 ^2 I$, $\mathcal{D}_X \coloneqq \lambda_X \mathrm{det} ( \breve{X}_k \breve{X}_k^\top ) ( I - \breve{X}_k^\top ( \breve{X}_k \breve{X}_k^\top )^{-1} \breve{X}_k )$, $\bm{l}_X \coloneqq  \Delta^\top W \bm{a}_k$, $I$ denotes an identity matrix, and $\breve{X}_k$ denotes a submatrix of $X$ \add{with the $k$-th row removed}. The subproblem to solve the optimization problem \eqref{eq:SOU_L2-VRDet_Q} is expressed as 
\begin{gather}
    \minimize_{\bm{a}_k}  q_{A,\mathrm{Tik}} \subjectto (\bm{a}_k)_j\ge 0\ (j=1,\ldots,N_j),\label{eq:SOU_L2-VRDet_subproblem_A}  \\
    \minimize_{\bm{x}_k}  q_{X,\mathrm{Tik}} \subjectto (\bm{x}_k)_l\ge 0\ (l=1,\ldots,N_j), \label{eq:SOU_L2-VRDet_subproblem_X} \\
    q_{A,\mathrm{Tik}}\coloneqq \frac{1}{2} \bm{a}_k^\top ( \mathcal{L}_A + \mathcal{T}_A ) \bm{a}_k - \bm{l}_A^\top \bm{a}_k,\\
    q_{X,\mathrm{Tik}}\coloneqq \frac{1}{2} \bm{x}_k^\top ( \mathcal{L}_X + \mathcal{D}_X ) \bm{x}_k - \bm{l}_X^\top \bm{x}_k.
\end{gather}

Given that \eqref{eq:SOU_L2-VRDet_subproblem_A} and \eqref{eq:SOU_L2-VRDet_subproblem_X} are optimization problems with non-negative constraints for differentiable objective functions \eqref{eq:nonneg_opt_prob}, they can be solved \add{then} using the proximal gradient method. In general, for the following optimization problem:
\begin{gather}
    \minimize_{\bm{w}} q(\bm{w}) \subjectto w_j\ge 0\ (j=1,\ldots,N_j), \label{eq:SOU_L2-VRDet_subproblems} \\
    q(\bm{w})\coloneqq\frac{1}{2} \bm{w}^\top  \mathcal{L} \bm{w} - \bm{l}^\top \bm{w},
\end{gather}
we have $\nabla q(\bm{w})=\mathcal{L} \bm{w} - \bm{l}$; thus, the update formula of the proximal gradient method for \eqref{eq:SOU_L2-VRDet_subproblems} is written as 
\begin{align}
    \bm{w}_{i+1} &= \max \{\bm{w}_i - \gamma(\mathcal{L} \bm{w}_i - \bm{l}),\bm{0}\}.
\end{align}
\citet{kawahara2020global} performed optimization by solving the subproblems \eqref{eq:SOU_L2-VRDet_subproblem_A} and \eqref{eq:SOU_L2-VRDet_subproblem_X} at each $k$. The Fast Iterative Shrinkage-Thresholding Algorithm (FISTA) \citep{beck2009fastSIAM}, which is a proximal gradient method using Nesterov's acceleration method \citep{nesterov2003introductory}, is used to solve each subproblem. Nesterov's acceleration method increases the convergence speed although the function value does not necessarily decrease monotonically. A method to solve this involves using the restart method \citep{o2015adaptive}, which restarts Nesterov's acceleration method when the function value does not decrease. Hence, the optimization algorithm for solving \eqref{eq:SOU_L2-VRDet_Q} is expressed as follows:
\begin{algorithm}[H]
    \caption{Spin-Orbit unmixing with Tikhonov regularization}
    \label{alg:SOU_with_Tikhonov}
    \begin{algorithmic}
    \STATE Initialization: non-negative matrices $A_0$ and $X_0$
    \FOR{$n$ in $(1,N_\mathrm{try})$}
    \FOR{$k$ in $(1,N_k)$}
    \STATE Update $\bm{x}_k$ using FISTA with the restart method
    \STATE Update $\bm{a}_k$ using FISTA with the restart method
    \ENDFOR
    \ENDFOR
    \end{algorithmic}
\end{algorithm}

%%%%%%%%%%%%
\section{Evaluation of Inferred Solutions} \label{sec:L1TSV_evaluate}

\begin{figure*}
\centering 
\includegraphics[width=1.0\textwidth ]{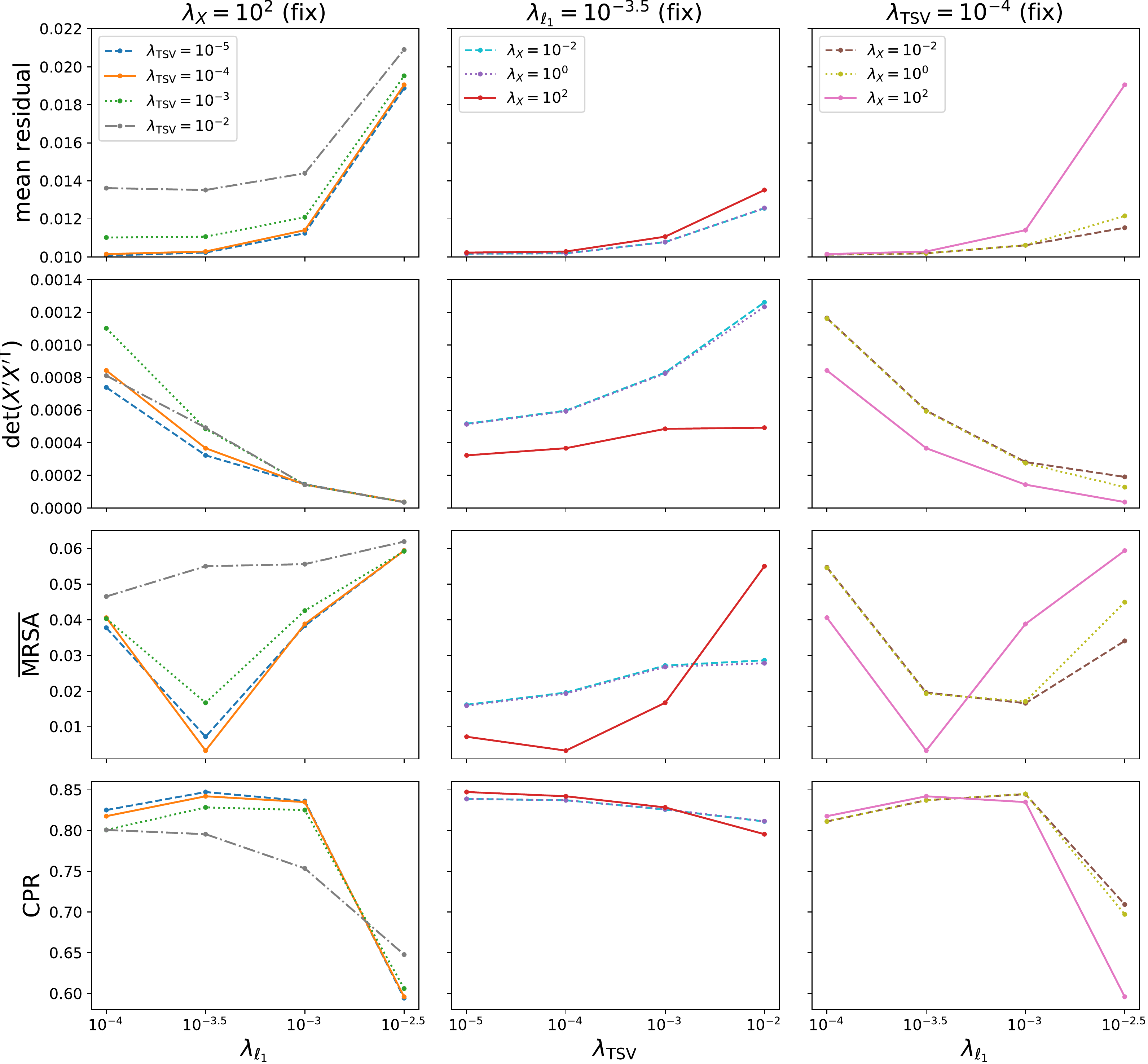}
\caption[]{Evaluation measures for selecting regularization parameters in spin-orbit unmixing with the $\ell_1$-norm and TSV regularization. The first row indicates mean residuals, second row indicates $\det(X'X'^\top)$, third row indicates $\overline{\mathrm{MRSA}}$, and fourth row indicates CPR. $\lambda_X=10^2$, $\lambda_{\ell_1}=10^{-3.5}$, and $\lambda_\mathrm{TSV}=10^{-4}$ are fixed in the first, second, and third columns, respectively, and their values are calculated by changing the other parameters.}
\label{fig:model_evaluate}
\end{figure*}

We present the evaluation measures used as criteria to select parameters $\lambda_{\ell_1}$, $\lambda_\mathrm{TSV}$, and $\lambda_X$ in the optimization problem \eqref{eq:SOU_L1TSV-VRDet_Q}. The first is the mean residual.
\begin{align}
    \mbox{mean residual}\coloneqq \frac{1}{\overline{D}}\sqrt{\frac{\|D-W\hat{A}\hat{X}\|_\mathrm{F}^2}{N_iN_l}}, \ \ \  \left(\overline{D}\coloneqq \frac{1}{N_i N_l}\sum_i \sum_l D_{il}\right)
\end{align}
which denotes the difference between the inferred model and observed data. The first row in Figure~\ref{fig:model_evaluate} shows mean residuals for each parameter. 

The second row of Figure~\ref{fig:model_evaluate} shows $\det(X'X'^\top)$, which corresponds to the volume of a simplex with each column vector of the normalized spectrum $X'$, where $X'$ is defined as:
\begin{align}
    X'_{kl}\coloneqq \frac{\hat{X}_{kl}}{\sum_{l=1}^{N_l} \hat{X}_{kl}}\ (k=1,\ldots,N_k,\ l=1,\ldots,N_l)
\end{align}
Given that the mean residuals \add{tend to} increase as $\det(X'X'^\top)$ decreases from Figure~\ref{fig:model_evaluate}, a trade-off exists between these two values in spin-orbit unmixing with the $\ell_1$-norm and TSV regularization. This relationship is also observed in spin-orbit unmixing with Tikhonov regularization \citep{kawahara2020global}.

\begin{figure*}
\centering 
\includegraphics[width=1.0\textwidth ]{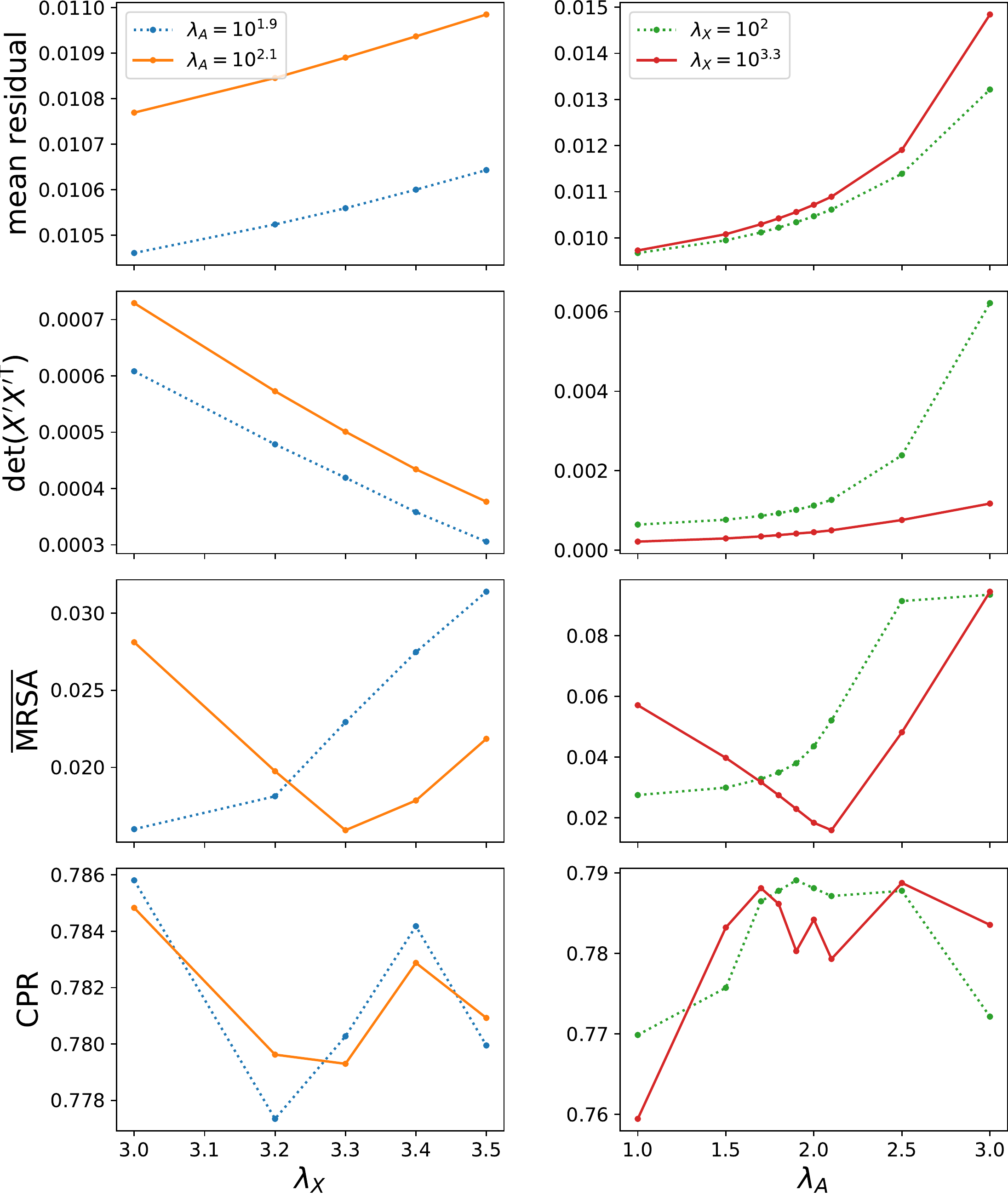}
\caption[]{Same as Figure~~\ref{fig:model_evaluate} but for with trace norm. $\lambda_A=10^{2.1}$ and $\lambda_X=10^{3.3}$ are fixed in the first and second columns, respectively, and their values are calculated by changing the other parameter. }
\label{fig:model_Trace_evaluate}
\end{figure*}

The Mean-Removed Spectral Angle (MRSA) is defined as an evaluation measure that directly compares the inferred spectra with the true values:
\begin{align}
    \MRSA (\bm{x},\bm{y})\coloneqq \frac{1}{\pi}\cos^{-1} \left( \frac{\left(\bm{x}-\overline{x}\bm{1}\right)^\top \left(\bm{y}-\overline{y}\bm{1}\right)}{\left\|\bm{x}-\overline{x}\bm{1}\right\|_2 \left\|\bm{y}-\overline{y}\bm{1}\right\|_2} \right).\label{eq:MRSA}
\end{align}
Because $\pi\MRSA (\bm{x},\bm{y})$ is the angle between $\left(\bm{x}-\overline{x}\bm{1}\right)/\left\|\bm{x}-\overline{x}\bm{1}\right\|_2$ and $\left(\bm{y}-\overline{y}\bm{1}\right)/\left\|\bm{y}-\overline{y}\bm{1}\right\|_2$, we obtain that $0\le\pi\MRSA (\bm{x},\bm{y})\le\pi$. Thus we get
\begin{align}
    0\le\MRSA (\bm{x},\bm{y})\le1.
\end{align}
Specifically, if $\bm{x}=\bm{y}$, $\MRSA (\bm{x},\bm{y})=0$. We use the average of the inferred spectra at each endmember then compared with the true spectra as the evaluation measure of the model:
\begin{align}
    \overline{\mathrm{MRSA}}\coloneqq \frac{1}{N_k}\sum_k^{N_k}\MRSA(\bm{x}_k,\bm{x}_{k,\mathrm{true}})
\end{align}
The third row of Figure~\ref{fig:model_evaluate} shows $\overline{\mathrm{MRSA}}$ for each parameter.

Furthermore, we consider the Correct Pixel Rate (CPR) as an evaluation measure that directly compares the surface distribution with the true value. The classification of the $j$-th pixel of the inferred surface distribution is
\begin{align}
    c_j(A)\coloneqq\argmax_{k\in\{1,\ldots,N_k\}} \left(\bm{a}_k\right)_j,
\end{align}
namely, $c_j(A)=k$ denotes the assignment of the $j$-th pixel of the surface distribution to the $k$-th end component. We then define CPR as  
\begin{align}
    \mathrm{CPR}\coloneqq \frac{1}{N_j} \#\{j\mid c_j(\hat{A})=c_j(A_\mathrm{true})\}.
\end{align}
The fourth row of Figure~\ref{fig:model_evaluate} shows CPR for each parameter.

We select a regularization parameter that resulted in smaller mean residuals, $\det(X'X'^\top)$ and $\overline{\mathrm{MRSA}}$, and a larger CPR. Given the large rate of change in the value of $\overline{\mathrm{MRSA}}$, as shown in Figure~\ref{fig:model_evaluate}, we measured the value of each evaluation measure by changing the parameter around the local minimum value of $\overline{\mathrm{MRSA}}$. Furthermore, $\lambda_{\ell_1}=10^{-3.5}$, $\lambda_\mathrm{TSV}=10^{-4}$, and $\lambda_X=10^2$ when $\overline{\mathrm{MRSA}}$ exhibits a local minimum. At these points, the mean residual and $\det(X'X'^\top)$ values are sufficiently small, and the CPR is sufficiently large. Thus, we adopt $\lambda_{\ell_1}=10^{-3.5}$, $\lambda_\mathrm{TSV}=10^{-4}$, and $\lambda_X=10^2$ under the assumption that the rating scale changes independently for each regularization parameter. Figure~\ref{fig:model_Trace_evaluate} is same as Figure~\ref{fig:model_evaluate} but except with trace norm. \add{We adopt $\lambda_{A}=10^{2.1}$ and $\lambda_X=10^{3.3}$.}

%%%%%%%%%%%%
\section{Theory of Convex Optimization} \label{chap:mathematics}

In this section, we review the theory of convex optimization, which is the basis of spin-orbit unmixing.

\subsection{Basis of a Convex Function}

\begin{dfn}[Convex function, {\citet[p.\ 23]{rockafellar1970convex}}]\label{dfn:convex}
Let $f: \mathbb{R}^n \rightarrow \mathbb{R}\cup\{\pm\infty\}$ be a function. $f$ is said to be convex if 
\begin{align}
     f( \alpha \bm{x} + ( 1-\alpha ) \bm{y} ) \le \alpha f(\bm{x})+(1-\alpha)f(\bm{y})
\end{align}
for any $\bm{x}, \bm{y} \in \mathbb{R}^d$ and any $\alpha \in [0, 1]$.
\end{dfn}

\begin{dfn}[Effective domain, {\citet[p.\ 23]{rockafellar1970convex}}] 
Let $f: \mathbb{R}^n \rightarrow \mathbb{R}\cup\{\pm\infty\}$ be a convex function. The effective domain of $f$ is as follows:
\begin{align}
    \dom f \coloneqq \{ \bm{x} \in \mathbb{R}^n \mid f(\bm{x}) < +\infty \}.
\end{align}
\end{dfn}

\begin{dfn}[Proper convex function, {\citet[p.\ 24]{rockafellar1970convex}}]\label{dfn:proper_convex}
Let $f: \mathbb{R}^n \rightarrow \mathbb{R}\cup\{\pm\infty\}$ be a convex function. $f$ is said to be proper if 
\begin{align}
    \begin{cases}
    \mathit{for\ any\ }\bm{x} \in\mathbb{R}^n , f(\bm{x})>-\infty \\
    \mathit{there\ exists\ } \bm{x} \in\mathbb{R}^n \ \mathit{such\ that\ } \ f(\bm{x})<+\infty,
    \end{cases}
\end{align}
that is,
\begin{align}
    \begin{cases}
    f:\mathbb{R}^n \rightarrow \mathbb{R}\cup\{+\infty\} \\
    \dom f \neq \emptyset .
    \end{cases}
\end{align}
\end{dfn}

\begin{dfn}[$\mu$-strongly convex function, {\citet[p.\ 565 Definition~12.58]{rockafellar2009variational}}]\label{dfn:strongly_convex}
Let $f: \mathbb{R}^n \rightarrow \mathbb{R}\cup\{+\infty\}$ be a proper convex function. $f$ is said to be $\mu$-strongly convex if there exists $\mu>0$ such that for any $\bm{x}, \bm{y} \in \dom f$ and any $\alpha \in [0, 1]$,
\begin{align}
    \frac{\mu}{2}\alpha (1-\alpha) \|\bm{x}-\bm{y}\|_2^2 +f( \alpha \bm{x} + ( 1-\alpha ) \bm{y} ) \le \alpha f(\bm{x})+(1-\alpha)f(\bm{y}).
\end{align}
Hence, a ($\mu$-)strongly convex function is strictly convex.
\end{dfn}

\begin{thm}[{\citet[p.\ 58 Theorem~2.42]{fukushima2001}}]\label{cor:unique_min_strong_exist}
Let $f: \mathbb{R}^n \rightarrow \mathbb{R}\cup\{+\infty\}$ be a ($\mu$-)strongly convex function. There uniquely exists a minimum value of $f$.
\end{thm}

\begin{thm}[{\citet[p.\ 565]{rockafellar2009variational}}]\label{thm:strongly_convex_equiv}
Let $f: \mathbb{R}^n \rightarrow \mathbb{R}\cup\{+\infty\}$ be a proper convex function. The following statements are equivalent.
\begin{itemize}
    \item[\rm{(i)}] $f$ is a $\mu$-strongly convex function.
    \vspace{2mm}
    \item[\rm{(ii)}] $\tilde{f} \coloneqq f(\bm{x})-{\displaystyle \frac{\mu}{2}} \|\bm{x}\|_2^2$ is a proper convex function.
    \vspace{2mm}
\end{itemize}
\end{thm}
\begin{proof}
For any $\bm{x},\bm{y}\in\dom f$ and any $ \alpha\in [0,1]$, we have
\begin{align}
    \tilde{f}(\alpha\bm{x}+(1-\alpha)\bm{y})
    &=f(\alpha\bm{x}+(1-\alpha)\bm{y})-\frac{\mu}{2} \|\alpha\bm{x}+(1-\alpha)\bm{y}\|_2^2 \\
    \begin{split}
    &=\alpha\tilde{f}(\bm{x}) +(1-\alpha)\tilde{f}(\bm{y})\\ 
    &- \left\{ \alpha f(\bm{x}) +(1-\alpha)f(\bm{y})- \frac{\mu}{2}\alpha(1-\alpha)\|\bm{x}-\bm{y}\|_2^2 -f(\alpha\bm{x}+(1-\alpha)\bm{y}) \right\}.
    \end{split}
\end{align}
\begin{itemize}
\setlength{\leftskip}{6mm}
    \item[\rm{(i)}$\Rightarrow$\rm{(ii)}] Suppose $f$ is $\mu$-strongly convex. Since 
    \begin{align}
        \alpha f(\bm{x}) +(1-\alpha)f(\bm{y})- \frac{\mu}{2}\alpha(1-\alpha)\|\bm{x}-\bm{y}\|_2^2 -f(\alpha\bm{x}+(1-\alpha)\bm{y}) \ge 0
    \end{align}
    by Definition~\ref{dfn:strongly_convex}, we obtain
    \begin{align}
        \tilde{f}(\alpha\bm{x}+(1-\alpha)\bm{y}) \le \alpha\tilde{f}(\bm{x}) +(1-\alpha)\tilde{f}(\bm{y}).
    \end{align}
    Therefore, $\tilde{f}$ is convex. Furthermore, by the definition of $\tilde{f}$, we obtain the following:
    \begin{align}
        \begin{cases}
        \tilde{f}:\mathbb{R}^n \rightarrow \mathbb{R}\cup\{+\infty\} \\
        \dom \tilde{f} \neq \emptyset .
        \end{cases}
    \end{align}
    Hence, $\tilde{f}$ is proper convex.
    
    \item[\rm{(i)}$\Leftarrow$\rm{(ii)}] Suppose $\tilde{f}$ is proper convex. Since
    \begin{align}
        \tilde{f}(\alpha\bm{x}+(1-\alpha)\bm{y}) \le \alpha\tilde{f}(\bm{x}) +(1-\alpha)\tilde{f}(\bm{y}).
    \end{align}
    by Definition~\ref{dfn:convex}, we obtain the following:
    \begin{align}
        \alpha f(\bm{x}) +(1-\alpha)f(\bm{y})- \frac{\mu}{2}\alpha(1-\alpha)\|\bm{x}-\bm{y}\|_2^2 -f(\alpha\bm{x}+(1-\alpha)\bm{y}) \ge 0,
    \end{align}
    which is equivalent to Definition~\ref{dfn:strongly_convex}. Hence, $f$ is $\mu$-strongly convex.
\end{itemize}
\end{proof}

\begin{prp}\label{prp:convex_sum}
Let $f: \mathbb{R}^n \rightarrow \mathbb{R}\cup\{+\infty\}$ be a convex function. The following statements are true.
\begin{itemize}
    \item[\rm{(i)}] If $g$ is a convex function, then $f+g$ is a convex function.
    \item[\rm{(ii)}] If $g$ is a $\mu$-strongly convex function and $\dom f\cap\dom g\neq\emptyset$, then $f+g$ is $\mu$-strongly convex.
\end{itemize}
\end{prp}
\begin{proof}
\begin{itemize}
    \item[\rm{(i)}] Suppose $g$ is convex. For any $\bm{x},\bm{y}\in\dom (f+g)=\dom f\cap\dom g$ and any $\alpha\in[0,1]$, we have
    \begin{align}
        &(f+g)( \alpha \bm{x} + ( 1-\alpha ) \bm{y} ) \\
        &=f( \alpha \bm{x} + ( 1-\alpha ) \bm{y} )+g( \alpha \bm{x} + ( 1-\alpha ) \bm{y} ) \\
        &\le \alpha f(\bm{x})+(1-\alpha)f(\bm{y}) + \alpha g(\bm{x})+(1-\alpha)g(\bm{y}) \ \ (\mathrm{convexity\ of\ } f \mathrm{\ and\ } g ) \\
        &= \alpha (f+g)(\bm{x})+(1-\alpha)(f+g)(\bm{y}).
    \end{align}
    Hence, $f+g$ is convex.
    \item[\rm{(ii)}] Suppose $g$ is $\mu$-strongly convex. The function $\tilde{g}(\bm{x})\coloneqq g(\bm{x})-(\mu /2) \|\bm{x}\|_2^2$ is proper convex according to Theorem~\ref{thm:strongly_convex_equiv}. Given that $\dom f\cap\dom g\neq\emptyset$, $f+\tilde{g}$ is proper convex based on (i) of Proposition~\ref{prp:convex_sum} and Definition~\ref{dfn:proper_convex}. Moreover, we obtain the following: 
    \begin{align}
        (f+\tilde{g})(\bm{x})&=f(\bm{x})+ g(\bm{x})-\frac{\mu}{2} \|\bm{x}\|_2^2 \\
        &=(f+g)(\bm{x})-\frac{\mu}{2} \|\bm{x}\|_2^2.
    \end{align}
    Hence, $f+g$ is $\mu$-strongly convex by Theorem~\ref{thm:strongly_convex_equiv}.
\end{itemize}
\end{proof}

\begin{dfn}[Closed convex function, {\citet[p.\ 52]{rockafellar1970convex}, \citet[p.\ 51]{fukushima2001}}]\label{dfn:closed_func}
Let $f:\mathbb{R}^n\rightarrow\mathbb{R}\cup\{\pm\infty\}$ be a convex function. Function $f$ is said to be closed if $\{\bm{x}\in\mathbb{R}^n\mid f(\bm{x})\le c\}$ is a closed set for any $c\in\mathbb{R}$.
\end{dfn}

%%%%%% differentiable convex function

\begin{dfn}[Conjugate function, {\citet[p.\ 104]{rockafellar1970convex}}]\label{dfn:conjugate_func}
Let $f: \mathbb{R}^n \rightarrow \mathbb{R}\cup\{+\infty\}$ be a proper convex function. The conjugate function $f^* :\mathbb{R}^n \rightarrow \mathbb{R}\cup\{+\infty\}$ of $f$ is defined as follows:
\begin{align}
    f^* (\bm{x})\coloneqq\sup_{\bm{y}\in\mathbb{R}^n} \left( \bm{x}^\top\bm{y}-f(\bm{y}) \right).
\end{align}
\end{dfn}

\begin{thm}[{\citet[p.\ 104]{rockafellar1970convex}}]\label{thm:biconjugate_func}
Let $f: \mathbb{R}^n \rightarrow \mathbb{R}\cup\{+\infty\}$ be a closed proper convex function. For function $f^{**}$, which is a conjugate function of $f^*$, $f^{**}=f$ is true.
\end{thm}

\begin{dfn}[Smooth function, {\citet[p.\ 14]{kanamori2016}}]\label{dfn:smooth_function}
Let $f: \mathbb{R}^n \rightarrow \mathbb{R}$ be a differentiable function. $f$ is said to be smooth if there exists $\gamma>0$ such that for any $\bm{x}, \bm{y} \in \mathbb{R}^d$,
\begin{align}
    \|\nabla f(\bm{x})-\nabla f(\bm{y})\|_2\le\gamma\|\bm{x}-\bm{y}\|_2.
\end{align}
\end{dfn}

\begin{thm}[{\citet[p.\ 566 Proposition~12.60]{rockafellar2009variational}}]\label{thm:strong_smooth}
Let $f: \mathbb{R}^n \rightarrow \mathbb{R}\cup\{+\infty\}$ be a closed proper convex function. For any $\mu>0$, the following statements are equivalent.
\begin{itemize}
    \item[\rm{(i)}] $f$ is $\mu$-strongly convex function.
    \item[\rm{(ii)}] $f^*$ is $(1/\mu)$-smooth function.
\end{itemize}
\end{thm}

\begin{thm}[{\citet[p.\ 242 Theorem~25.1]{rockafellar1970convex}}]\label{thm:grad_conv_func}
Let $f: \mathbb{R}^n \rightarrow \mathbb{R}\cup\{+\infty\}$ be a function such that $\dom f$ is an open convex set. When $f$ is differentiable in $\dom f$, the following statements are equivalent: 
\begin{itemize}
    \item[\rm{(i)}] $f$ is convex function.
    \item[\rm{(ii)}] For any $\bm{x}, \bm{y} \in \dom f $, $ f(\bm{y})-f(\bm{x})\ge \nabla f(\bm{x})^\top (\bm{y}-\bm{x}).$
\end{itemize}
\end{thm}

\begin{prp}\label{prp:grad_conjugate_func_equiv}
Let $f: \mathbb{R}^n \rightarrow \mathbb{R}\cup\{+\infty\}$ be a differentiable closed proper convex function. The following statements are equivalent:
\begin{itemize}
    \item[\rm{(i)}] $\bm{g}=\nabla f(\bm{x}).$
    \item[\rm{(ii)}] $\bm{x}=\nabla f^*(\bm{g}).$
    \item[\rm{(iii)}] $f(\bm{x})+f^*(\bm{g})=\bm{g}^\top \bm{x}.$
\end{itemize}
\end{prp}
\begin{proof}
\begin{align}
    &\bm{g}=\nabla f(\bm{x}) \nonumber \\
    &\Leftrightarrow f(\bm{y})-f(\bm{x})\ge \bm{g}^\top (\bm{y}-\bm{x})\ \mathrm{for\ any\ } \bm{y}\in\mathbb{R}^n \ (\mathrm{Theorem~\ref{thm:grad_conv_func}}) \\
    &\Leftrightarrow \bm{g}^\top \bm{x}-f(\bm{x}) \ge \max_{\bm{y}\in\mathbb{R}^n}  \left( \bm{g}^\top\bm{y}-f(\bm{y}) \right) = f^*(\bm{g}) \ (\mathrm{Definition~\ref{dfn:conjugate_func}}) 
\end{align}
Since $\bm{g}^\top \bm{x}-f(\bm{x}) \le f^*(\bm{g})$ for any $\bm{x},\bm{g}\in\mathbb{R}^n$ (Definition~\ref{dfn:conjugate_func}), we have
\begin{align}
    \bm{g}=\nabla f(\bm{x}) \Leftrightarrow \bm{g}^\top \bm{x}-f(\bm{x}) = f^*(\bm{g}).
\end{align}
Hence, (i)$\Leftrightarrow$(iii) is true. As $f=f^{**}$ by Theorem~\ref{thm:biconjugate_func}, we obtain
\begin{align}
    &f(\bm{x})+f^*(\bm{g})=\bm{g}^\top \bm{x} \nonumber \\
    &\Leftrightarrow f^*(\bm{g})+f^{**}(\bm{x})=\bm{x}^\top \bm{g} \\
    &\Leftrightarrow \bm{x}=\nabla f^*(\bm{g}). \ (\mathrm{using\ (i)}\Leftrightarrow\mathrm{(iii)\ for\ } f^*)
\end{align}
Therefore, (ii)$\Leftrightarrow$(iii) is true.
\end{proof}

\begin{cor}\label{cor:grad_conjugate_func}
Let $f: \mathbb{R}^n \rightarrow \mathbb{R}\cup\{+\infty\}$ be a differentiable closed proper convex function. Then, we obtain the following:
\begin{align}
    \nabla f^* (\bm{x})=\argmax_{\bm{y}\in\mathbb{R}^n} \left( \bm{x}^\top\bm{y}-f(\bm{y}) \right).
\end{align}
\end{cor}
\begin{proof}
Suppose $\hat{\bm{y}}=\mathrm{argmax}_{\bm{y}\in\mathbb{R}^n} \left( \bm{x}^\top\bm{y}-f(\bm{y}) \right)$. Using $f^*(\bm{x})=\bm{x}^\top\hat{\bm{y}}-f(\hat{\bm{y}})$ by the definition of a conjugate function and Proposition~\ref{prp:grad_conjugate_func_equiv}, we obtain $\hat{\bm{y}}=\nabla f^* (\bm{x})$.
\end{proof}

%%%%%%%%%%%%%%%%%%%%%%%
\subsection{Application for Optimization Problems}

\begin{dfn}[Indicator function, {\citet[p.\ 28]{rockafellar1970convex}}]\label{dfn:indicator_func}
Let $S\subset\mathbb{R}^n$ be a set. The indicator function  $\delta_S$ is defined as follows:
\begin{align}
\delta_S ( \bm{x} ) \coloneqq 
    \begin{cases}  
        0 & ( \bm{x} \in S ) \\
        \infty & ( \bm{x} \notin S ).
    \end{cases}
\end{align}
\end{dfn}

\begin{prp}[{\citet[p.\ 60 Theorem~2.43]{fukushima2001}}]\label{prp:indicator_proper}
If $S$ is a non-empty convex set, then $\delta_S$ is a proper convex function. If $S$ is also closed, then $\delta_S$ is a closed proper convex function.
\end{prp}
\begin{proof}
Given that $S$ is a convex set, we obtain the following:
\begin{align}
\alpha\bm{u}+(1-\alpha)\bm{v}\in S
\end{align}
for any $\bm{u},\bm{v}\in S$ and $\alpha\in[0,1]$. We consider an arbitrary $\bm{x},\bm{y}\in\mathbb{R}^n$. If $\bm{x},\bm{y}\in S$, we obtain $\alpha\bm{x}+(1-\alpha)\bm{y}\in S$ for any $\alpha\in[0,1]$. Thus, we obtain
\begin{align}
    0=\delta_S\left( \alpha\bm{x}+(1-\alpha)\bm{y} \right) \le \alpha\delta_S(\bm{x})+(1-\alpha)\delta_S(\bm{y})=0.
\end{align}
When at least one of $\bm{x}$ or $\bm{y}$ is not included in $S$, the function value becomes $\infty$. This also satisfies the definition of a convex function (Definition~\ref{dfn:convex}). Therefore, $\delta_S$ is a convex function. Furthermore, $\delta_S$ is a proper convex function based on the non-emptiness of $S$ and  definition of $\delta_S$. Using Definition~\ref{dfn:closed_func}, the following statement is true:
\begin{align}
    \delta_S \mathrm{\ is\ a\ closed\ function} &\Leftrightarrow \{\bm{x}\in\mathbb{R}^n\mid \delta_S(\bm{x})\le c\} \mbox{ is a closed set for any } c\in\mathbb{R} \\
    &\Leftrightarrow \{ \bm{x}\in\mathbb{R}^n\mid\bm{x}\in S\} \mbox{ is a closed set} \\
    &\Leftrightarrow S \mbox{ is a closed set}.
\end{align}
Therefore, if $S$ is a non-empty closed convex set, $\delta_S$ is a closed proper convex function.
\end{proof}

\begin{dfn}[Proximal Operator, {\citet[p.\ 284]{moreau1965}}]\label{dfn:prox_ope}
Let $\psi: \mathbb{R}^n \rightarrow \mathbb{R}\cup\{+\infty\}$ be a closed proper convex function. The proximal operator of $\psi$ is defined as:
\begin{align}
    \prox \left(\bm{x}\mid\psi\right) \coloneqq \argmin_{ \bm{w} \in \mathbb{R}^n} \left( \psi( \bm{w} ) + \frac{1}{2} \| \bm{w} - \bm{x} \|_2^2 \right).
\end{align}
\end{dfn}

\begin{prp}[{\citet[p.\ 284]{moreau1965}}]\label{prp:prox_unique}
Let $\psi: \mathbb{R}^n \rightarrow \mathbb{R}\cup\{+\infty\}$ be a proper convex function. Then, the value of $\prox \left( \bm{x} \mid \psi \right)$ is unique.
\end{prp}
\begin{proof}
Given that function $\|\cdot\|_2^2$ is a strongly convex function, function $\psi( \bm{w} ) + (1/2) \| \bm{w} - \bm{x} \|_2^2$ is also a strongly convex function by Proposition~\ref{prp:convex_sum}. Hence, the value of $\prox \left( \bm{x} \mid \psi \right)$ is unique based on Theorem~\ref{cor:unique_min_strong_exist}.
\end{proof}

%=============================
%=============================
%=============================

\bibliography{bibtex}{}
\bibliographystyle{aasjournal}

\end{document}